\documentclass[pre,singlecolumn,noshowpacs,preprintnumbers,amsmath,amssymb]{revtex4}

\usepackage{mathrsfs}
\usepackage[dvipdfmx]{graphicx}
\usepackage{dcolumn}
\usepackage{bm}
\usepackage{color}

\newcommand{\mcL}{\mathcal{L}}
\newcommand{\tl}{\tilde}
\newcommand{\ra}{\rangle}
\newcommand{\la}{\langle}
\newcommand{\tb}{\textbf}

\newcommand{\hxi}{\hat{\xi}}
\newcommand{\ha}{\hat{a}}
\newcommand{\hb}{\hat{b}}
\newcommand{\hz}{\hat{z}}

\newcommand{\hp}{\hat{p}}
\newcommand{\hDp}{\Delta \hat{p}}
\newcommand{\Dp}{\Delta p}

\newcommand{\bGamma}{\bm{\Gamma}}
\newcommand{\hGamma}{\hat{\bGamma}}
\newcommand{\htau}{\hat{\tau}}
\newcommand{\sgn}{\mathrm{sgn}}

\newcommand{\heta}{\hat{\eta}}
\newcommand{\hzeta}{\hat{\zeta}}
\newcommand{\ve}{\varepsilon}

\newcommand{\bz}{{\bar{z}}}
\newcommand{\hr}{\hat{r}}
\newcommand{\mrG}{\mathrm{G}}
\newcommand{\mrP}{\mathrm{P}}
\newcommand{\mrR}{\mathrm{R}}

\newcommand{\mrT}{\mathrm{T}}
\newcommand{\mrB}{\mathrm{B}}
\newcommand{\mrA}{\mathrm{A}}
\newcommand{\mrCUT}{\mathrm{cut}}

\newcommand{\CM}{\mathrm{CM}}
\newcommand{\MF}{\mathrm{MF}}

\newcommand{\zCM}{z_{\CM}}
\newcommand{\hzCM}{\hat{z}_{\CM}}
\newcommand{\pst}{\mathrm{pst}}
\newcommand{\tr}{\tilde{r}}
\newcommand{\erfc}{\mathrm{erfc}}

\newcommand{\EMA}{\mathrm{EMA}}

\newcommand{\bmp}{\bm{p}}
\newcommand{\bmq}{\bm{q}}
\newcommand{\bmP}{\bm{P}}
\newcommand{\bmQ}{\bm{Q}}

\newcommand{\tpartial}{\tilde{\partial}}

\begin{document}

\title{Kinetic Theory for Finance Brownian Motion from Microscopic Dynamics}

\author{Kiyoshi Kanazawa$^{1,2}$, Takumi Sueshige$^{2}$, Hideki Takayasu$^{1,3}$, and Misako Takayasu$^{1,2}$}

\affiliation{
	$^1$Institute of Innovative Research, Tokyo Institute of Technology, 4259 Nagatsuta-cho, Midori-ku, Yokohama, 226-8502, Japan\\
	$^2$Department of Mathematical and Computing Sciences, Graduate School of Information Science and Engineering, Tokyo Institute of Technology, 4259 Nagatsuta-cho, Midori-ku, Yokohama, 226-8502, Japan\\
	$^3$Sony Computer Science Laboratories, 3-14-13 Higashi-Gotanda, Shinagawa-ku, Tokyo, 141-0022, Japan
}
\date{\today}

\begin{abstract}
	Recent technological development has enabled researchers to study social phenomena scientifically in detail 
	and financial markets has particularly attracted physicists since the Brownian motion has played the key role as in physics. 
	In our previous report (arXiv:1703.06739; to appear in Phys. Rev. Lett.), 
	we have presented a microscopic model of trend-following high-frequency traders (HFTs) and its theoretical relation to the dynamics of financial Brownian motion, 
	directly supported by a data analysis of tracking trajectories of individual HFTs in a financial market. 
	Here we show the mathematical foundation for the HFT model paralleling to the traditional kinetic theory in statistical physics. 
	We first derive the time-evolution equation for the phase-space distribution for the HFT model exactly, 
	which corresponds to the Liouville equation in conventional analytical mechanics.
	By a systematic reduction of the Liouville equation for the HFT model, the Bogoliubov-Born-Green-Kirkwood-Yvon hierarchal equations are derived for financial Brownian motion. 
	We then derive the Boltzmann-like and Langevin-like equations for the order-book and the price dynamics by making the assumption of molecular chaos. 
	The qualitative behavior of the model is asymptotically studied by solving the Boltzmann-like and Langevin-like equations for the large number of HFTs,
	which is numerically validated through the Monte-Carlo simulation. 
	Our kinetic description highlights the parallel mathematical structure between the financial Brownian motion and the physical Brownian motion.  
\end{abstract}
\pacs{??}

\maketitle

\section{Introduction}
	The goal of statistical physics is to reveal macroscopic behavior of physical systems from their microscopic setups,
	and has been partially achieved in equilibrium and nonequilibrium statistical mechanics~\cite{KuboB}. 
	For example, kinetic theory has provided a mathematically rigid foundation for various non-equilibrium systems, 
	such as dilute molecular gas, Brownian motion, granular gas, active matter, traffic flows, neural networks, and social dynamics~\cite{Chapman1970,Broeck2004,Broeck2006,Brilliantov,Bertin2006,Helbing,Nishinari,Prigogine,Cai2004,Buice2013,Pareschi}. 
	The fundamental equations of kinetic theory (i.e., the Boltzmann and Langevin equations) were historically introduced on the basis of phenomenological arguments 
	within the frameworks of non-linear master equations and stochastic processes~\cite{Resibois1977,GardinerB}.  
	Furthermore, their systematic derivations were mathematically developed from analytical mechanics by Bogoliubov-Born-Green-Kirkwood-Yvon (BBGKY) 
	and van Kampen~\cite{Resibois1977,McDonald,VanKampen,Spohn1980}. 
	
	Inspired by these successes, physicists have attempted to apply statistical physic approaches even to social science beyond material science. 
	In particular, financial markets have attracted physicists as an interdisciplinary area~\cite{Mantegna1999,Slanina2014} since they exhibit quite similar phenomena to physics, represented by the Brownian motion. 
	It is noteworthy that the concept of the Brownian motion was historically first invented by Bachelier in finance~\cite{Bachelier1900} before the famous work by Einstein in physics~\cite{Einstein1905}. 
	After the work by Bachelier, various characters of Brownian motions in finance and their differences from physical Brownian motions have been found by both theoretical and data analyses. 
	On the level of price time series, the power-law behavior of price movements has been reported empirically~\cite{Mantegna1995,Lux1996,Plerou1999,Guillaume1997,Longin1996}. 
	Such universal characters have been summarized as the stylized facts~\cite{Slanina2014} and have been theoretically studied 
	by time-series models~\cite{Slanina2014,JDHamilton,Engle1982,PUCK2006} and agent-based models~\cite{Kyle1985,Takayasu1992,Bak1997,Lux1999,SatoTakayasu1998,Yamada2007,Yamada2009,Yamada2010}.
	In addition, characters of order books (i.e., current distributions of quoted prices) are studied 
	by both empirical analysis and order-book models~\cite{Slanina2014,Maslov2000,Daniels2003,Smith2003,Bouchaud2002,Farmer2005,Toth2011,Donier2015}. 
	For example, the zero-intelligence order-book models~\cite{Maslov2000,Daniels2003,Smith2003,Bouchaud2002,Farmer2005,Toth2011,Donier2015} have been investigated from various viewpoints, 
	such as power-law price movement statistics~\cite{Maslov2000}, order-book profile~\cite{Bouchaud2002}, and market impact by large meta orders~\cite{Toth2011,Donier2015}. 
	The collective motion of the full order book was further found by analyzing the layered structure of the order book~\cite{Yura2014,Kanazawa2017}, which was a key to generalize the fluctuation-dissipation relation to financial Brownian motion. 
	To date, however, the modeling of individual traders' dynamics based on direct microscopic evidence has not been fully studied, which was a crucial obstacle to apply the statistical mechanics from microscopic dynamics. 
	To fully apply statistical mechanics to financial systems, it is expected necessary to establish the microscopic dynamical model of traders based on microscopic evidence 
	and to develop a non-equilibrium statistical mechanics for such non-Hamiltonian many-body systems. 
	
	Recently, an extension of the kinetic framework for financial Brownian motion has been proposed by studying high-frequency data including traders identifiers (IDs)~\cite{Kanazawa2017}. 
	The dynamics of high-frequency traders (HFTs) were directly analyzed by tracking trajectories of the individuals, 
	and a microscopic model of trend-following HFTs have been established showing agreeing with empirical analyses of microscopic trajectories. 
	On the basis of the ``equation of motions" for the HFTs, the Boltzmann-like and Langevin-like equations are finally derived for the mesoscopic and macroscopic dynamics, respectively. 
	This framework is shown consistent with empirical findings, such as HFTs' trend-following, average order book, price movement, and layered order-book structure. 
	However, the mathematical argument therein was rather heuristic similarly to the original derivation of the conventional Boltzmann and Langevin equations. 
	Considering the traditional stream of kinetic theory, a mathematical derivation beyond heuristics is necessary for the financial Brownian motion paralleling to the works by BBGKY and van Kampen. 
	
	In this paper, we show the mathematical foundation for the financial Brownian motion in the parallel mathematics in kinetic theory. 
	For the trend-following HFT model~\cite{Kanazawa2017}, we first define the phase space and the corresponding phase-space distribution (PSD) according to analytical mechanics~\cite{McDonald,Evans2008}. 
	We then exactly derive the time-evolution equation for the PSD, which corresponds to the Liouville equation in analytical mechanics. 
	The many-body dynamics for the PSD are reduced into few-body dynamics for reduced PSD according to the reduction method by BBGKY. 
	By assuming the molecular chaos, we obtain the non-linear Boltzmann equation for the order-book profile and the master-Boltzmann equation for the market price dynamics. 
	We also present their perturbative solutions for large number of HFTs to study the dynamical behavior of this system for all hierarchies. 
	The validity of our framework is finally examined by Monte Carlo simulation. 

	This paper is organized as follows: 
	In Sec.~\ref{sec:Review_Kinetic}, we briefly review the mathematical structure of the standard kinetic theory before proceeding to our work. 
	In Sec.~\ref{sec:setup}, we describe the detail of the trend-following HFTs model as the microscopic setups. 
	In Sec.~\ref{sec:result_micro}, the microscopic dynamics of the model are exactly formulated in terms of the Liouville equation and the corresponding BBGKY hierarchal equation. 
	In Sec.~\ref{sec:result_meso}, the financial Boltzmann equation is derived as the mesoscopic description of this financial system. 
	In Sec.~\ref{sec:result_macro}, the macroscopic behavior is analyzed by deriving the financial Langevin equation. 
	In Sec.~\ref{sec:discussion}, implications of our theory are discussed for several related topics. 
	We conclude this paper in Sec.~\ref{sec:conclusion} with some remarks. 

\section{Brief Review of Conventional Kinetic Theory for Brownian Motion}\label{sec:Review_Kinetic}
	\begin{figure*}
		\centering
		\includegraphics[width = 180mm]{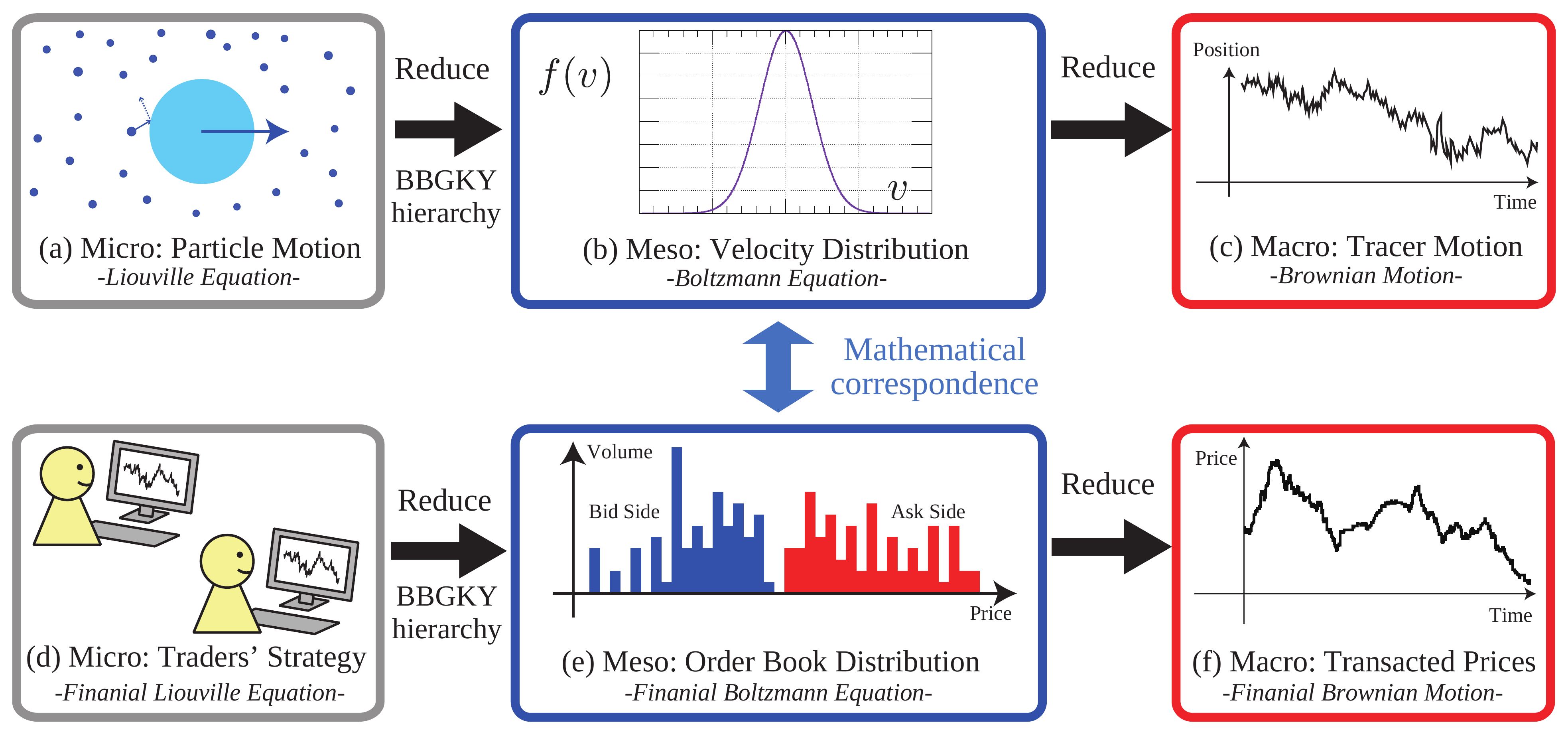}
		\caption	{
							(a--c)~Hierarchal description of the conventional Brownian motion in kinetic theory (Fig.~a).  
							Microscopic setup for the Brownian motions. Gas particles and a massive tracer interact with each other,
							where the dynamics are described by the Liouville equation~{(\ref{eq:Liouville_cnvnt})}. 
							As the mesoscopic description (Fig.~b), the full-dynamics are reduced to the one-body distribution $\phi^{(1)}$ for the gas particles, which are governed by the Boltzmann equation~{(\ref{eq:conventional_BE})}. 
							The macroscopic dynamics of the tracer (Fig.~c) are described by the master-Boltzmann equation~{(\ref{eq:conventional_BLE})}, or the Langevin equation~{(\ref{eq:Langevin_cnvnt})} asymptotically for large system size $M\to \infty$. 
							(d--f)~Hierarchal structure of financial markets parallel to molecular kinetic theory. 
							In the microscopic hierarchy (Fig.~d), each traders make decisions to submit or cancel orders. 
							The dynamics of the traders correspond to those of molecules in kinetic theory. 
							In the mesoscopic hierarchy (Fig.~e), the information on traders identifiers is lost by coarse-graining. 
							We thus obtain the dynamics of the order book (i.e., the quoted price distribution). 
							The order-book profile corresponds to the velocity distribution in the conventional kinetic theory. 
							In the macroscopic hierarchy (Fig.~f), the dynamics of the market price movement is finally deduced by the coarse-graining, 
							which exhibits the anomalous random walks. 
							The market price dynamics corresponds to those of the Brownian motion in kinetic theory. 
						}
		\label{fig:Hierarchy_Brownian}
	\end{figure*}
	Before proceeding to the core part of our work, we here briefly review the scenario of conventional kinetic theory for Brownian motion to convey our essential idea for generalization toward financial systems. 
	Let us consider the Hamiltonian dynamics of $N$ gas particles of mass $m$ and a tracer particle of mass $M$ with the hard-core interaction in a hard-core box of volume $V$ (see Fig.~{\ref{fig:Hierarchy_Brownian}a} for a schematic). 
	The momentum and position of the $i$th gas particle are denoted by $\bmp_i\equiv (p_{i;x},p_{i;y},p_{i;z})$ and $\bmq_i\equiv (q_{i;x},q_{i;y},q_{i;z})$ for $1\leq i \leq N$,
	and those of the tracer are denoted by $\bmP=\bmp_0$ and $\bmQ=\bmq_0$. 
	The dynamics of this system are described by the equation of motions,
	\begin{equation}
			\frac{d\bmq_i}{dt}=\frac{\bmp_i}{m_i}, \>\>\> \frac{d\bmp_i}{dt} = \sum_{j\neq i}\bm{F}_{ij}\label{eq:Newtonain_cnvt}
	\end{equation}
	with interaction force $\bm{F}_{ij}$ between particles $i$ and $j$ for $0\leq i,j,\leq N$ ($m_i = M$ for $i=0$ and $m_i=m$ otherwise). 
	
	\subsection{Liouville equation}
	In analytical mechanics, the phase space is defined as $\mathcal{S}\equiv \prod_{i=0}^{N} (-\infty,\infty)^6$. 
	The state of the system can be designated as the phase point defined by $\bGamma\equiv (\bmP,\bmQ;\bmp_1,\bmq_1;\dots;\bmp_N,\bmq_N) \in \mathcal{S}$, 
	and the corresponding PSD is denoted by $P_t(\bGamma)$. The time evolution of PSD is described by the Liouville equation,
	\begin{equation}
		\frac{\partial P_t(\bGamma)}{\partial t} = \mcL P_t(\bGamma)\label{eq:Liouville_cnvnt}
	\end{equation}
	with the Liouville operator $\mcL$~\footnote{In the presence of the hard-core interaction, 
	the Liouville operator $\mcL$ is non-local and is technically called the pseudo-Liouville operator~\cite{Resibois1977,Ernst1969,Beijeren1979,KanazawaTheses}.}
	(see Refs.~\cite{Resibois1977,McDonald,Evans2008,Ernst1969,Beijeren1979,KanazawaTheses} for the details). 
	This equation is exactly equivalent to the equation of motions~{(\ref{eq:Newtonain_cnvt})} mathematically, 
	and is the fundamental equation for the microscopic description (Fig.~\ref{fig:Hierarchy_Brownian}a). 
	This equation is however not analytically solvable as it fully addresses the original many-body dynamics without any approximation.

	\subsection{BBGKY hierarchy and Boltzmann equation}\label{sec:BBGKY_cnvnt}
	To focus on the one-body dynamics of a gas particle or the tracer, let us introduce the reduced PSDs,
	\begin{align*}
		\phi^{(1)}_t(\bmp_1,\bmq_1) \equiv \int P_t (\bGamma)\prod_{i=0,i\geq 2}d\bmp_i d\bmq_i, \>\>\>
		\phi^{(2)}_t(\bmp_1,\bmq_1,\bmp_2,\bmq_2) \equiv \int P_t (\bGamma)\prod_{i=0,i\geq 3}d\bmp_i d\bmq_i,\>\>\> \\
		P^{(\mathrm{T})}_t(\bmP,\bmQ) \equiv \int P_t (\bGamma)\prod_{i\geq 1}d\bmp_i d\bmq_i, \>\>\>
		P^{(\mathrm{TG})}_t(\bmP,\bmQ,\bmp_1,\bmq_1) \equiv \int P_t (\bGamma)\prod_{i\geq 2}d\bmp_i d\bmq_i.
	\end{align*}
	On the assumption of binary interaction, we can exactly derive hierarchies of PSDs, such that
	\begin{align}
		\frac{\partial \phi^{(1)}_t}{\partial t} 				&= \mcL^{(1)}\phi^{(1)}_t + \mcL^{(2)}\phi^{(2)}_t + \frac{1}{N}\mcL^{(\mathrm{TG})} P^{(\mathrm{TG})}_t\label{eq:BBGKY_BE}\\
		\frac{\partial P^{(\mathrm{T})}_t}{\partial t} 	&= \mcL^{(\mathrm{T})}P^{(\mathrm{T})}_t + \mcL^{(\mathrm{TG})} P^{(\mathrm{TG})}_t\label{eq:BBGKY_BLE}
	\end{align}
	with one-body Liouville operators $\mcL^{(1)}, \mcL^{(\mathrm{T})}$ and two-body collision operators $\mcL^{(2)}, \mcL^{(\mathrm{TG})}$. 
	These equations are exact but not closed in terms of $\phi^{(1)}_t$ and $P^{(\mathrm{T})}_t$. 
	
	To obtain analytical solutions, a further approximation is necessary. 
	The standard approximation in kinetic theory is a mean-field approximation, called molecular chaos, 
	\begin{align}
		\phi^{(2)}(\bmp_1,\bmq_1,\bmp_2,\bmq_2) \approx \phi^{(1)}(\bmp_1,\bmq_1)\phi^{(1)}(\bmp_2,\bmq_2), \label{eq:MlclrChs_conventional1}
	\end{align}
	which is mathematically shown asymptotically exact for dilute gas in the thermodynamic limit $N,V \to \infty$ (called the Boltzmann-Grad limit~\cite{Cercignani1994}).
	We then obtain the closed dynamical equation for $\phi^{(1)}$ as 
	\begin{align}
		\frac{\partial \phi^{(1)}}{\partial t} 				&\approx \mcL^{(1)}\phi^{(1)} + \mcL^{(2)}\left(\phi^{(1)}\phi^{(1)}\right) \label{eq:conventional_BE}
	\end{align}
	which is the fundamental equation for the mesoscopic description (Fig.~\ref{fig:Hierarchy_Brownian}b). 
	The steady solution for $\phi^{(1)}$ of the non-linear Boltzmann equation~{(\ref{eq:conventional_BE})} is then given by the celebrated Maxwell-Boltzmann distribution. 
	
	\subsection{Langevin equation}
	The stochastic dynamics for the macroscopic variables $(\bmP,\bmQ)$ can be also obtained within kinetic theory.
	By applying molecular chaos for $P^{(\mathrm{TG})}(\bmP,\bmQ,\bmp_1,\bmq_1)$ as 
	\begin{equation}
		P^{(\mathrm{TG})}(\bmP,\bmQ,\bmp_1,\bmq_1) \approx P^{(\mathrm{T})}(\bmP,\bmQ)\phi^{(1)}(\bmp_1,\bmq_1),\label{eq:MlclrChs_conventional2}
	\end{equation}
	we obtain the master-Boltzmann equation (or the linear Boltzmann equation) 
	\begin{equation}
		\frac{\partial P^{(\mathrm{T})}}{\partial t} 	\approx \mcL^{(\mathrm{T})}P^{(\mathrm{T})} + \mcL^{(\mathrm{TG})} \left(P^{(\mathrm{T})}\phi^{(1)}\right), \label{eq:conventional_BLE}
	\end{equation}
	which belongs to the linear-master equations in the Markov process and describes the dynamics of the tracer particle. 
	Equation~{(\ref{eq:conventional_BLE})} can be further approximated as the Fokker-Planck equation within the system size expansion~\cite{VanKampen}.
	One can thus deduce the Langevin equation for the tracer as the macroscopic description of the Brownian motion (Fig.~\ref{fig:Hierarchy_Brownian}c),
	\begin{equation}
		\frac{d\bmP}{dt} \approx -\frac{\gamma}{M} \bmP +\sqrt{2\gamma T}\bm{\xi}^{\mathrm{G}}\label{eq:Langevin_cnvnt}
	\end{equation}
	with viscous coefficient $\gamma$, temperature of the gas $T$, and the white Gaussian noise $\bm{\xi}^{\mathrm{G}}$ with unit variance. 
	
	The above formulation shows the systematic connection from the microscopic Newtonian dynamics to the mesoscopic dynamics and macroscopic dynamics. 
	This methodology is shown valid even for non-equilibrium systems when the gas is sufficiently dilute 
	(see Refs.~\cite{Broeck2004,Broeck2006,Brilliantov,Bertin2006,Helbing,Nishinari,Prigogine,Pareschi} for its application to various nonequilibrium systems),
	and is one of the most successful formulations in statistical physics. 

	\subsection{Idea to generalize kinetic theory toward finance}
	Here, let us remark our idea to generalize the framework toward financial Brownian motion. 
	Financial markets have a quite similar hierarchal structure to the conventional Brownian motion (see Fig.~\ref{fig:Hierarchy_Brownian}d--f for a schematic): 
	In the microscopic hierarchy, individual traders make decisions to buy or sell currencies at a certain price (Fig.~\ref{fig:Hierarchy_Brownian}d). 
	In the mesoscopic hierarchy, the dynamics are coarse-grained into the order-book dynamics with removal of traders' IDs (Fig.~\ref{fig:Hierarchy_Brownian}e). 
	In the macroscopic hierarchy, the dynamics are reduced to the price dynamics (Fig.~\ref{fig:Hierarchy_Brownian}f).
	One can notice that these hierarchies directly correspond to those in kinetic theory; 
	traders, order book, and price correspond to molecules, velocity distribution, and Brownian particle, respectively. 
	In this sense, the financial markets have a similar hierarchal structure to that in kinetic theory. 
	From the next section, we present a parallel mathematical framework for the description of financial markets from microscopic dynamics. 

\section{Microscopic Setup}\label{sec:setup}
	In this section, the dynamics of the trend-following HFT model in Ref.~\cite{Kanazawa2017} is mathematically formulated within the many-body stochastic processes with collisions 
	on the basis of microscopic empirical evidences. 
	
	\subsection{Notation}
		\begin{figure*}
		\centering
		\includegraphics[width=180mm]{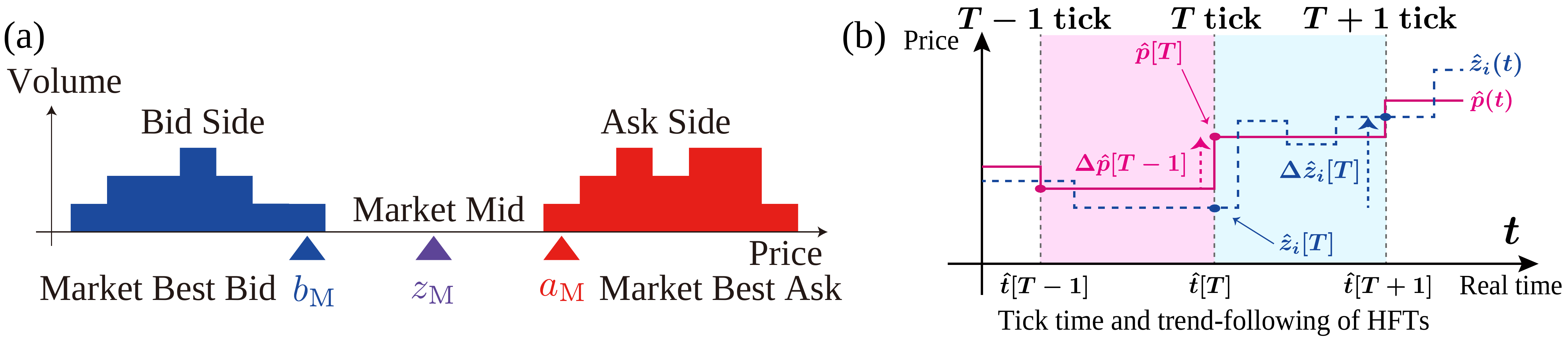}
		\caption	{
							(a)~Notation of the market best bid $\hb_{\mathrm{M}}$ and ask $\ha_{\mathrm{M}}$ prices. 
							The market mid price is also defined by $\hz_{\mathrm{M}}\equiv (\hb_{\mathrm{M}}+\ha_{\mathrm{A}})/2$.   
							(b)~Schematic of the tick time $T$, incremented every transaction. 
							For the trend-following analysis of individual traders~\cite{Kanazawa2017}, the correlation was studied between future movement of HFT's quoted mid price $\Delta \hz_i[T]$ and historical price movement $\hDp[T-1]$.
						}
		\label{fig:notation}
	\end{figure*}
	We here briefly explain the notation in this paper. 
	Any stochastic variable accompanies the hat symbol such as $\hat A$ to stress its difference to non-stochastic real numbers such as $A$. 
	For example, the probability distribution function (PDF) of a stochastic variable $\hat A(t)$ at real time $t$ is denoted by $P(A,t)\equiv P(\hat A(t)=A)$ with a non-stochastic real number $A$
	(i.e., the probability of $\hat A(t)\in [A,A+dA)$ is given by $P(A,t)dA$).  
	The complementary cumulative distribution function (CDF) is also defined as $P(\geq A,t)\equiv \int_A^\infty P(A',t)dA'$. 
	To simplify the notation, arguments in functions are sometimes abbreviated without mention if they are obvious.  
	The ensemble average of any stochastic quantity $\hat A(t)$ is denoted by $\la \hat A(t)\ra\equiv \int_{-\infty}^\infty AP(A,t)dA$. 
	
	We next explain the terminology for the order book for the whole market (Fig.~\ref{fig:notation}a). 
	The highest bid (lowest ask) quoted price among all the traders is called the market best bid (ask) price $\hb_{\mathrm{M}}$ ($\ha_{\mathrm{M}}$). 
	The average of the market best bid and ask prices is called the market mid price $\hz_{\mathrm{M}} \equiv (\hb_{\mathrm{M}}+\ha_{\mathrm{M}})/2$. 
	The difference between the market best bid and ask prices is called the market spread. 
	The market transacted price means the price at which a transaction occurs in the market. 
	In this paper, the market price (mathematically denoted by $\hp$) means the market transacted price for short. 
	
	As for a single trader, the highest bid (lowest ask) quoted price by a single trader is called the best bid (ask) price of the trader (denoted by $\hb_i$ ($\ha_i$) for the $i$th trader). 
	The average of the best bid and ask prices of the trader is called the mid price of the trader (denoted by $\hz_i$). 
	Also, the difference between the best bid and ask prices of the trader is called the {\it buy-sell spread} of the trader (denoted by $\hat{L}_i\equiv \ha_i-\hb_i$), which is different from the market spread. 

	There are two types of time in this paper. One is the real time $t$ and the other is the tick time $T$ (Fig.~\ref{fig:notation}b). 
	The tick time $T$ is defined as a discrete time incremented by every market transaction and corresponds to the real time as a stochastic variable, such as $t=\hat{t}[T]$. 
	Here the square brackets for the function argument (e.g., $\hat{A}[T]$) means that the stochastic variable $\hat{A}(t)$ is measured according to the tick time $T$ (i.e., $\hat{A}[T]\equiv \hat{A}(\hat{t}[T])$),
	highlighting the differences to that measured according to the real time $t$ (e.g., $\hat{A}(t)$ with the round brackets).

	\subsection{Characters of real HFTs}\label{sec:Char_RealHFTs}
		\begin{figure*}
			\centering
			\includegraphics[width=180mm]{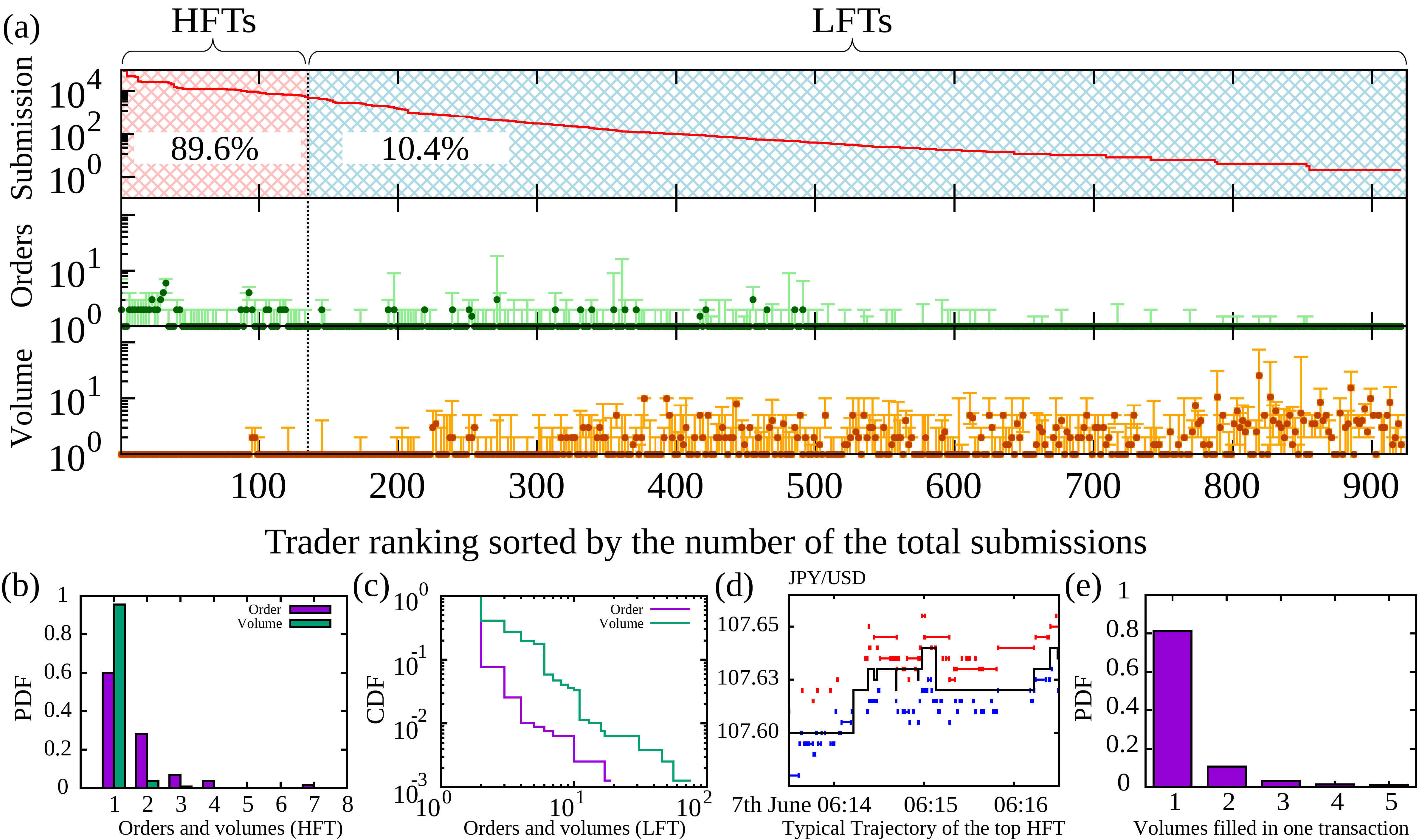}
			\caption	{
								(a)~The number of submissions, typical number of orders, typical volumes designated in one order, depending on the ranking of the trader. 
									For this figure, we studied representative numbers every two traders for anonymization. 
										(Top) We sorted the traders by their total number of submissions to define their rankings. 
										The top 135 traders were defined as HFTs, while the remaining 788 traders were defined as LFTs in this paper. 
										We plotted the average of their total submissions for every two traders. 
										(Center) We studied the number of total orders in the bid (ask) side at every bid (ask) order submission
										and take its median, first and third quartiles every two traders. 
										(Bottom) We studied volumes designated in one order at every order submission and take its median, first and third quartiles for every two traders. 
								(b)~Probability distribution function (PDF) for the numbers of orders maintained by a single HFT for one side (purple) and
									  volumes designated in a single order of HFTs (green). 
									  For this figure, we studied medians as representative numbers every single HFT.
								(c)~Complementary cumulative distribution function (CDF) for the numbers of orders maintained for one side (purple) by a single LFT and
									  volumes designated in a single order of LFTs (green). 
									  For this figure, we studied medians as representative numbers every single LFT.
								(d)~Typical trajectories of the top HFT, continuously maintaining both sides as key liquidity providers. 
								(e)~PDF for volumes filled in a single transaction. The percentage of one-to-one transaction is 81.5\% of all transactions. 
									  Transactions within 5 volumes occupy 98.2\%. 
							}
			\label{fig:BasicProp}
		\end{figure*}
		Here we describe the characters of real HFTs on the basis of high-frequency data analysis of a foreign exchange (FX) market. 
		We analyzed the order-book data including anonymized trader IDs and anonymized bank codes in Electronic Broking Services (EBS) from the 5th 18:00 to the 10th 22:00 GMT June 2016. 
		EBS is an interbank FX market and is one of the biggest financial platforms in the world. 
		The minimum volume unit for transaction was one million US dollars (USD) for the FX market between the USD and the Japanese Yen (JPY). 
		We particularly focus on HFTs, who frequently submit or cancel their orders according to algorithms.  
		As reported in our previous work~\cite{Kanazawa2017}, HFTs have several characters quite different from low frequency traders (LFTs). 
		For this paper, an HFT is defined as a trader who submitted more than 2500 time during the week, similarly to a previous research~\cite{EBS_Schmit}. 
		With this definition, the number of HFTs was 135 during this week, while the total number of traders submitting limit orders was 922~\footnote{
		In Ref.~\cite{Kanazawa2017}, 134 traders were defined as HFTs with one trader excluded whose order lifetime is extremely short.},
		and 89.6\% of all the orders in this market were submitted by the HFTs. 
		Here we summarize the reported characters with several additional evidence: 
		
		\renewcommand{\theenumi}{($\alpha$\arabic{enumi})}
		\begin{enumerate}
			\item 	\tb{Small number of live orders and volume}: 
					HFTs typically maintain a few live orders, less than ten (see Fig.~\ref{fig:BasicProp}a and b). 
					Furthermore, a single order submitted by HFTs typically implies one unit volume of the currency. 
					These characters are in contrast to those of LFTs, who sometimes submit a large amount of volumes by a single order 
					(see Fig.~\ref{fig:BasicProp}a and c for the fat-tailed distributions of the number of orders or volumes for LFTs). 
					\label{item:ordervolume}
					
			\item 	\tb{Liquidity providers}: 
					Typical HFTs plays the role of key liquidity providers (or market makers) and 
					have the obligation to maintain continuous two-way quotes during their liquidity hours according to the EBS rulebook~\cite{EBSRule} (see Fig.~\ref{fig:BasicProp}d for a typical trajectory of the top HFT). 
					The balance between the ask and bid order book is kept statistically symmetric to some extent, seemingly thanks to the liquidity providers. 
					\label{item:KLP}
					
			\item 	\tb{Frequent price modification}: 
					Typical HFTs frequently modify their quoted prices by successive submission and cancellation of orders (see Fig.~\ref{fig:BasicProp}d for a typical trajectories of the top HFT). 
					The lifetime of orders were typically within seconds for the top HFT, while the typical transaction interval was $9.3$ seconds in our dataset. 
					In addition, 94.4\% of the submissions by all the HFTs were canceled finally without transactions. 
					\label{item:pricemod}
					
			\item 	\tb{Trend-following property}: 
					HFTs tend to follow the market trends. 
					We here denote the best bid and ask quoted price of the $i$th trader and the market price at the $T$ tick time by $\hb_i(\hat{t}[T])\equiv \hb_i[T]$, $\ha_i(\hat{t}[T])\equiv \ha_i[T]$, and $\hp(\hat{t}[T])\equiv \hp[T]$, respectively (see Fig.~\ref{fig:notation}b). 
					We also denote the mid quoted price of the $i$the trader by $\hz_i[T]\equiv (\hb_i[T]+\ha_i[T])/2$. 
					According to Ref.~\cite{Kanazawa2017}, the future price movement of the $i$th HFT $\Delta \hz_i[T]\equiv \hz[T+1]-\hz[T]$ statistically obeys 
					\begin{equation}
						\la \Delta \hz_i [T] \ra_{\hDp[T-1]=\Dp} \approx c_i\tanh \frac{\Dp}{\Dp_i^*}
					\end{equation}
					conditionally on the historical price movement $\hDp[T-1]\equiv \hp[T] - \hp[T-1]=\Dp$ with characteristic constants $c_i$ and $\Dp_i^*$. 
					The constant $c_i$ characterizes the strength of trend-following of the $i$th trader, whereas $\Dp^*_i$ characterizes the saturation threshold for the trader's reaction to market trends. 
					Here the ensemble average $\la \dots \ra_{\hDp[T-1]=\Dp}$ is taken for active traders $\Delta z_i[T]\neq 0$ on the condition that the previous price movement is given by $\hDp[T-1]=\Dp$ with a non-stochastic real number $\Delta p$. 
					In the following, we introduce short hand notation for the conditional ensemble average as $\la ...\ra_{\hDp[T-1]=\Dp}=\la ...\ra_{\Dp}$.  
					In addition, the variance of the HFT's future price movement is independent of historical market trends as 
					\begin{equation}
						V_{\Delta p}\left( \Delta \hz_i[T]\right) \approx \sigma^2_i
					\end{equation}
					with variance $V_{\Delta p}\left(\Delta \hz_i[T]\right) \equiv \la \left(\Delta \hz_i[T] - \la \Delta \hz_i[T]\ra_{\Dp}\right)^2\ra_{\Dp}$ and constant $\sigma_i^2$ independent of $\Dp$. 
					\label{item:trendfollowing}
		\end{enumerate}
		\renewcommand{\theenumi}{\arabic{enumi}}
		We also note that the one-to-one transaction is the basic interaction among traders in this market. 
		The percentage of the one-to-one transaction was indeed 81.5\% in our dataset (see Fig.~\ref{fig:BasicProp}e for more detailed evidence). 
		On the basis of the above empirical results, the trend-following HFT model was proposed in Ref.~\cite{Kanazawa2017} as the corresponding minimal microscopic model as reviewed in the next section. 

	\subsection{Theoretical Model}
			On the basis of the above HFT's characters, let us consider the microscopic model of HFTs according to Ref.~\cite{Kanazawa2017}, within the framework of many-body stochastic systems with collisions. 
			\begin{figure*}
				\centering
				\includegraphics[width=170mm]{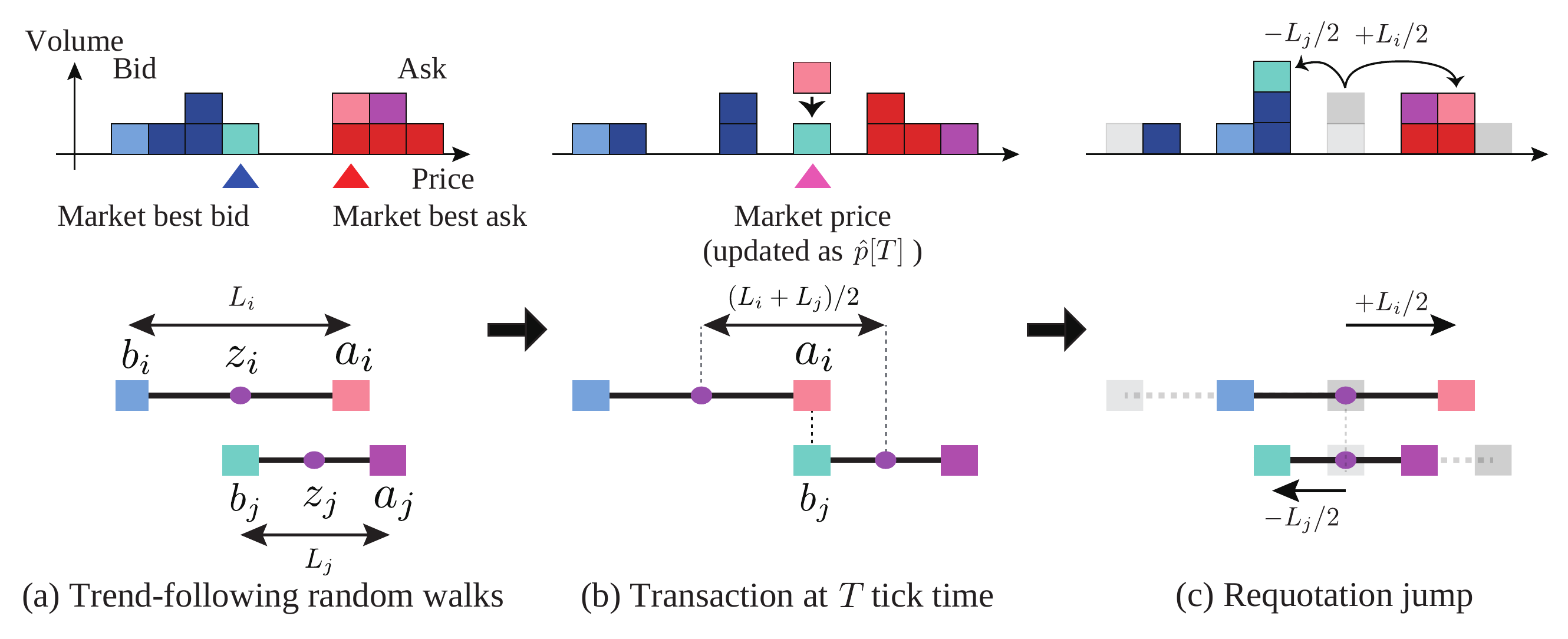}
				\caption	{
									Schematics of the dynamics of the trend-following HFT model. 
									(a) Traders maintain two-sided quotes with constant buy-sell spreads $L_i$ and $L_j$ for traders $i$ and $j$. 
									Their mid-prices moves according to deterministic trend-following and random movement. 
									(b) When bid and ask prices coincide, transaction occurs at that price. 
									(c) Both traders requote their bid and ask prices at a distance from the market price after transaction. 
								}
				\label{fig:ModelDynamics}
			\end{figure*}
			
		\subsubsection{State variables}
			Let us consider a market composed of $N$ HFTs quoting both bid and ask prices $\{\hb_i\}_i$ and $\{\ha_i\}_i$ at all the time with the unit volume, 
			where the index $i$ identifies the individual trader ($1\leq i\leq N$). 
			This assumption is consistent with the empirical HFT's characters~{\ref{item:ordervolume}} and~{\ref{item:KLP}}. 
			For simplicity, the difference between the best bid and ask prices of a single trader (called {\it buy-sell spread} $L_i$) is assumed to be time-constant unique to the trader (see Fig.~\ref{fig:ModelDynamics}a): 
			\begin{equation}
				L_i\equiv \hb_i-\ha_i=\mathrm{const}.
			\end{equation}
			On this assumption, the dynamics of individual HFTs are uniquely characterized by the mid price of HFTs as $\hz_i \equiv (\hb_i+\ha_i)/2$. 
			The maximum and minimum values of the buy-sell spread among traders are denoted by $L_{\max}$ and $L_{\min}$, respectively. 
			According to Ref.~\cite{Kanazawa2017}, the buy-sell distribution $\rho_L$ is directly measured to obey 
			the $\gamma$-distribution, such that 
			\begin{equation}
				\rho_L \equiv \frac{1}{N}\sum_{i=1}^N \delta (L-L_i) \approx \frac{L^\alpha }{\alpha !L^{*(\alpha+1)}}e^{-L/L^*}
			\end{equation}
			with decay length $L^*$ and empirical exponent $\alpha\approx 3$. 
		
		\subsubsection{Trend-following random walks}
			HFTs have a tendency to maintain continuous two-sided quotes by frequently modifying their prices (i.e., successive cancellation and submission of limit orders), as required by the market rule~\cite{EBSRule}.
			This implies that the mid-price trajectory of an HFT can be modeled as a continuous random trajectory (i.e., the characters~{\ref{item:KLP}} and~{\ref{item:pricemod}}). 
			Remarkably, there is a mathematical theorem guaranteeing that the It\^o processes (i.e., SDEs driven by the white Gaussian noise) are the only Markov processes with continuous sample trajectory~\cite{GardinerB}. 
			As a minimal model satisfying all the characters of real HFTs~{\ref{item:ordervolume}--\ref{item:trendfollowing}}, the dynamics of the HFTs are modeled within the It\^o processes as 
			\begin{equation}
				\frac{d\hz_i}{dt} = c\tanh \frac{\Delta \hat p}{\Delta p^*} + \sigma \heta_i^{\mrR}\label{eq:TF_RW_wo_Trnsct}
			\end{equation}
			in the absence of transactions (Fig.~\ref{fig:ModelDynamics}a) by taking into account the empirical trend-following properties~{\ref{item:trendfollowing}}. 
			Here $c$ and $\Dp^*$ are constants characterizing the strength and threshold of trend-following effect 
			and $\heta_i^{\mrR}$ is the white Gaussian noise with unit variance. 
			The presence of the trend-following effect in Eq.~{(\ref{eq:TF_RW_wo_Trnsct})} is the character of our HFT model,
			which induces the collective motion of limit orders~\cite{Kanazawa2017}. 
			The trend-following effect triggers translational motion of the full order book, which was crucial to reproduce the layered structure of the order book reported in Ref.~\cite{Yura2014}. 
		
		\subsubsection{Transaction rule}
			When the best bid and ask prices coincide, there occurs an transaction (see Fig.~\ref{fig:ModelDynamics}b). 
			The transaction condition (i.e., the condition of price matching) is mathematically given by 
			\begin{equation}
				\hb_j = \ha_i \label{eq:transaction_occurs}
			\end{equation}
			for $i\neq j$. In the following, we assume that the index $i$ is an integer always different from another integer $j$. 
			At the instance of transaction $\hb_i = \ha_j$, let us assume that the traders requote their prices simultaneously (see Fig.~\ref{fig:ModelDynamics}c) such that 
			\begin{equation}
				\hb^{\pst}_j = \hb_j - \frac{L_j}{2}, \>\>\>
				\ha^{\pst}_i = \ha_i + \frac{L_i}{2}, 
				\label{eq:requotation}
			\end{equation}
			where $\hb^{\pst}_i$ and $\ha^{\pst}_i$ are post-transactional bid and ask prices after transaction for between traders $i$ and $j$, respectively. 
			By introducing the mid-price of the individual traders as $\hz_i\equiv (\hb_i+\ha_i)/2$, the transaction rule is rewritten as 
			\begin{equation}
				\hz_i-\hz_j=\frac{L_i+L_j}{2} \Longrightarrow \hz^{\pst}_i = \hz_i - \frac{L_i}{2}, \>\>\> \hz^{\pst}_j = \hz_j + \frac{L_j}{2}. \label{eq:execution_rule}
			\end{equation}

			We here define the market price $\hp(t)$ and the previous price movement $\hDp(t)$ at time $t$.
			$\hp(t)$ is the market price at the previous transaction; 
			$\hDp(t)$ is the price movement by the previous transaction. 
			They are updated after transactions under the following post-transaction rule (Fig.~\ref{fig:ModelDynamics}b and c): 
			\begin{equation}
				|\hz_i-\hz_j|=\frac{L_i+L_j}{2}
				\>\>\>\Longrightarrow\>\>\>
				\hp^{\pst} = \hz_i-\frac{L_i}{2}\sgn(\hz_i-\hz_j),\>\>\>
				\hDp^{\pst} = \hz_i-\frac{L_i}{2}\sgn(\hz_i-\hz_j) - \hp\label{eq:execution_rule_price}
			\end{equation}
			with signature function $\sgn(x)$ defined by $\sgn(x) = x/|x|$ for $x\neq 0$ and $\sgn(0)=0$.

	\subsection{Complete model dynamics}
		\begin{figure*}
			\centering
			\includegraphics[width=175mm]{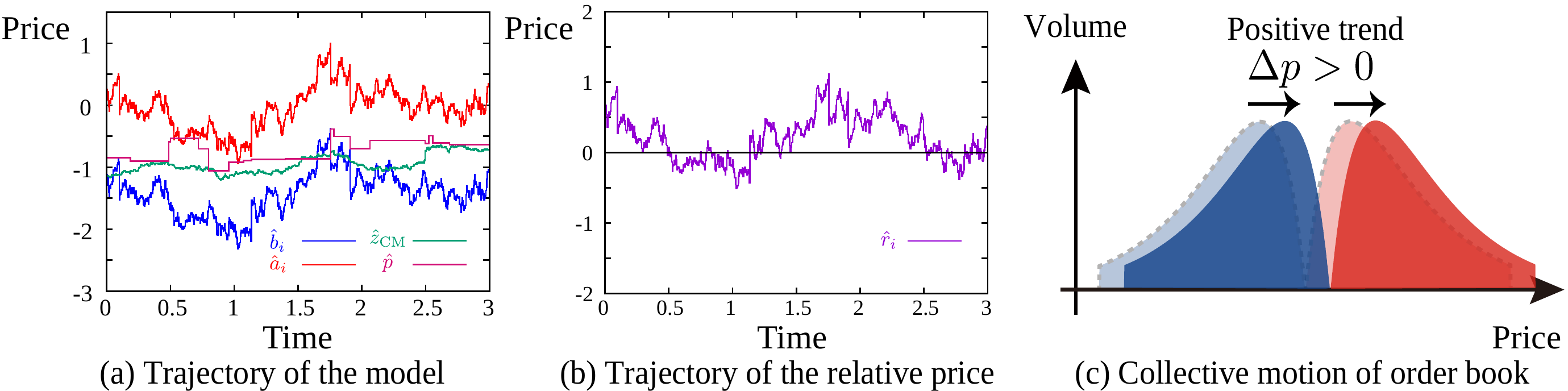}
			\caption	{
								(a)~Sample trajectory of the HFT model, showing the bid $\hb_i$ and ask $\ha_i$ quoted prices of $i$th HFT, the CM $\hzCM$, and the market transaction price $\hp$. 
								(b)~Sample trajectory of the relative price $\hr_i$ from the CM $\hzCM$, showing that $\hr_i$ stationarily fluctuates around zero. 
								(c)~Collective motion of the order book, showing herding behavior of traders. This collective motion is minimally implemented as trend-following in this HFT model. 
							}
			\label{fig:modeltrajectory}
		\end{figure*}
		We here specify the complete dynamics of the quoted prices $\{\hz_i(t)\}_i$ within the framework of stochastic processes with collisions. 
		When the previous price movement is $\hDp$, we assume that traders' quoted prices are described by the trend-following random walks:  
		\begin{equation}
			\frac{d\hz_i}{dt} = c\tanh \frac{\Delta \hat p}{\Delta p^*} + \sigma \heta_i^{\mrR} + \heta_i^{\mrT}, \>\>\>\>\>
			\heta_i^{\mrT}\equiv \sum_{k=1}^\infty \sum_{j}\Delta z_{ij}\delta(t-\htau_{k;ij}), \label{eq:Dealer1}
		\end{equation}
		where $\heta_i^{\mrT}$ is requotation jump term and $\htau_{k;ij}$ is the $k$th transaction time between traders $i$ and $j$ satisfying 
		\begin{equation}
			|\hz_i(\htau_{k;ij})-\hz_j(\htau_{k;ij})| = \frac{L_i+L_j}{2}, \>\>\> \Delta z_{ij}\equiv -\frac{L_i}{2}\sgn (\hz_i-\hz_j).\label{eq:collisions_condition}
		\end{equation}
		The requotation jump $\heta_i^{\mrT}$ corresponds to collisions in molecular kinetic theory. 
		The price-matching condition~{(\ref{eq:transaction_occurs})} and the requotation rule~{(\ref{eq:requotation})} correspond to 
		the contact condition and the momentum exchange rule in standard kinetic theory for hard-sphere gases, respectively. 
		The summary of the model parameters is presented in the Table~\ref{table:parameters} with their dimensions.
		A sample trajectory of this model is depicted in Fig.~\ref{fig:modeltrajectory}a. 
		We note that this model is a generalization of the previous theoretical model in Refs.~\cite{Takayasu1992,SatoTakayasu1998,Yamada2007,Yamada2009,Yamada2010} on the basis of the above empirical facts~{\ref{item:ordervolume}--\ref{item:trendfollowing}} on HFTs. 
		\begin{table}[b]
			\centering
			\begin{tabular}{|c||c|c|}\hline 
				Parameter & Meaning & Dimension \\ \hline \hline
				$N$ & Number of traders & dimensionless \\ \hline
				$\{L_i\}_{1\leq i\leq N}$ & Buy-sell spreads of traders & price \\ \hline
				$c$ & Strength of trend-following & price/time \\ \hline 
				$\Delta p^*$ & Saturation for trend-following & price \\ \hline
				$\sigma^2$ & Variance of random noise & price$^2$/time \\ \hline
			\end{tabular}
			\caption	{
								Summary of the model parameters and their dimensions. 
							}
			\label{table:parameters}
		\end{table}

		The dynamics of the price $\hp$ and the previous price movement $\hDp$ can be specified within the framework of stochastic processes. 
		Since $\hp$ and $\hDp$ are updated at the instance of transactions, 
		their dynamics synchronizes with collision time $\htau_{k;ij}$. 
		Considering the transaction rule for prices~{(\ref{eq:execution_rule_price})}, their concrete dynamical equations are thus given by
		\begin{align}
			\frac{d\hp}{dt}		= \sum_{k=1}^\infty\sum_{i,j}^{i<j} \left(\hp_{ij}^{\pst} - \hp\right)\delta(t-\htau_{k;ij}), \>\>\>\>\>
			\frac{d\hDp}{dt}	= \sum_{k=1}^\infty\sum_{i,j}^{i<j} \left(\hDp_{ij}^{\pst} - \hDp\right)\delta(t-\htau_{k;ij})\label{eq:eq_motion_price}
		\end{align}
		with the price after collision $\hp^{\pst}_{ij}\equiv \hz_i-(L_i/2)\sgn(\hz_i-\hz_j)$ and the price movement after collision $\hDp^{\pst}_{ij}\equiv \hp^{\pst}_{ij} - \hp$. 
		In this paper, the It\^o convention is used for the multiplication to $\delta$-functions. 
		
	\subsection{Slow variable}
		Introduction of slow variables is the key for reduction of the complex dynamics in general (e.g., the center of mass (CM) of the Brownian particle~\cite{VanKampen} and the slaving principles in synergetics~\cite{HakenB}). 
		Here we introduce the CM of the quoted prices as the slow variable of this system (Fig.~\ref{fig:modeltrajectory}a). 
		The definition of the CM and its dynamics are given by
		\begin{equation}
			\hzCM \equiv \frac{1}{N}\sum_{i=1}^N\hz_i,\>\>\> \frac{d\hzCM}{dt} = c\tanh \frac{\Delta \hat p}{\Delta p^*} + \bar{\eta} \label{eq:eq_motion_relative}
		\end{equation}
		with $\bar{\eta} \equiv (\sigma/N)\sum_{i=1}^N\heta^{\mrR}_i + (1/N)\sum_{i=1}^N\heta^{\mrT}_i$. 
		The CM $\hzCM$ characterizes the macroscopic dynamics of this system. 
		As will be shown in Sec.~\ref{sec:WeakTrendCase}, indeed, the diffusion coefficient of the CM turns out to be proportional to $N^{-1}$ for the weak trend-following case, 
		implying that the selection of $\hzCM$ is reasonable as a slow variable. 
		
		Another motivation to introduce the CM is to define the relative price from the CM such that
		\begin{equation}
			\hr_i \equiv \hz_i-\hzCM,
		\end{equation}
		since the relative price $\hr_i$ has better mathematical characters than $\hz_i$. 
		For example, the relative price $\hr_i$ fluctuates around zero (see Fig.~\ref{fig:modeltrajectory}b for the dynamics in the comoving frame of CM) and has the stationary distribution,
		while the original variable $\hz_i$ diffuses to infinity for a long time and has no stationary distribution. 

	\subsection{Difference to other order-book models}
		One of the unique characters of the HFT model is the collective motion of order book due to trend-following. 
		As shown in Ref.~\cite{Yura2014}, the order book has the layered structure in the sense that the difference in volumes of bid (ask) order book near best price has positive (negative) correlation with price movements. 
		This implies that the order book exhibits the translational motion like inertia in physics (Fig.~\ref{fig:modeltrajectory}c), and thus movements of HFTs are not independent of each other like herding behavior. 
		This collective motion has not been implemented in conventional order-book models, which are based on independent Poisson processes for order submission and cancellation, 
		and is minimally implemented in our HFT model as trend-following for the consistency with the layered order-book structure~\cite{Kanazawa2017}. 
		
\section{Main Result 1: Microscopic Description}\label{sec:result_micro}
		As the main results of this paper, the analytical solutions to the trend-following HFT model are presented by developing the mathematical technique of kinetic theory. 
		We first introduce the phase space for the HFT model in the standard manner of analytical mechanics, and derive the dynamical equation for the PSD, which we call the financial Liouville equation. 
		We next derive the hierarchy for the reduced distributions similarly to the BBGKY hierarchy in molecular kinetic theory,
		which is the theoretical key to understand the financial system systematically as shown in Secs.~\ref{sec:result_meso} and~\ref{sec:result_macro}.
		
	\subsection{Phase space and phase-space distribution}
		Here first we introduce the phase space for the HFT model according to the standard manner of analytical mechanics. 
		Let us introduce a vector $\hGamma \equiv (\hat z_1,\dots \hat z_N;\hzCM, \hp,\hDp)$,
		which corresponds to a phase point in the phase space $\mathcal{S} \equiv \prod_{i=1}^{N+3} (-\infty,\infty)$ as $\hGamma \in \mathcal{S}$. 
		Equations~{(\ref{eq:Dealer1})}, ~{(\ref{eq:eq_motion_price})}, and~{(\ref{eq:eq_motion_relative})} are the complete set of dynamical equations for the phase point, 
		corresponding to the Newtonian equations of motions in conventional mechanics.	
		Also, let us define the PSD function $P_t(\bGamma)$.
		Using the PSD, the probability is given by $P_t(\bGamma)d\bGamma$ where the phase point $\bGamma$ exists at the time $t$ 
		in the volume element $d\bGamma \equiv \prod_{i=1}^N [z_i,z_i+dz_i)\times [z_{\CM},z_{\CM}+dz_{\CM})\times [p,p+dp)\times [\Dp,\Dp+d\Dp)$. 
	
	\subsection{Financial Liouville Equation}
		\begin{figure*}
			\centering
			\includegraphics[width=150mm]{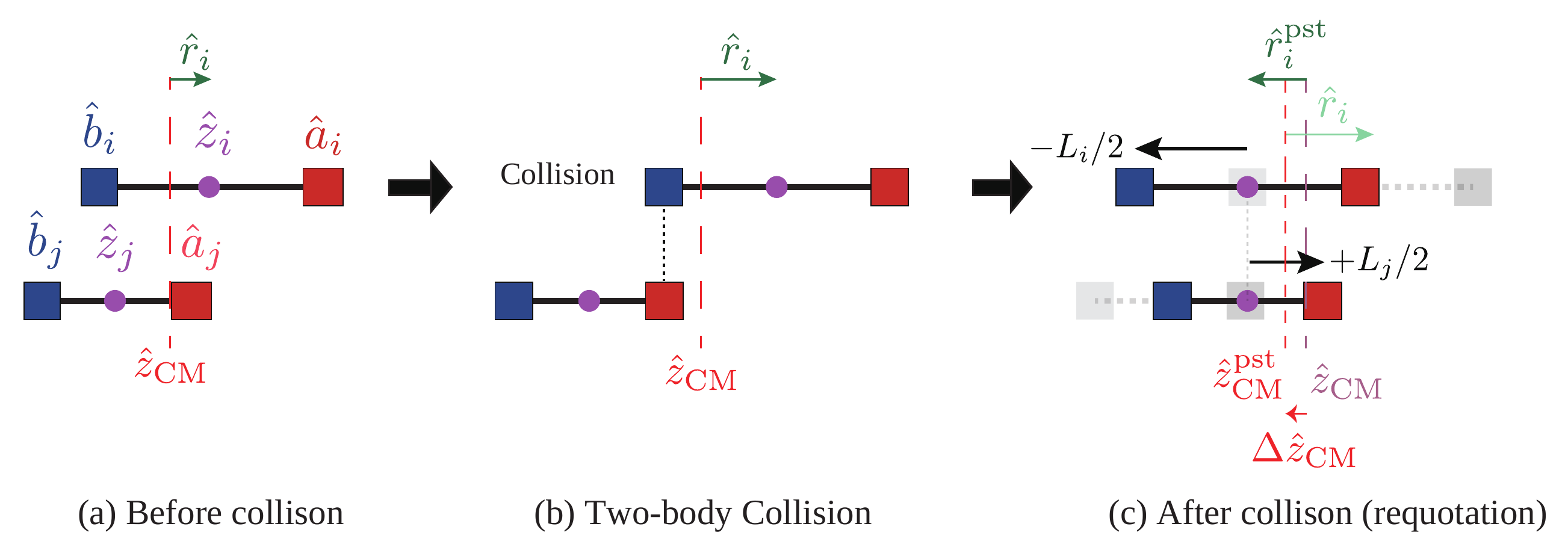}
			\caption		{
										Schematic of the two-body collision. 
										When the prices match between the traders $i$ and $j$, they requote their prices far from the market price. 
										Note that the CM also moves through a distance of $\Delta \hz_{\CM}=-(L_i-L_j)/2N$ during this requotation. 
								}
		\end{figure*}
		As the first main result in this paper, we present the Liouville equation for the trend-following trader model~{(\ref{eq:Dealer1})}--(\ref{eq:eq_motion_relative}) 
		as the dynamical equation for the PSD. 
		The dynamical equation for the PSD is given by 
		\begin{equation}
			\frac{\partial P_t(\bGamma)}{\partial t} = \mcL^{\mathrm{a}}P_t(\bGamma) + \mcL^{\mathrm{c}}P_t(\bGamma),\label{eq:pseudo_Liouville}
		\end{equation}
		where the advective and diffusive Liovuille operator $\mcL^{\mathrm{a}}$ and the binary collision Liouville operator $\mcL^{\mathrm{c}}$ are defined by
		\begin{subequations}
		\begin{align}
			\mcL^{\mathrm{a}}P_t &\equiv \sum_{i=1}^N\left[-c\tanh \frac{\Delta p}{\Delta p^*} \left\{\partial_i + \frac{1}{N}\partial_{\CM}\right\}
				+\frac{\sigma^{2}}{2}\left\{\partial_i +\frac{1}{N}\partial_{\CM}\right\}^2\right] P_t(\bGamma),\\
			\mcL^{\mathrm{c}}P_t &\equiv \sum_{i,j}\frac{\sigma^{2}}{2}\left\{\delta(z_i-z_j)\delta(p-z_i)\int d\Dp'|\tpartial_{ij}|P_t(\bGamma-\Delta \bGamma'_{ij}) - \delta\left(z_i-z_j-\frac{L_i+L_j}{2}\right)|\tpartial_{ij}|P_t(\bGamma)\right\}.
		\end{align}
		\end{subequations}
		Here we have introduced the symmetric absolute derivative $|\tpartial_{ij}|f\equiv |\partial_{i}f|+|\partial_{j}f|$ for an arbitrary function $f(z_i,z_j)$ 
		and abbreviated derivatives $\partial_{i}\equiv \partial/\partial z_i$ and $\partial_{\CM}\equiv \partial/\partial z_{\CM}$ (see Appendix.~\ref{sec_app:Liouville_der} for the detailed derivation). 
		We have also introduced a difference vector:
		\begin{equation}
			\Delta \bGamma'_{ij} \equiv \left(0,\dots,-\frac{L_i}{2},\dots,+\frac{L_j}{2},\dots,0;\Delta z_{\CM},\Dp,\Delta p-\Delta p'\right)
		\end{equation}
		with movement of the CM $\Delta z_{\CM}\equiv -(L_i-L_j)/2N$. 
		This is the first main result in this paper. 
		The advective and diffusive Liovuille operator $\mcL^{\mathrm{a}}$ describes the continuous dynamics of the system in the absence of transactions, 
		while the binary collision Liouville operator $\mcL^{\mathrm{c}}$ describes the discontinuous dynamics in the presence of transactions. 
		Equation~{(\ref{eq:pseudo_Liouville})} formally corresponds to the Liouville equation~{(\ref{eq:Liouville_cnvnt})} in molecular kinetic theory, 
		and is called the financial Liouville equation in this paper. 
		The financial Liouville equation completely characterizes the microscopic dynamics of all traders (Fig.~\ref{fig:Hierarchy_Brownian}d). 
		
	\subsection{Financial BBGKY Hierarchy}
		The financial Liouville equation~{(\ref{eq:pseudo_Liouville})} is exact but cannot be solved analytically. 
		We therefore reduce Eq.~{(\ref{eq:pseudo_Liouville})} toward a simplified dynamical equation for a one-body distribution in the parallel method to molecular kinetic theory. 
		According to the standard method in the kinetic theory, the Boltzmann equation, a closed dynamical equation for the one-body distribution, 
		can be derived by systematically reducing the Liouville equation in the parallel method to BBGKY (see Sec.~\ref{sec:BBGKY_cnvnt}). 
		We here present the lowest-order equation of reduced distributions for the trend-following HFT model in the parallel calculation in kinetic theory. 
		We first introduce the relative price from the CM as $r_i\equiv z_i-z_{\CM}$. 
		We also define the one-body, two-body and three-body reduced distribution functions for the relative price: 
		\begin{subequations}\label{eq:def:123bodydists}
		\begin{align}
			&P^i_t(r_i) \equiv \int P_t(\bGamma )dz_{\CM}dpd\Dp\prod_{l\neq i}^N dr_l, \>\>\>
			P^{ij}_t(r_i,r_j) \equiv \int P_t(\bGamma)dz_{\CM}dpd\Dp\prod_{l\neq i,j}^N dr_l, \\
			&P^{ijk}_t(r_i,r_j,r_k) \equiv \int P_t(\bGamma)dz_{\CM}dpd\Dp\prod_{l\neq i,j,k}^N dr_l.
		\end{align}
		\end{subequations}
		We then obtain the lowest-order hierarchal equation for the one-body distribution as 
		\begin{align}
		\frac{\partial P_t^i(r_i)}{\partial t} &= \mcL^{(i)}P_t^{i}(r_i)+ \sum_{j\neq j}\mcL^{(ij)}P^{ij}_t(r_i,r_j) + \sum_{j,k\neq i}\mcL^{(ijk)}P_t^{ijk}(r_i,r_j,r_k).\label{eq:BBGKY_main}
		\end{align}
		with one-body, two-body, and three-body Liouville operators $\mcL^{(i)}$, $\mcL^{(ij)}$, $\mcL^{(ijk)}$ defined by
		\begin{subequations}
		\begin{align}
			\mcL^{(i)} P_t^{i} &\equiv \frac{\tilde{\sigma}^{2}}{2}\frac{\partial^2 P_t^{i}(r_i)}{\partial r_i^2}\\
			\mcL^{(ij)} P_t^{ij} &\equiv \sum_{s=\pm 1}\frac{\sigma^{2}}{2}\left[|\tpartial_{ij}|P^{ij}_t(r_i-\Delta r_{ij;s},r_j+\Delta r_{ji;s})\big|_{r_i=r_j} - |\tpartial_{ij}|P_t^{ij}(r_i,r_j)\big|_{r_i-r_j=s(L_i+L_j)/2}\right]\\
			\mcL^{(ijk)} P_t^{ij} &\equiv \sum_{s=\pm 1} \sum_{j,k\neq i}\frac{\sigma^2}{2}\int dr_j  \left[|\tpartial_{jk}|P^{ijk}_t\left(r_i -\Delta r_{jk;s}^{(1)},r_j,r_k\right) - |\tpartial_{jk}|P_t^{ijk}(r_i,r_j,r_k)\right]\bigg|_{r_j-r_k=s(L_j+L_k)/2},\label{eq:op_BBGKY_3body}
		\end{align}
		\end{subequations}
		effective variance $\tilde{\sigma}^{2} \equiv \sigma^{2}(1-1/N)$, and jump size $\Delta r_{ij;s}\equiv \Delta r_{ij;s}^{(0)}+\Delta r_{ij;s}^{(1)}$, 
		\begin{equation}
			\Delta r_{ij;s}^{(0)} \equiv -\frac{sL_i}{2}, \>\>\>
			\Delta r_{ij;s}^{(1)} \equiv \frac{s(L_i-L_j)}{2N}.
		\end{equation}
		Here $\Delta r_{ij;s}^{(1)}$ indirectly originates from the movement of the CM during requotation. 
		The detailed derivation of Eq.~{(\ref{eq:BBGKY_main})} is described in Appendix.~\ref{sec_app:BBGKY_der}.
		Equation~{(\ref{eq:BBGKY_main})} formally corresponds to the conventional BBGKY hierarchal equation~{(\ref{eq:BBGKY_BE})} for the mesoscopic description. 
		On the basis of Eq.~{(\ref{eq:BBGKY_main})}, the Boltzmann-type closed equation for the one-body distribution is derived in the next section. 
		
		We also derive the hierarchal equation for the macroscopic dynamics. 
		For the macroscopic variables $\bm{Z}\equiv (z_{\CM},p,\Delta p)$, we here define the reduced distributions:
		\begin{equation}
			P_t(\bm{Z})\equiv P_t(z_{\CM},p,\Delta p) \equiv \int P_t(\bGamma) \prod_{k=1}dz_k, \>\>\>
			P_t^{ij}(z_i,z_j;\bm{Z})\equiv P_t^{ij}(z_i,z_j;z_{\CM},p,\Delta p) \equiv \int  P_t(\bGamma)\prod_{k\neq i,j}dz_k.
		\end{equation}
		We then obtain the hierarchal equation for the macroscopic dynamics,
		\begin{align}
			\frac{\partial P_t(\bm{Z})}{\partial t} &= \mcL_{\mathrm{CM}}^{\mathrm{a}} P_t(\bm{Z}) + \sum_{i,j}\mcL_{\mathrm{CM}}^{\mathrm{c};ij}P_t^{ij}(z_i,z_j;\bm{Z})\label{eq:BBGKY_main_2}
		\end{align}
		with advective and diffusive Liouville operator $\mcL_{\mathrm{CM}}^{\mathrm{a}}$ and collision Liouville operator $\mcL_{\mathrm{CM}}^{\mathrm{a};ij}$ between particles $i$ and $j$: 
		\begin{subequations}
		\begin{align}
			\mcL_{\mathrm{CM}}^{\mathrm{a}} P_t =& \left[-c\tanh \frac{\Delta p}{\Delta p^*} \partial_{\CM}+\frac{\sigma^{2}}{2N}\partial_{\CM}^2\right] P_t(z_{\CM},p,\Delta p) \\
			\mcL_{\mathrm{CM}}^{\mathrm{c};ij} P_t =& \frac{\sigma^{2}}{2}\bigg[\! \int \!\! |\tpartial_{ij}|P_t^{ij}\left(\!z_i\!+\!\frac{L_i}{2},z_j\!-\!\frac{L_j}{2};z_{\CM}\!+\!\frac{L_i-L_j}{2N},p\!-\!\Dp,\Delta p'\right)\bigg|_{z_i=z_j=p} \!\!\!\!\!\!\!\!\!\!\!\!\!d\Dp'
										\!-\! \int\!\! |\tpartial_{ij}|P_t^{ij}\bigg|_{z_i=z_j+(L_i+L_j)/2} \!\!\!\!\!\!\!dz_j\bigg].
		\end{align}
		\end{subequations}
		Equation~{(\ref{eq:BBGKY_main_2})}  formally corresponds to the lowest-order conventional BBGKY hierarchal equation~{(\ref{eq:conventional_BLE})} for the macroscopic description. 
		Using this hierarchal equation~{(\ref{eq:BBGKY_main_2})}, a closed master-Boltzmann equation is derived for the macroscopic variables in the next section. 
		
		The set of Eqs.~{(\ref{eq:BBGKY_main})} and~{(\ref{eq:BBGKY_main_2})} is the second main result in this paper.
		Equation~{(\ref{eq:BBGKY_main})} connects the microscopic description (Fig.~\ref{fig:Hierarchy_Brownian}d) to the mesoscopic description (Fig.~\ref{fig:Hierarchy_Brownian}e),
		and Eq.~{(\ref{eq:BBGKY_main_2})} connects the mesoscopic description (Fig.~\ref{fig:Hierarchy_Brownian}e) to the macroscopic description (Fig.~\ref{fig:Hierarchy_Brownian}f). 
		Their detailed derivation is presented in Appendix.~\ref{sec_app:BBGKY_der}. 
		These equations are derived in a parallel calculation to the conventional BBGKY hierarchal equations~{(\ref{eq:BBGKY_BE})} and~{(\ref{eq:conventional_BLE})}, 
		and are called the financial BBGKY hierarchal equations in this paper. 
		Similarly to the conventional BBGKY hierarchal equations~{(\ref{eq:BBGKY_BE})} and~{(\ref{eq:conventional_BLE})}, our hierarchal equations~{(\ref{eq:BBGKY_main})} and~{(\ref{eq:BBGKY_main_2})} are exact but are not closed:
		the dynamics of low-order distributions are driven by those of higher-order distributions.
		Appropriate approximations are necessary to derive closed equations, such as the molecular chaos, which will be studied in the next section.  
		
		\subsubsection*{Remark on the three-body collision term.}
		\begin{figure*}
			\centering
			\includegraphics[width=150mm]{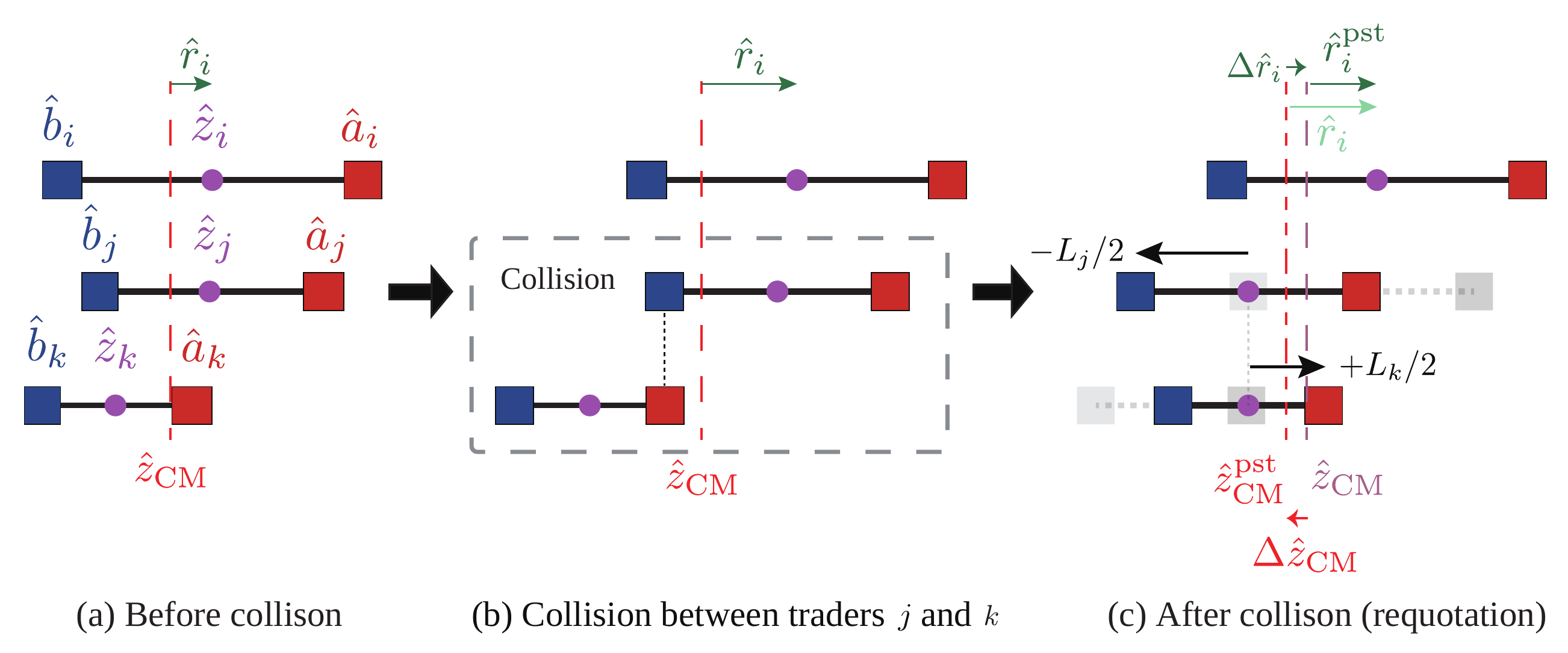}
			\caption	{
								Schematic of the three-body collision term $\mcL^{(ijk)}$. 
								Let us assume that there is a collision between the traders $i$ and $j$. 
								Because of the assumption of the binary interaction, the mid price $\hz_i$ of the $i$th trader does not move during this collision. 
								On the other hand, the CM of this system $\hz_{\CM}$ moves through a short distance of $\Delta \hz_{\CM}\equiv \hz_{\CM}^{\pst}-\hz_{\CM}=-s(L_j-L_k)/2N$ because of the requotation. 
								The relative price $\hr_i$ of the $i$th trader indirectly moves through a short distance of $\Delta \hr_i\equiv \hr_i^{\pst}-\hr_i=\Delta r_{ij;s}^{(1)}$.  
							}
			\label{fig:ThreeBodyCollision}
		\end{figure*}
		We here remark the emergence of the three-body collision term $\mcL^{(ijk)}$ in the BBGKY hierarchy~{(\ref{eq:BBGKY_main})}, which is slightly different from the conventional BBGKY hierarchy~{(\ref{eq:BBGKY_BE})}. 
		This term appears because our kinetic theory is formulated on the basis of the relative price $\hr_i$. 
		To understand this point, let us consider the movement of the relative price $\hr_i$ of the $i$th trader during collision between traders $j$ and $k$ (see Fig.~\ref{fig:ThreeBodyCollision} for a schematic of three-body collision). 
		While the mid price $\hz_i$ of the $i$th trader does not move during the collision between traders $j$ and $k$, the CM of this system $\hz_{\CM}$ moves through a distance of $\Delta \hz_{\CM}\equiv \hz_{\CM}^{\pst}-\hz_{\CM}=-s(L_j-L_k)/2N$. 
		The relative price $\hr_i$ thus moves indirectly through a distance of $\Delta \hr_i\equiv \hr_i^{\pst}-\hr_i=-\Delta \hz_{\CM}=\Delta r_{jk;s}^{(1)}$,
		which appears in the three-body collision operator~{(\ref{eq:op_BBGKY_3body})}. 
		This effect is intuitively small for the large $N$ limit and is finally shown irrelevant to the leading-order (LO) and next-leading-order (NLO) approximations as discussed later.

\section{Main Result 2: Mesoscopic description}\label{sec:result_meso}
	From microscopic dynamics, we have derived the BBGKY hierarchal equation~{(\ref{eq:BBGKY_main})} for the mesoscopic description of the HFT model 
	in a parallel manner to the conventional BBGKY hierarchal equation~{(\ref{eq:BBGKY_BE})}. 
	Here we proceed to derive the closed mean-field model for the mesoscopic description, which will be finally shown useful to understand the order-book profile systematically. 
	
	\subsection{Financial Boltzmann Equation}
		We here derive a closed equation for the one-body distribution function by assuming a mean-field approximation. 
		The one-body and two-body distribution functions $\phi^L_t(r)$ and $\phi^{LL'}_t(r,r')$ are introduced conditional on the traders' spreads $L$ and $L'$, 
		satisfying $\phi^{L_i}_t(r)= P_t^i(r)$ and $\phi^{L_iL_j}_t(r,r')= P_t^{ij}(r,r')$. 
		Let us approximately truncate the two-body correlation as 
		\begin{equation}
			\phi^{LL'}_t(r,r') \approx \phi^L_t(r)\phi^{L'}_t(r'),\label{eq:mlclr_chaos_MF_BE}
		\end{equation}
		which corresponds to molecular chaos~{(\ref{eq:MlclrChs_conventional1})}, the standard approximation in the conventional kinetic theory. 
		The validity of this approximation will be numerically evaluated in Sec.~\ref{sec:BE_solution}. 
		A closed mean-field equation for the one-body distribution $\phi^L_t(r)$ is thus obtained as
		\begin{equation}
			\frac{\partial \phi^L_t}{\partial t} \approx \frac{\sigma^{2}}{2}\frac{\partial^2 \phi^L_t}{\partial r^2}
			+ N\!\!\sum_{s=\pm1}\int_{L_{\min}}^{L_{\max}} \!\!\!dL'\rho_{L'}\left[J_{t;s}^{LL'}(r+sL/2)-J_{t;s}^{LL'}(r)\right], \>\>\>
			J_{t;s}^{LL'}(r) = \frac{\sigma^2}{2}|\tpartial_{rr'}|\phi^{L}_t(r)\phi^{L'}_t(r') \big|_{r-r'=s(L+L')/2}.\label{eq:BoltzmannEq}
		\end{equation}
		with mean-field probability flux $J_{t;s}^{LL'}(r)$ for $s=\pm 1$. 
		The systematic derivation of this equation is the third main result in this paper (see Appendix.~\ref{sec:app:FBE_der} for the detail). 
		Equation~{(\ref{eq:BoltzmannEq})} is a closed equation for the one-body distribution function, 
		and corresponds to the Boltzmann equation in molecular kinetic theory (see Fig.~\ref{fig:Hierarchy_Brownian}b). 
		Equation~{(\ref{eq:BoltzmannEq})} is therefore called the financial Boltzmann equation in this paper. 
		Here the dummy variable $s=+1$ ($s=-1$) implies the transactions as a bidder (an asker), and the integrals on the right-hand side (rhs) correspond to the collision integrals in the standard Boltzmann equation~{(\ref{eq:conventional_BE})}. 
		Remarkably, Eq.~{(\ref{eq:BoltzmannEq})} is derived from a systematic calculation from the Liouville equation~{(\ref{eq:pseudo_Liouville})},
		whereas it was originally introduced with a rather heuristic discussion in our previous paper~\cite{Kanazawa2017}.
		
	\subsection{Solution}\label{sec:BE_solution}
		\begin{figure*}
			\centering
			\includegraphics[width=160mm]{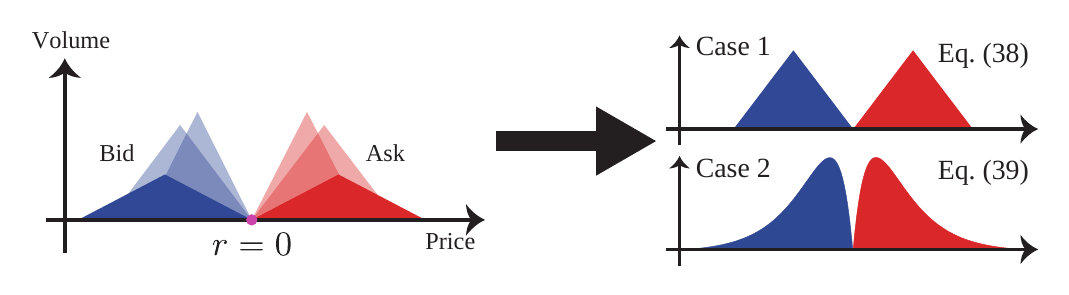}
			\caption	{
							Average order-book profile is given by the superposition~{(\ref{eq:MF-avg_orderbook})} of the tent function~{(\ref{eq:limit_triangular})}. 
							For the $\delta$-distributed spread (Case 1), the profile is the tent function~{(\ref{eq:avg_order_book_1})}. 
							For the $\gamma$-distributed spread (Case 2), the profile obeys Eq.~{(\ref{eq:avg_order_book_2})}.
						}
		\end{figure*}
		Let us focus on the steady solution of Eq.~{(\ref{eq:BoltzmannEq})}. 
		Equation~{(\ref{eq:BoltzmannEq})} can be analytically solved for $N\to \infty$ on an appropriate boundary condition (See Appendix.~\ref{sec_app:Boundary_Condition} for the detail) for the steady state.
		The LO steady solution is given by the tent function: 
		\begin{equation}
				\psi^L(r) \equiv \lim_{t\to \infty}\lim_{N\to 0}\phi^L_t(r) = \frac{4}{L^2}\max \left\{ \frac{L}{2}-|r|, 0 \right\}.\label{eq:limit_triangular}
		\end{equation}
		The average order-book profile for the ask side $f_{\mathrm{A}}(r)$ is given by the superposition of the tent function: 
		\begin{equation}
			f_{\mathrm{A}}(r) = \int_{L_{\min}}^{L_{\max}} dL\rho_L\psi^L(r-L/2).\label{eq:MF-avg_orderbook}
		\end{equation}
		
		We note that the average order-book profile has a symmetry, such that $f_{\mathrm{B}}(r) = f_{\mathrm{A}}(-r)$ for the average bid order-book $f_{\mathrm{B}}(r)$. 
		We also note that the NLO correction~{(\ref{eq:app:subleading_order_FBE})} can be obtained as shown in Appendix.~\ref{sec:app:subleading}. 
		Though the LO solution~{(\ref{eq:limit_triangular})} is sufficient to understand the average order-book profile, 
		the NLO solution~{(\ref{eq:app:subleading_order_FBE})} is necessary to understand the dynamics of the financial Langevin equation, as shown in Sec.~\ref{sec:result_macro}.

\begin{figure*}
			\centering
			\includegraphics[height=45mm]{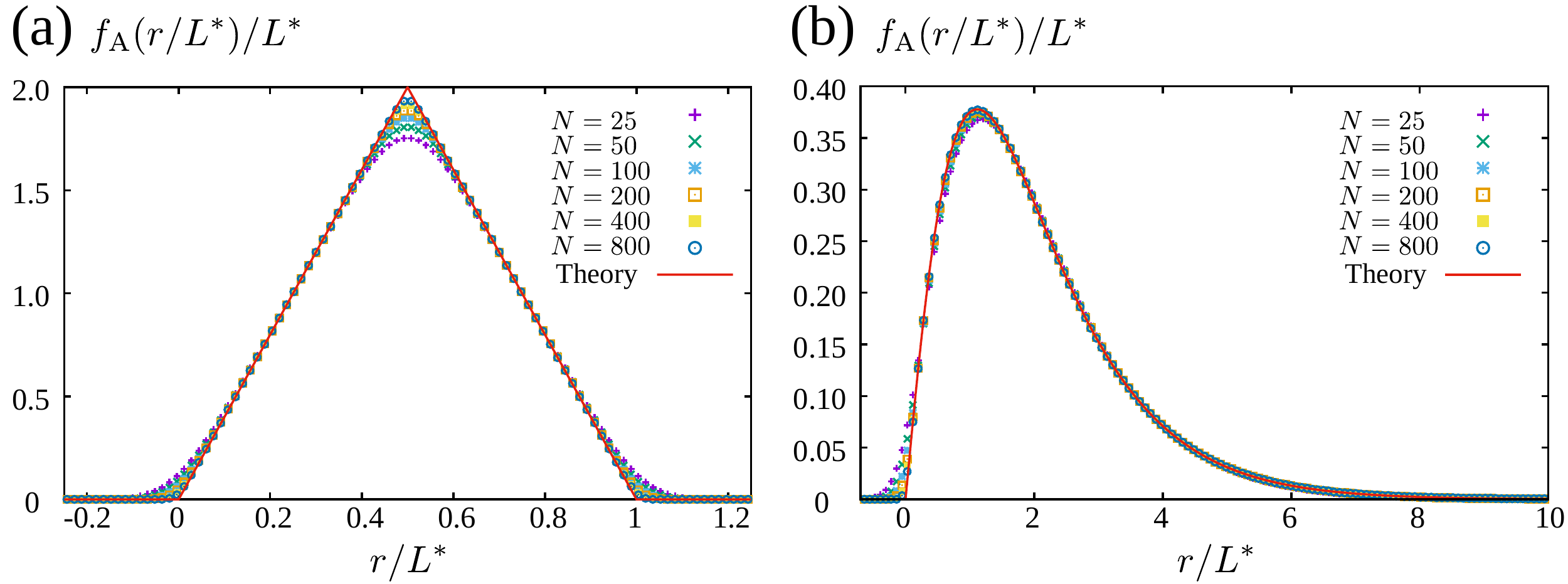}
			\caption	{
							(a) Numerical average order-book profile for the $\delta$-distributed spread $\rho_L = \delta(L-L^*)$, showing the agreement with the theoretical formula~{(\ref{eq:avg_order_book_1})} for $N\to \infty$. 
							(b) Numerical average order-book profile for the $\gamma$-distributed spread $\rho_L = L^{3}e^{-L/L^*}/(6L^{*4})$, showing the agreement with the theoretical formula~{(\ref{eq:avg_order_book_2})} for $N\to \infty$. 
						}
			\label{fig:OrderBook}
		\end{figure*}

	\paragraph*{Numerical comparison 1: $\delta$-distributed spread.}
		We here study the theoretical order-book profiles for two concrete examples with numerical validation (see Appendix.~\ref{sec:app:MonteCarlo} for the detailed implementation). 
		Let us first consider the case of a single spread $L^*$. 
		The corresponding average order-book profile is given by the tent function 
		\begin{equation}
			\rho_L = \delta(L-L^*) \Longrightarrow 			f_{\mathrm{A}}(r) =  \psi^{L^*}(r-L^*/2) = 
									\frac{4}{L^{2*}}\max\left\{\frac{L^*}{2}-\left|r-\frac{L^*}{2}\right| , 0 \right\}.\label{eq:avg_order_book_1}
		\end{equation}
		We have numerically examined the validity of this formula in Fig.~{\ref{fig:OrderBook}a}, which shows the numerical agreement with our formula~{(\ref{eq:avg_order_book_1})}. 
		The LO solution~{(\ref{eq:avg_order_book_1})} works quite well for the description of the order-book profile, and 
		the numerical convergence in Fig.~{\ref{fig:OrderBook}a} implies that Eq.~{(\ref{eq:avg_order_book_1})} might be exactly valid for $N\to \infty$.

	\paragraph*{Numerical comparison 2: $\gamma$-distributed spread. }
		The formula~{(\ref{eq:MF-avg_orderbook})} works well even for $L_{\min}\to 0$ and $L_{\max}\to \infty$ when the integrals converge. 
		As an example, let us consider the case where the spread obeys the $\gamma$-distribution 
		\begin{equation}
			\rho_L = \frac{L^{3}e^{-L/L^*}}{6L^{*4}} \>\>\> \Longleftarrow \>\>\>
			f_{\mathrm{A}}(r) = \frac{1}{L^*}\psi\left(\frac{r}{L^*}\right), \>\>\>
			\psi(r) \equiv \frac{4}{3}e^{-\frac{3r}{2}}\left[\left(2+r\right)\sinh \frac{r}{2}-\frac{r}{2}e^{-\frac{r}{2}}\right],\label{eq:avg_order_book_2}
		\end{equation}
		which was empirically validated through single-trajectory analysis of individual traders in our previous work~\cite{Kanazawa2017}. 
		We have numerically examined the validity of this formula in Fig.~{\ref{fig:OrderBook}b}, which shows the numerical agreement with our formula~{(\ref{eq:avg_order_book_2})}. 
		The numerical convergence in Fig.~{\ref{fig:OrderBook}b} implies that the LO solution~{(\ref{eq:avg_order_book_2})} might be also exact for $N\to \infty$. 

\section{Main Result 3: Macroscopic Description}\label{sec:result_macro}
	In this section, we derive the stochastic equations for the macroscopic dynamics of this system from the BBGKY hierarchal equation~{(\ref{eq:BBGKY_main_2})} 
	in the parallel method to the master-Boltzmann equation~{(\ref{eq:conventional_BLE})} for physical Brownian motions. 

	\subsection{Master-Boltzmann Equation for Financial Brownian Motion}
		On the basis of the financial BBGKY hierarchy~{(\ref{eq:BBGKY_main_2})} for the macroscopic dynamics, 
		we derive a closed dynamical equation for the macroscopic variables $\bm{Z}\equiv (z_{\CM},p,\Dp)$. 
		Here we first make the assumption of molecular chaos,
		\begin{equation}
			P_t^{ij}(z_i,z_j;\bm{Z})\approx \phi^{L_i}_t(z_i-z_{\CM})\phi^{L_j}_t(z_j-z_{\CM})P_t (\bm{Z}).\label{eq:mlclr_chs_macro3}
		\end{equation}
		Using the NLO solution~{(\ref{eq:app:subleading_order_FBE})}, 
		we deduce a closed master-Boltzmann equation for the macroscopic dynamics (see Appendix.~\ref{sec:app:MB_macro} for the detailed calculation): 
		\begin{align}
			\frac{\partial P_t(\bm{Z})}{\partial t} \approx (\mathcal{L}_{\CM}^{\mathrm{a}}+\mathcal{L}_{\CM}^{\mathrm{c};\MF})P_t(\bm{Z})\label{eq:main_MB_macro}
		\end{align}
		where the mean-field collision Liouville operators for the macroscopic variables $\mathcal{L}_{\CM}^{\mathrm{c};\MF}$ is defined by 
		\begin{equation}
			\mathcal{L}_{\CM}^{\mathrm{c};\MF} P_t \equiv \frac{1}{\tau^*}\left[\mathcal{N}\left(p-z_{\CM};\frac{L^{*2}_{\rho}}{4N}\right)\int d\Delta p'dyw_N(y)P_t\left(z_{\CM}-y,p-\Dp,\Delta p'\right)- P_t(\bm{Z})\right]
			\label{eq:LinearBL_macro}
		\end{equation}
		with $1/L^{*2}_\rho\equiv \int dL\rho_L/L^2$, Gaussian distribution $\mathcal{N}(x;\sigma^2)$, jump size distribution $w_N(y)$, and mean transaction interval $\tau^*$ defined by
		\begin{equation}
			\mathcal{N}\left(x;\sigma^2\right) \equiv \frac{e^{-x^2/2\sigma^2}}{\sqrt{2\pi\sigma^2}}, \>\>\>
			w_N(y) \equiv \int_{-\infty}^\infty \frac{2NL^{*4}_{\rho}dL}{L^2(L+2Ny)^2}\rho_L\rho(L+2Ny), \>\>\>
			\tau^* \simeq \frac{L_\rho^{*2}}{2N\sigma^{2}}
		\end{equation}
		by assuming $\rho_L$ is zero for $L \not \in [L_{\min},L_{\max}]$.
		Note that Eq.~{(\ref{eq:LinearBL_macro})} is a master equation (or the differential form of the Chapman-Kolmogorov equation~\cite{GardinerB}) 
		and is equivalent to a set of stochastic differential equations (SDEs) (see Eq.~{(\ref{eq:set:SDE_macro})} in Appendix.~\ref{sec:app:MB_macro}). 
			
	\subsection{Financial Langevin Equation}
		We have derived the stochastic dynamics for the three macroscopic variable $\hat{\bm{Z}}=(\hz_{\CM},\hp,\hDp)$ as the master equation~{(\ref{eq:LinearBL_macro})} (or equivalently SDEs~{(\ref{eq:set:SDE_macro})}) in the continuous time $t$. 
		We next simplify the dynamics~{(\ref{eq:LinearBL_macro})} of the three macroscopic variables into that of a single macroscopic variable $\hDp$ in the tick time $T$. 
		In the tick time $T$...., the dynamical equation for the price movement $\hDp$ is given by 
		\begin{equation}
			\hDp[T+1] =\underbrace{c\hat{\tau}[T]  \tanh \frac{\hDp[T]}{\Delta p^*}}_{\mbox{Trend-following}} + \underbrace{\Delta \hxi[T]}_{\mbox{Zigzag}}  + \underbrace{\hzeta[T]}_{\mbox{Random}},\label{eq:Financial_Brownian_Motion}
		\end{equation}
		where $\hat{\tau}[T]\equiv \hat{t}[T+1]-\hat{t}[T]$ is time interval between transaction, 
		$\Delta \hxi[T]$ is the zigzag noise of order $N^{-1/2}$,
		and $\hzeta[T]$ is a random noise of order $N^{-1}$ (see Appendix.~\ref{sec:app:F_Langevin_der} for the detail). 
		The systematic derivation of Eq.~{(\ref{eq:Financial_Brownian_Motion})} is the fourth main result of this paper. 
		Equation~{(\ref{eq:Financial_Brownian_Motion})} corresponds to the conventional Langevin equation~{(\ref{eq:Langevin_cnvnt})},
		and is thus called the financial Langevin equation in this paper. 
		
		Within the mean-field approximation, we can specify all the statistics of the random noise terms from analytics. 
		The time interval $\hat{\tau}[T]$ is given by the exponential random number with mean interval $\tau^*$,
		\begin{equation}
			P(\tau) = \frac{1}{\tau^*}e^{-\tau/\tau^*}, \>\>\>\tau^*=\frac{L_\rho^{*2}}{2N\sigma^{2}}. \label{eq:MF_timeinterval}
		\end{equation}
		The zigzag noise $\Delta \hxi[T]$ is defined by the difference of two Gaussian random numbers as 
		\begin{equation}
			\Delta \hxi[T] \approx \sqrt{\frac{L^{*2}_{\rho}}{4N}}\left(\hxi[T]-\hxi[T-1]\right) = O(N^{-1/2}),
		\end{equation}
		where $\hxi[T]$ is a discrete-time white Gaussian noise with unit variance. 
		The random noise term $\hzeta[T]$ is specified as 
		\begin{equation}
			\hzeta[T] \approx \sqrt{\frac{\sigma^{2}\hat{\tau}(T)}{N}}\hat {\mu}[T] + \frac{1}{N}\hat{\nu}[T] = O(N^{-1}),
		\end{equation}
		where  $\hat{\mu}[T]$ is a discrete-time white Gaussian noise with unit variance and $\hat{\nu}[T]$ is a discrete-time white noise term obeying $P(\nu)=\tilde{w}(\nu)$ with an $N$-independent distribution $\tilde{w}(\nu)=w_N(\nu/N)/N$.
		
		We next discuss the interpretation of each term on the rhs of Eq.~{(\ref{eq:Financial_Brownian_Motion})}. 
		The trend-following term induces the collective motion of the order book and thus keeps the price movement in the same direction for a certain time-interval similarly to the inertia in physics. 
		On the other hand, the zigzag noise term exhibits one-tick negative autocorrelation, such that 
		\begin{equation}
			C_{\Delta \hxi}[K]\equiv \frac{\la \Delta \hxi[T+K]\Delta \hxi[T]\ra}{\la \Delta \hxi[T]^2 \ra}
			\approx 	\begin{cases}
							1		& (K=0) \cr
							-1/2	& (K=1) \cr
							0		& (K\geq 2)
						\end{cases},
		\end{equation}
		and has the effect to change the price movement direction alternately. 
		In this sense, the trend-following term and the zigzag noise have the opposite effect to each other;
		the balance between their strengths is crucial for the qualitative behavior of the market price dynamics. 
		The random noise term $\hzeta[T]$ originates from the slow dynamics of the CM:
		$(\sigma^2\hat{\tau[T]}/N)^{1/2}\hat{\mu}[T]$ is the diffusion term of the CM during a transaction time-interval $\hat{\tau}[T]$ 
		and $\hat{\nu}[T]/N$ is the movement term of the CM by requotation jumps of traders after transaction. 

	\subsection{Solution}\label{sec:solution:Langevin}
\begin{figure*}
			\centering
			\includegraphics[width=180mm]{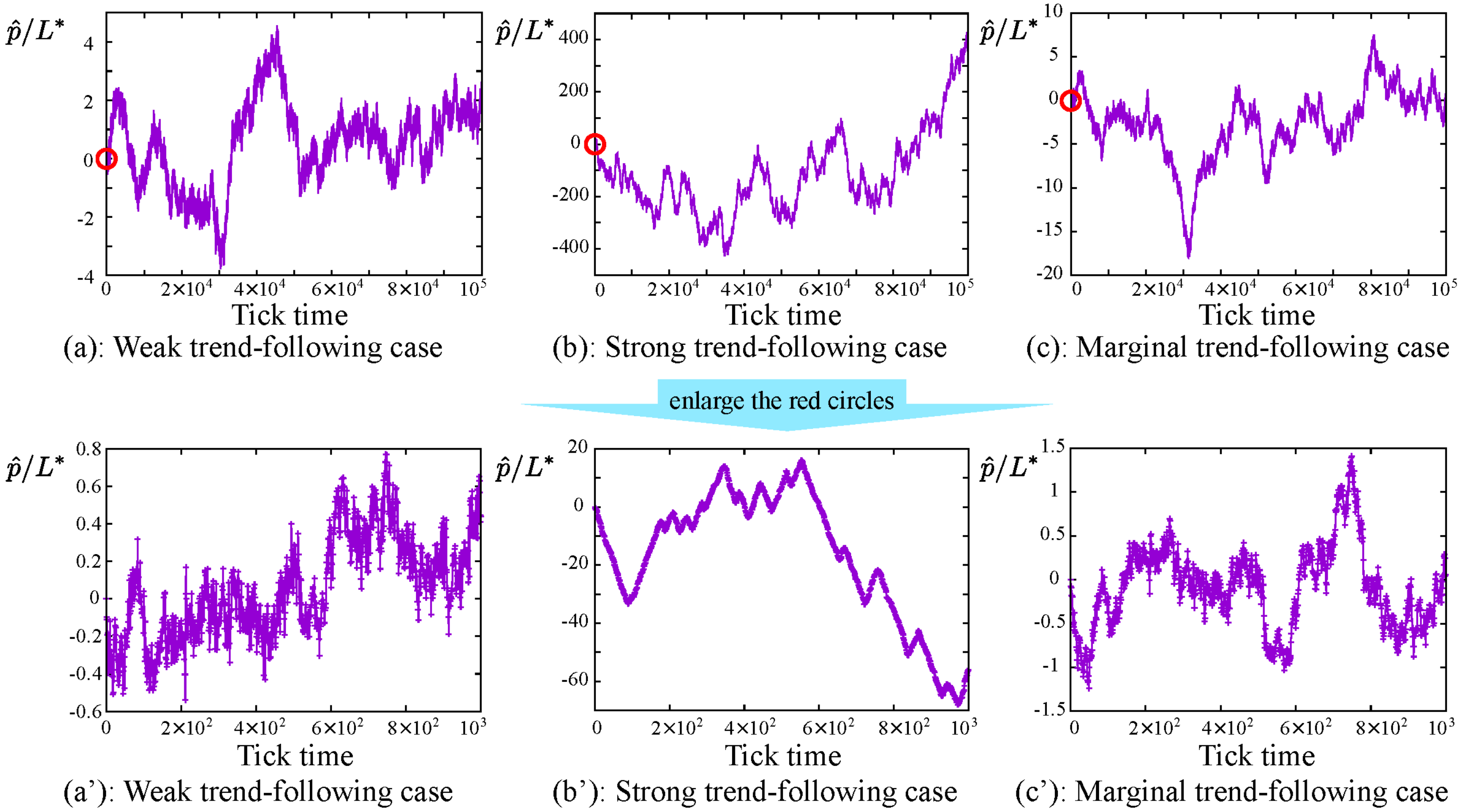}
			\caption	{
							(a--c)~Sample trajectories are plotted for (a)~the weak trend-following case $\tl{c}=0$, (b)~the strong trend-following case $(\tl{c},\Delta \tl{p}^*)=(2.0,0.1)$, and
							(c)~the marginal trend-following case $(\tl{c},\Delta \tl{p}^*)=(0.5,2.5)$ for $N=100$, $10^5$ ticks, and the $\gamma$-distributed spread. 
							All parameters are shared except for the trend-following parameters $(\tl{c},\Delta \tl{p}^*)$. 
							As can be seen from the figures, all trajectories seem to be the normal diffusion in the long timescale. 
							(a'--c')~The sample trajectories are enlarged 100 times (in the circles in Fig.~{a--c}),
							where the character of each trajectory can be seen. 
							(a')~The price trajectory exhibits a zigzag behavior in the absence of trend-following. 
							(b')~The price keeps moving toward the same direction for a certain tick period because of the strong trend-following. 
							(c')~The price trajectory exhibits both zigzag behavior and trend-following because both effects are in balance. 
						}
			\label{fig:trajectories}
		\end{figure*}

		The macroscopic dynamics of the price strongly depends on the balance between the strength of trend-following effect and that of the zigzag noise. 
		Here we present the solutions of the financial Langevin equation depending on the strength of trend-following with the dimensional analysis.  
		The price movement originating from trend-following behavior is estimated to be $c\tau^*$ (of price dimension). 
		On the other hand, the amplitude of the zigzag noise is estimated to be $L^{*}_\rho/\sqrt{2N}$ (of price dimension). 
		Their balance is thus characterized by the dimensionless parameter $\tl{c}$ defined by
		\begin{equation}
			\tl{c}\equiv \frac{c\tau^*}{L^{*}_\rho/\sqrt{2N}} = \frac{cL^*_\rho}{\sigma^2\sqrt{2N}}. 
		\end{equation}
		Another dimensionless control parameter is the ratio $\Delta \tl{p}^*$ between the average movement by the trend-following $c\tau^*$  (of price dimension) and the saturation threshold against the market trend $\Delta p^*$ (of price dimension): 
		\begin{equation}
			\Delta \tl{p}^* \equiv \frac{\Delta p^*}{c\tau^*}. 
		\end{equation}
		The set of dimensionless parameters $(\tl{c},\Delta \tl{p}^*)$ governs the qualitative dynamics of the market price. 
		For consistency with the empirical report~\cite{Kanazawa2017}, we focus on the case of $\Delta \tl{p}^* \lesssim 1$ in this section, whereby the saturation of the hyperbolic function is valid. 
		(see Sec.~{\ref{sec:DiscussLinearTrend}} for the discussion on the case with $\Delta \tl{p}^* \gg 1$). 
		Here we introduce three classifications in terms of the strength of trend-following: 
		\begin{enumerate}
			\item 	Weak trend-following case: $\tl{c}\ll 1$
			\item 	Strong trend-following case: $\tl{c}\gg 1$
			\item 	Marginal trend-following case: $\tl{c}\sim 1$
		\end{enumerate}
		Sample trajectories are plotted in Fig.~{\ref{fig:trajectories}} to highlight the character of each case:
		For the weak trend-following case (Fig.~{\ref{fig:trajectories}}a), the price tends to move upward and downward alternatively every tick because of the zigzag noise $\Delta \hxi$. 
		For the strong trend-following case (Fig.~{\ref{fig:trajectories}}b), the unidirectional movement of price is kept for a certain time period. 
		For the marginal trend-following case (Fig.~{\ref{fig:trajectories}}c), both zigzag and unidirectional movements randomly appear because both effects are in balance. 
		As will be shown later in detail, the marginal case may be the most realistic, at least in our dataset. 
		We next study these qualitative characters through statistical analysis of price time series within the mean-field approximation. 

		\subsubsection{Weak trend-following case}\label{sec:WeakTrendCase}
			For the weak trend-following case $\tl{c}\ll 1$, the trend-following effect is negligible compared with the zigzag noise: $|c\hat{\tau}[T]\tanh(\hDp[T]/\Dp^*)|\ll |\Delta\hxi[T]|$. 
			The master equation~{(\ref{eq:LinearBL_macro})} can then analytically solved in continuous time $t$. 
			By applying the system size expansion~\cite{VanKampen} (see Appendix.~\ref{sec:app:Diffusion_WT} for derivation), we obtain the diffusion equation for the CM
			\begin{equation}
				\frac{\partial P_t(z_{\CM})}{\partial t} = D(N)\frac{\partial^2 P_t(z_{\CM})}{\partial z_{\CM}^2},\>\>\> D(N)\equiv \frac{\sigma^2}{2N}\left(1 + \frac{2\alpha_2}{L_{\rho}^{*2}}\right).\label{eq:diffusion_WT}
			\end{equation}
			with the renormalized diffusion coefficient $D(N)$ up to the order of $N^{-1}$ and the second-order Kramers-Moyal coefficient $\alpha_2\equiv\int_{-\infty}^\infty dyy^2\tl{w}(y)$. 
			The diffusion constant $D(N)$ decays for $N\to \infty$, which implies that the dynamics of the CM become slower as the number of the traders increases. 
			Given that the dynamics of price $\hp$ coincides with that of the CM $\hz_{\CM}$ for a long timescale,  
			the diffusion of the price is also shown normal for a long timescale with the same diffusion coefficient $D(N)$ in the real time $t$. 
			The mean square displacement (MSD) based on real time $t$ is thus analytically obtained as 
			\begin{equation}
				\mathrm{MSD}(t) \equiv \la\left[\hp(t)-\hp(0)\right]^2\ra \sim 2D(N)t,\label{eq:MSD_RT}
			\end{equation}
			showing the normal diffusion for a long time. 
			
			We also study price movement at one-tick precision. 
			For the weak trend-following case, the only relevant term in Eq.~{(\ref{eq:Financial_Brownian_Motion})} is the zigzag noise $\Delta \hxi(T)$ for a short timescale. 
			Price movement $\hDp$ then obeys the Gaussian distribution
			\begin{equation}
				P(\Delta p) \approx \mathcal{N}\left(\Delta p;\frac{L_\rho^{*2}}{2N}\right), \>\>\> \la\hDp^2\ra\approx\frac{L_{\rho}^{*2}}{2N}.\label{eq:Dp_WT}
			\end{equation}
			The autocorrelation function of the price movement $\hDp$ is also given by
			\begin{equation}
				C_{\hDp}[K]\equiv \frac{\la \hDp[T+K]\hDp[T]\ra}{\la \hDp[T]^2 \ra}
				\approx C_{\Delta \hxi}[K]
				\approx 
						\begin{cases}
							1		& (K=0) \cr
							-1/2	& (K=1) \cr
							0		& (K\geq 2)
						\end{cases}.\label{eq:weak_trend_ngtv_corr}
			\end{equation}
			Interestingly, this property is consistent with an empirical fact that price movements typically exhibit zigzag behavior for a short timescale, which is reflected in the one-tick strong negative autocorrelation of the price movement. 
		
			Here we discuss the origin of the strong negative correlation in terms of price movement. 
			Remarkably, only the random noise $\hzeta[T]$ is dominant for long time whereas only the zigzag noise $\Delta \hxi[T]$ is dominant for a short timescale.  
			For $K\gg N$, indeed, we obtain
			\begin{equation}
				\hp [T+K] - \hp[T] = \sum_{i=0}^{K-1} \left(\Delta \hxi[T+i]+\hzeta [T+i]\right) = \underbrace{\sqrt{\frac{L^{*2}_{\rho}}{4N}}\left(\hxi[T+K-1]-\hxi[T-1]\right)}_{O(N^{-1/2})} + \underbrace{\sum_{i=0}^{K-1}\hzeta[T+i]}_{O(N^{-1}K^{1/2})}, \label{eq:discuss_orgn_ngtv_corr}
			\end{equation}
			which implies that the contribution by the zigzag noise $\hxi[T]$ is negligible compared with that of the random noise $\hzeta[T]$ (i.e., $\sum_{i=0}^{K-1}\hzeta[T+i]=O(N^{-1}K^{1/2}) \gg O(N^{-1/2})$). 
			Considering that the random noise $\hzeta[T]$ originates from the diffusion of the CM, Eq.~{(\ref{eq:discuss_orgn_ngtv_corr})} means that the macroscopic behavior of price is governed by the slow dynamics of the CM. 
			Even though the price movement at one-tick precision is much larger than that of the CM, such movement is irrelevant to the macroscopic dynamics of the whole system. 
			This is the origin of the strong negative correlation for price movement in this model with weak trend-following. 
			To relieve such negative correlation, stronger trend-following is necessary to induce the collective motion of the order book as discussed in Ref.~\cite{Kanazawa2017}. 
			We note that similar slow diffusion is observed in the conventional zero-intelligence order-book models~\cite{Maslov2000,Daniels2003,Smith2003}, with which the trend-following effect is not incorporated likewise. 
			
			We also note that the negative correlation~{(\ref{eq:weak_trend_ngtv_corr})} is also related to the slow diffusion of price for a short timescale. 
			Indeed, the MSD is given by
			\begin{equation}
				\mathrm{MSD}[K] = \la\left(\hp[T+K]-\hp[T]\right)^2\ra \approx \frac{L_{\rho}^{*2}}{2N} + 2D(N)\tau^* K\label{eq:MSD_Tick}
			\end{equation}
			within the mean-field approximation. 
			This formula implies that the MSD is almost constant (i.e., no diffusion) for a short timescale $K\ll N$ while it is asymptotically linear (i.e., the normal diffusion) for a long timescale $K\gg N$. 
			
			\paragraph*{Numerical comparison.}
				\begin{figure*}
					\centering
					\includegraphics[width=180mm]{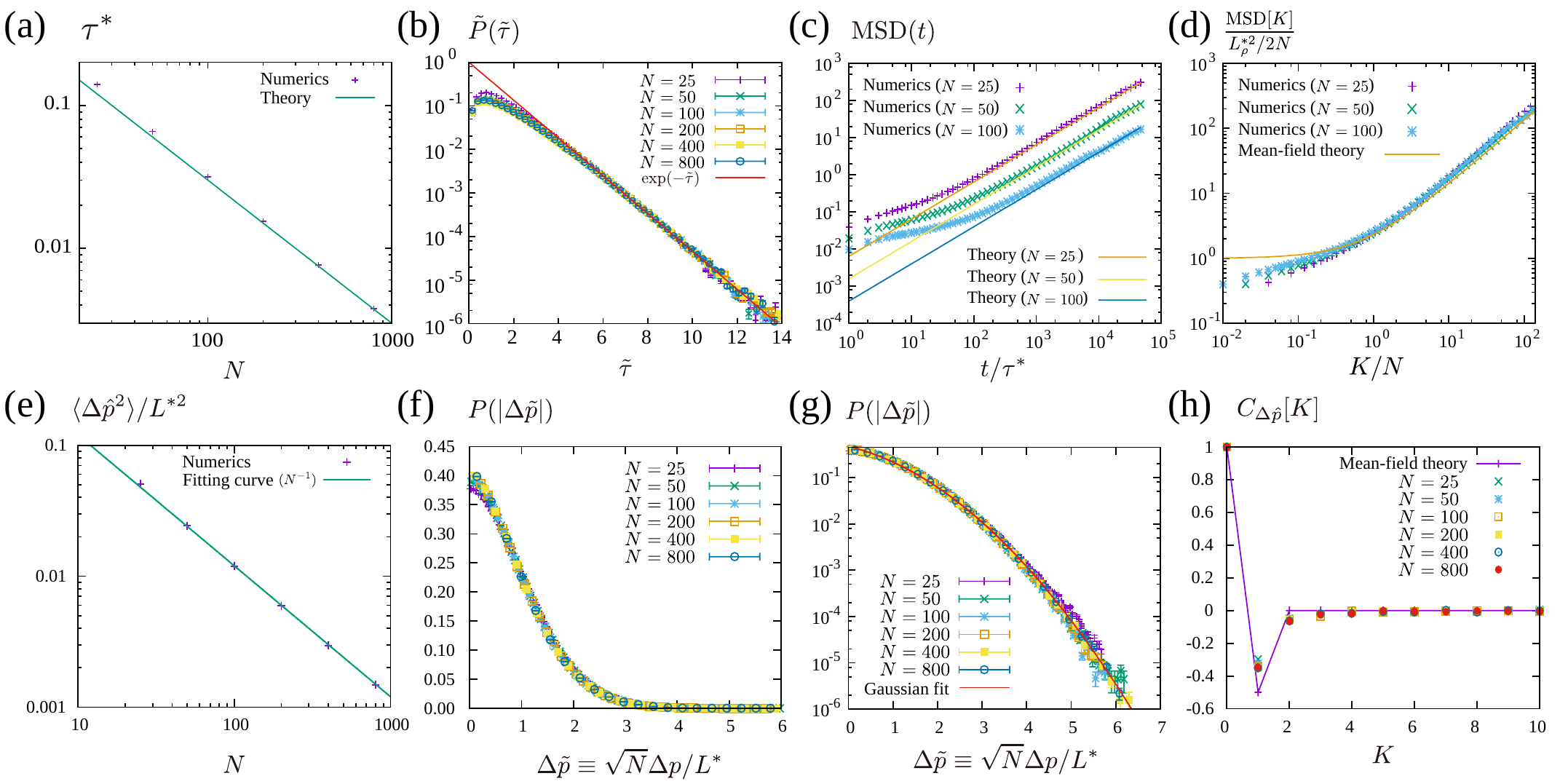}
					\caption	{
											Numerical study in the absence of trend-following $c=0$. 
											(a)~Numerical mean transaction intervals $\tau^*$ and the theoretical line for various $N$.   
											(b)~Transaction interval distribution for various $N$ with an exponential guideline by scaling the horizontal and vertical axes. 
											The scaled interval is given by $\tl{\tau}=c_\tau\hat{\tau}/\tau^*$, where the fitting parameter $c_\tau$ for the decay time was estimated by the least square method for the tail as 
											$c_{\tau}=1.34$, $1.49$, $1.56$, $1.58$, $1.62$, and $1.59$ for $N=25$, $50$, $100$, $200$, $400$, and $800$, respectively.
											(c)~MSD plot based on real time $t$ for various $N$ with the theoretical lines, showing the normal diffusion for large $t$ but the slow diffusion $t\sim \tau^*$
											(d)~MSD plot based on tick time $K$ for various $N$ with the theoretical line, showing the normal diffusion for large $K$ but the slow diffusion for small $K$. 
											(e)~Variance of price difference $\hDp$ for various number of traders $N$ with a fitting curve of power-law exponent $N^{-1}$. 
											(f)~Plot of the peak of the PDF $P(\Delta \tl{p})$ for the scaled price movement $\Delta \tl{p}\equiv \sqrt{N}\Delta p/L^*$. 
											(g)~Log-plot of the tail of the PDF $P(\Delta \tl{p})$ with a Gaussian fitting curve $h(\Delta \tl{p})$. 
											(h)~Auto-correlation function $C_{\hDp}[K]$ with tick time $K$, showing the negative correlation at $K=1$. 
									}
					\label{fig:WeakTrend}
				\end{figure*}
				Here we examine the validity of our formulas through comparison with numerical results for the $\gamma$-distributed spread (see Appendix.~\ref{sec:app:MonteCarlo} for the implementation). 
				
				\subparagraph{Transaction interval.}
				We first check the statistics of the time-interval between transactions $\hat{\tau}$. 
				The mean transaction interval $\tau^*\equiv \la\hat{\tau}\ra$ is numerically plotted in Fig.~{\ref{fig:WeakTrend}a},
				showing the quantitative agreement with the theoretical prediction~{(\ref{eq:MF_timeinterval})} including the coefficient. 
				We also numerically plotted the probability distribution of $\hat{\tau}$ with scaling parameters for horizontal and vertical axes, qualitatively showing the exponential tail for large $\hat{\tau}$.
				Here, we have introduced a scaled transaction interval $\tilde{\tau}\equiv c_{\tau}\hat{\tau}/\tau^*$ and plotted the scaled probability distribution in Fig.~{\ref{fig:WeakTrend}b}
				\begin{equation}
					\tl{P}(\tilde{\tau}) \equiv \frac{\tau^* P(\tau)}{Z_{\tau}} \sim e^{-\tilde{\tau}} \>\>\>(\tilde{\tau} \to \infty),
				\end{equation}
				with scaling parameters for the horizontal and vertical axes $c_{\tau}$ and $Z_{\tau}$. 
				The coefficients $c_{\tau}$ and $Z_{\tau}$ were determined by the least-square method to fit the exponential tail for each $N$. 
				The numerical results imply the modification for the decay length $c_{\tau}\approx 1.6$, whereas the mean-field solution~{(\ref{eq:MF_timeinterval})} predicts $c_{\tau}=1$. 
				This means that the mean-field solution~{(\ref{eq:MF_timeinterval})} is not exact but is rather qualitatively correct for the probability distribution $P(\tau)$. 
				
				This factor modification $c_{\tau}\approx 1.6$ can be roughly understood from the viewpoint of the order statistics, as discussed in Ref.~\cite{Kanazawa2017}. 
				The mean-field approximation predicts the exponential interval distribution~{(\ref{eq:MF_timeinterval})}, 
				which means that the transaction obeys the exact Poisson process. 
				As the numerics shows, however, the transaction obeys the Poisson process not exactly but only asymptotically. 
				One candidate of its origin is that a transaction occurs as a pair of arrivals of both bid and ask quotes. 
				Let us assume that the arrival of a bid (ask) quote at the transaction price obeys the Poisson statistics as $P(\tau_{\mrB})=e^{-\tau_{\mrB}/\tau_{\mrB}^*}/\tau_{\mrB}^*$ $(P(\tau_{\mrA})=e^{-\tau_{\mrA}/\tau_{\mrA}^*}/\tau_{\mrA}^*)$.
				Any transaction is assumed to occurs when both bid and ask quotes arrive at the transaction price. 
				We then make an approximation that $\hat{\tau}\approx \max \{\hat{\tau}_{\mrB},\hat{\tau}_{\mrA}\}$ and $\tau_{\mrB}^*=\tau_{\mrA}^*$. 
				On the basis of the orders statistics~\cite{David2003}, we obtain
				\begin{equation}
					P(\tau) \approx 1 - (1-e^{-3\tau/2\tau^*})^2 \sim e^{-3\tau/2\tau^*},\label{eq:imprvd_MF}
				\end{equation}
				where the fitting parameter was determined by the consistency condition for the average interval as $\la \hat{\tau}\ra=\tau^*\Longleftrightarrow \tau^*_{\mrB}=3\tau^*/2$. 
				We thus obtain the modification factor $c_{\tau}=3/2$ as an approximation. 
				
				We note that the transaction interval is not under the influence of the trend-following effect. 
				The above statistical characters on transaction interval are therefore shared for any parameter set of $(\tl{c},\Delta \tl{p}^*)$. 
			
				\subparagraph{MSD.} 
				Our theoretical prediction on the MSD is numerically examined here for analyses based on both real time $t$ and tick time $K$. 
				We first numerically check the MSD~{(\ref{eq:MSD_RT})} based on real time $t$ in Fig.~{\ref{fig:WeakTrend}c}. 
				This figure shows the quantitative agreement with our theoretical formula~{(\ref{eq:MSD_RT})} without fitting parameters. 
				We also check the MSD based on tick time $K$ in Fig.~{\ref{fig:WeakTrend}d}, showing a quantitative agreement with the theoretical prediction~{(\ref{eq:MSD_Tick})} for $K\gg 1$. 
				For small $K\sim 1$, the agreement is not perfect between the numerical data and the theoretical line, 
				but the slowness of the diffusion is qualitatively observed as predicted in the mean-field solution~{(\ref{eq:MSD_Tick})}. 
			
				\subparagraph{Price movement.}
				The dependence of the variance of price movement is checked in Fig.~\ref{fig:WeakTrend}e on the number of traders $N$. 
				We numerically obtained $\la \hDp^2\ra \approx C_{\la \hDp^2\ra} (L^{*2}_{\rho}/2N)$
				with modification factor $C_{\la \hDp^2\ra} \approx 0.4$ and $L^{*2}_{\rho}=6L^{*2}$. 
				Though there is a discrepancy in terms of the factor $C_{\la \hDp^2\ra}$, the mean-field solution~{(\ref{eq:Dp_WT})} qualitatively works well for the variance of price movement. 
				We also checked the PDF $P(|\Delta \tl{p}|)$ of the scaled price movement $\Delta \tl{p}\equiv \sqrt{N}\Delta p/L^*$ (Fig.~\ref{fig:WeakTrend}f and g for the peak and tail of PDF, respectively). 
				In Fig.~\ref{fig:WeakTrend}g, we also show a Gaussian-type fitting curve $h(\Delta \tl{p})=\exp\left(-h^*_0-h_1^*\Delta \tl{p}-h_2^*\Delta \tl{p}^2\right)$ 
				for the tail with parameters $h_0^* = 0.75\pm 0.05$, $h_1^*=0.54\pm 0.04$, and $h_2^*=0.238\pm0.006$. 
				These figures suggests that the PDF of the price movement has the Gaussian tail,
				which is qualitatively consistent with the theoretical prediction~{(\ref{eq:Dp_WT})} ($h_1^*=0$ and $h_2^*=1/6$).
				
				\subparagraph{Autocorrelation.} 
				The autocorrelation function $C_{\hDp}[K]$ is checked in Fig.~\ref{fig:WeakTrend}h, which supports the qualitative consistency between the theory~{(\ref{eq:discuss_orgn_ngtv_corr})} and the numerical results in terms of the negative correlation at $K=1$ tick. 
				This negative correlation implies that the price time series exhibits zigzag behavior in the absence of the trend-following effect. 
				Indeed, the probability of $\hDp[T+1]\hDp[T]<0$ is theoretically $2/3=66.6...\%$ for the mean-field model (see Appendix.~\ref{sec:app:successive_sign_MF}), considerably higher than $50\%$ (i.e., the pure random walks). 
				This result is also qualitatively consistent with the numerical result (around $61\%$) as shown in Table~\ref{table:sign_WT}. 
				\begin{table}[b]
					\centering
						\begin{tabular}{|l||r|r|}\hline 
							Case &Same sign & Different sign \\ \hline \hline
							(a) Weak trend-following case & 0.389 & 0.611 \\ \hline
							(b) Strong trend-following case & 0.949 & 0.051 \\ \hline 
							(c) Marginal trend-following case & 0.480 & 0.520\\ \hline
							(d) Real price time series & 0.479 & 0.521 \\ \hline
						\end{tabular}
						\caption	{
											Table of the probabilities of the same successive sign and different sign for the price movement time series $\{\hDp[T]\}_{T}$. 
											We numerically obtained the probability that the next price movement $\hDp[T+1]$ has the same (different) sign as (from) that of the previous price movement $\hDp[T]$ for $N=100$. 
											(a) For the weak trend-following case $\tl{c}=0$,
											the probability of taking different sign is higher than that of taking same sign, implying the zigzag motion of the price movement. 
											(b) For the strong trend-following case $(\tl{c},\Delta \tl{p}^*)=(2.0,0.1)$, 
											the probability of taking the same successive sign is much higher than that of taking different sign, implying the ballistic motion of the price movement. 
											(c) For the marginal trend-following case $(\tl{c},\Delta \tl{p}^*)=(0.5,2.5)$,
											the probability of taking different sign is slightly higher than that of taking same sign. 
											(d) We also obtained the probabilities from the real price time series in our dataset, 
											showing that the probability of taking different sign is slightly higher than that of taking same sign. 
											For simplicity, we omitted zero, such as $\Delta p[T]=0$, during the data analysis of real price movement time series $\{\hDp[T]\}_{T}$. 
											This table implies that the marginal trend-following case is consistent with the real price time series and is the most realistic at least for stable markets. 
										}
						\label{table:sign_WT}
				\end{table}

		\subsubsection{Strong trend-following case}
			The strong trend-following case $\tl{c}\gg 1$ is also analytically tractable, 
			whereby the trend-following term is dominant such that $|c\hat{\tau}[T]\tanh(\hDp[T]/\Dp^*)|\gg |\max\{\Delta\hxi[T], \hzeta[T]\}|$. 
			Here we assume that the saturation threshold is sufficiently small such that $\Delta \tl{p}^*\ll 1$. 
			This condition simplifies the analysis below because the hyperbolic function can be approximated as the signature function, such that $\tanh(\hDp[T]/\Dp^*) \approx \sgn(\hDp[T])$. 
			Price movement is then governed by the first term on the rhs of Eq.~{(\ref{eq:Financial_Brownian_Motion})}, 
			which approximately leads the exponential distribution,
			\begin{equation}
				P(|\Dp|) \propto e^{-|\Dp|/\kappa}\label{eq:pricemovement_exp}
			\end{equation}
			with decay length $\kappa$ for $|\Dp|\to \infty$.  
			The decay length is given by the mean movement originating from the trend-following as $\kappa= c\tau^*$ within the mean-field approximation~{(\ref{eq:MF_timeinterval})}. 
			By applying the improved mean-field approximation~{(\ref{eq:imprvd_MF})}, more consistent coefficient $\kappa= 2c\tau^*/3$ is obtained with the numerical result below. 
			The trend-following effect plays similar roles to momentum inertia in physics, which are reflected in the autocorrelation function and the MSD plot as shown numerically in the next paragraph. 
			
			\paragraph*{Numerical comparison.}
				\begin{figure*}
					\centering
					\includegraphics[width=135mm]{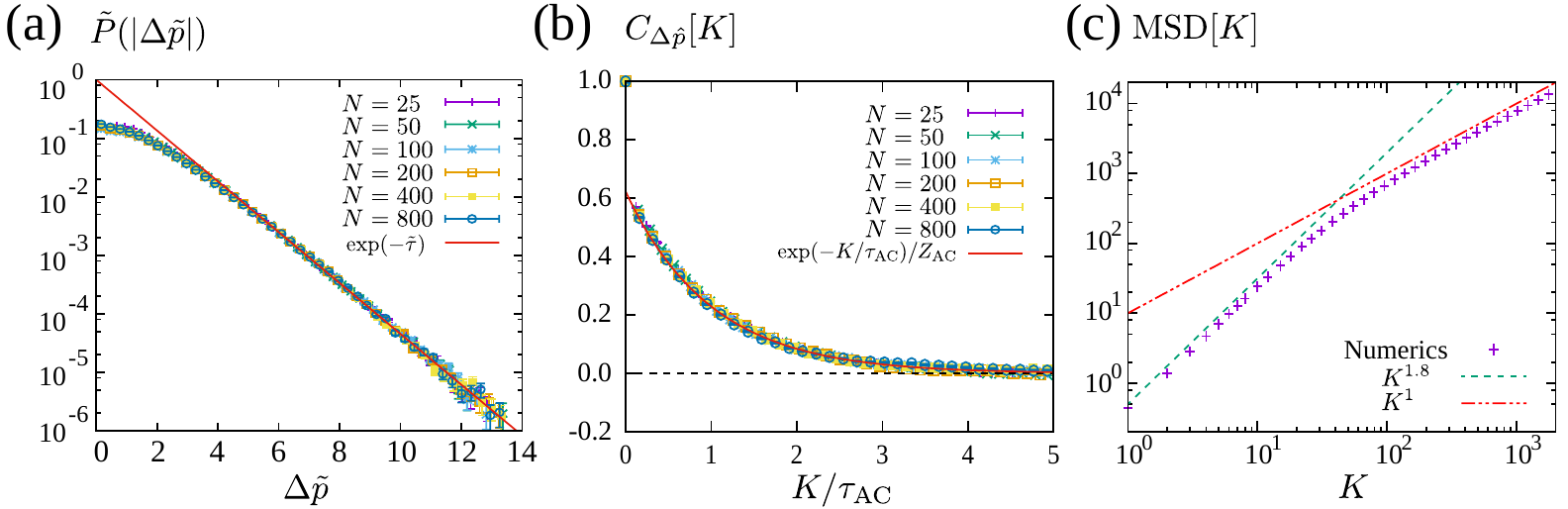}
					\caption	{
											Numerical study for the strong trend-following case $\tl{c}\gg 1$. 
											We adopted $(\tl{c}^*,\Delta \tl{p}^*)=(2.0,0.1)$ as the trend-following parameters. 
											(a)~Price movement distribution for various $N$ by scaling both horizontal and vertical axes, qualitatively showing the exponential law~{(\ref{eq:pricemovement_exp})}. 
											The least square method numerically estimates the decay length as $\kappa/c\tau^*=0.74$, $0.68$, $0.65$, $0.64$, $0.64$, and $0.64$ for $N=25$, $50$, $100$, $200$, $400$, and $800$, respectively.  
											(b)~Auto-correlation function $C_{\hDp}[K]$ with tick time $K$, showing the positive correlation with exponential decay. 
											The fitting parameters were estimated to be $Z_{\mathrm{AC}}=0.62$ and $\tau_{\mathrm{AC}}=16.4$, $14.2$, $13.1$, $12.8$, $12.2$, and $12.6$ for $N=25$,  $50$, $100$, $200$, $400$, and $800$, respectively. 
											(c)~Numerical MSD plot for $N=50$, showing a rapid diffusion of exponent $K^{1.8}$ (almost the ballistic motion of exponent $K^2$) for a short time and a normal diffusion of exponent $K^1$ for a long time. 
									}
					\label{fig:StrongTrend}
				\end{figure*}
				Numerical characters are studied here for the strong trend-following case under the parameter set $(\tl{c},\Delta \tl{p}^*)=(2.0,0.1)$. 
				We first study the price movement distribution $P(|\Delta p|)$. 
				In Fig.~{\ref{fig:StrongTrend}a}, the price movement distribution is plotted by scaling the horizontal and vertical axes, 
				\begin{equation}
					\tl{P}(|\Delta \tl{p}|) = \frac{\kappa P(|\Delta p|)}{Z_{\Delta \tl{p}}}\sim e^{-|\Delta \tl{p}|},\label{eq:Dp_ExpDist_Scaled}
				\end{equation}
				qualitatively showing the exponential tail for the scaled price movement $\Delta \tl{p}\equiv \Delta p/\kappa$. 
				Here the scaling parameters $\kappa$ and $Z_{\Delta \tl{p}}$ were determined by the least square method for the tail. 
				The mean-field solution~{(\ref{eq:MF_timeinterval})} and the improved mean-field solution~{(\ref{eq:imprvd_MF})} predicts $\kappa= c\tau^*$ and $\kappa= 2c\tau^*/3$, respectively.
				These theoretical predictions are qualitatively consistent with the numerical estimation $\kappa\approx 0.64c\tau^*$. 
				
				We next study the autocorrelation function $C_{\hDp}[K]$ of the price difference $\hDp$ based on tick time $K$ in Fig.~{\ref{fig:StrongTrend}b} by scaling the horizontal line. 
				For our parameter sets, the numerical result implies that the autocorrelation function can be written as 
				\begin{equation}
					C_{\hDp}[K] \approx 	\begin{cases}
																1 & (K=1)\\
																\frac{1}{Z_{\mathrm{AC}}}e^{-K/\tau_{\mathrm{AC}}}  & (K\geq 2)
															\end{cases}
				\end{equation}
				with fitting parameters $\tau_{\mathrm{AC}}$ and $Z_{\mathrm{AC}}$. 
				This autocorrelation suggests that the strong trend-following keeps unidirectional price movements for a certain time-interval. 
				Indeed, the probability of $\hDp[T]\hDp[T+1]>0$ is much higher than $50\%$ under this condition as shown in Table~\ref{table:sign_WT}. 
				In addition, the numerical MSD plot in Fig.~\ref{fig:StrongTrend}c shows the rapid diffusion (almost ballistic motion $K^2$) for a short time and the normal diffusion for a long time

		\subsubsection{Marginal case}
			The most complex case is the marginal case $\tl{c}\sim 1$, where both trend-following effect and zigzag noise contribute to the price movement as $|c\hat{\tau}[T]\tanh(\hDp[T]/\Dp^*)|\sim |\Delta\hxi[T]|$. 
			While both trend-following term $c\hat{\tau}\tanh(\hDp/\Delta p^*)$ and random noise term $\Delta \hxi$ are relevant on this condition, 
			the main contribution to the price movement tail originates from the trend-following term
			because the former yields the exponential tail while the latter yields the Gaussian tail. 
			We thus obtain the exponential tail~{(\ref{eq:pricemovement_exp})} for the price movement for the marginal case. 
			This theoretical conjecture is to be validated numerically below. 

			\paragraph*{Numerical comparison.}
			\begin{figure*}
				\centering
				\includegraphics[width=180mm]{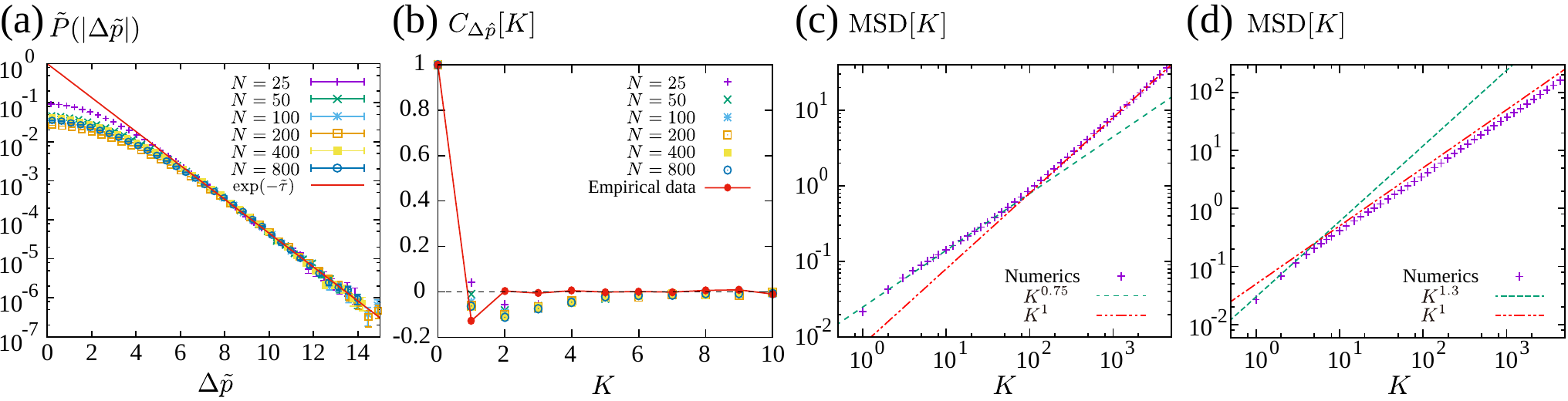}
				\caption	{
										Numerical study for the marginal case $\tl{c}\sim 1$. 
										(a)~The price movement distribution $\Delta p$ by scaling the horizontal and vertical axes. 
										(b)~Auto-correlation function $C_{\hDp}[K]$ based on tick time $K$ (points), showing the negative correlation around $K=1$. 
											This numerical result was consistent with the empirical result obtained from our dataset (solid line). 
										(c)~MSD plot under the parameters $(\tl{c},\Delta \tl{p}^*)=(0.5,2.5)$,
											showing a slightly slow diffusion. 
										(d)~MSD plot under the parameters $(\tl{c},\Delta \tl{p}^*)=(0.86,1.43)$,
											showing a slightly rapid diffusion with the Hurst exponent $H=0.65$. 
								}
				\label{fig:MarginalTrend}
			\end{figure*}	
				We studied the marginal case under the parameter set $(\tl{c},\Delta \tl{p}^*)=(0.5,2.5)$. 
				In Fig.~\ref{fig:MarginalTrend}a, we plot the price movement distribution by scaling both horizontal and vertical axes as Eq.~{(\ref{eq:Dp_ExpDist_Scaled})}. 
				We thus obtain the exponential-law tail~{(\ref{eq:Dp_ExpDist_Scaled})} for the price movement qualitatively. 
				
				In Fig.~\ref{fig:MarginalTrend}b, we also studied the autocorrelation function $C_{\hDp}[K]$ on tick time $K$ through both numerical simulation (points) and empirical data analysis (solid line) of the real time series. 
				This figure shows the slight negative correlation around $K=1$, which was qualitatively consistent with the empirical result in our dataset. 
				This result also implies that the price time series exhibits a slight zigzag behavior for a certain tick period. 
				This theoretical implication was validated by analyzing the probability of $\hDp[T]\hDp[T+1]<0$ as summarized in Table~\ref{table:sign_WT}. 
				The table~\ref{table:sign_WT} shows the quantitative consistency between the marginal trend-following case and the real price time series. 
				
				We also discuss the behavior of MSD in Fig.~\ref{fig:MarginalTrend}c and d, which shows both slow and rapid diffusions dependently on the parameters. 
				For example, we set the parameters $(\tl{c},\Delta \tl{p}^*)=(0.5,2.5)$ and $(\tl{c},\Delta \tl{p}^*)=(0.86,1.43)$ for Fig.~\ref{fig:MarginalTrend}c and d, respectively. 
				In Fig.~\ref{fig:MarginalTrend}c, the MSD plot exhibits a slightly slow diffusion for a short time and the normal diffusion for a long time. 
				In Fig.~\ref{fig:MarginalTrend}d, on the other hand, the MSD plot exhibits a slightly rapid diffusion with the Hurst exponent $H=0.65$ for a short time and the normal diffusion for a long time. 
				We thus conclude that our HFT model can reproduce a variety of diffusion by adjusting the trend-following parameters.

\section{Discussion}\label{sec:discussion}
	We here discuss implications of our theory to understand various topics intensively. 

		\subsection{Comparison with real dataset}
			\begin{table}[b]
				\centering
				\begin{tabular}{|l||c|c|c|c|}\hline 
					Case & $P(|\Delta p|)$ & $P(\tau)$ & $C_{\Delta p}[K]$ & Prob. of diff. sign \\ \hline \hline
					(a) Weak trend-following case & Gaussian & Exponential & Strongly negative at $K=1$ & around $60\%$ \\ \hline
					(b) Strong trend-following case & Exponential & Exponential & Strongly positive & less than $10\%$ \\ \hline 
					(c) Marginal trend-following case & Exponential & Exponential & Slightly negative around $K=1$ & around $52\%$\\ \hline
					(d) Empirical facts & Exponential & Exponential & Slightly negative around $K=1$ & around $52\%$\\ \hline
				\end{tabular}
				\caption	{
								Comparison between the empirical facts of the EBS market and our theoretical prediction. 
							}
				\label{table:comparison_summary}
			\end{table}
			Here we provide a detailed comparison between empirical facts and the above theoretical predictions as follows:
			As for the order-book profile $f_{\mathrm{A}}(r)$, the validity of the formula~{(\ref{eq:avg_order_book_2})} was examined by analyzing daily average order-book in Ref.~\cite{Kanazawa2017}. 
			The exponential-tail for time interval distribution $P(\tau)\sim e^{-\tau/\tau^*}$ was studied in Ref.~\cite{MTakayasu2002} by removing the non-stationary property of time series. 
			The price movement was reported to obey the exponential-law $P(|\Delta p|)\sim e^{-|\Delta p|/\kappa}$ in Ref.~\cite{Kanazawa2017} by removing the non-stationary property of time series. 
			The price time series tended to exhibit zigzag behaviors, which were reflected in 
			the negative autocorrelation function $C_{\Delta p}[K]$ around $K=1$ (see Fig.~\ref{fig:MarginalTrend}c) 
			and the probability of $\hDp[T]\hDp[T+1]<0$ (i.e., taking different signs) slightly over $50\%$ (see Table~\ref{table:sign_WT}). 
			All these characters are consistent with our theoretical prediction for the marginal trend-following case (see Table~\ref{table:comparison_summary} for the summary of the comparison). 
			The HFT model presented here can show precise agreements with these empirical facts. 
			Considering that the market was stable in our dataset, we concluded that our HFT model can describe the FX market well, at least during the stable period. 
			Description of unstable markets is out of scope of this paper and is a next interesting problem for future studies.

		\subsection{Validity of Mean-Field Approximation}\label{sec:DiscussionMeanField}
			We have numerically validated the mean-field theory. 
			The LO solution~{(\ref{eq:limit_triangular})} quantitatively describes the order-book profile~{(\ref{eq:MF-avg_orderbook})} with high precision 
			and the NLO solution~{(\ref{eq:app:subleading_order_FBE})} qualitatively describes the price movement~{(\ref{eq:Financial_Brownian_Motion})}. 
			Possible reasons are discussed here why the mean-field approximation works so well for the trend-following HFT model considering the common sense in physics.

			The mean-field approximation is expected invalid for low-dimensional physical systems because two-body correlations do not disappear between colliding pairs for a long time. 
			Colliding particles are not allowed to be separated far from each other because of the continuity of paths and the low-dimensional space geometry. 
			For one-dimensional Hamiltonian systems with hard-core interactions, for example, any particle successively collides against the fixed neighboring particles and two-body correlations then remain forever. 
			The mean-field approximation is therefore shown valid only for high-dimensional systems, at least for several concrete setups. 
			From this viewpoint, the precise agreement is not trivial between the mean-field solution~{(\ref{eq:MF-avg_orderbook})} and the numerical result. 

			In contrast, the continuity of the path is absent due to requotation jumps though our model is a one-dimensional system. 
			The transaction rule~{(\ref{eq:execution_rule})} compulsorily separates the transaction pairs after their collision,
			because of which there is no restriction on the combination of possible transaction pairs. 
			In the $N\to \infty$ limit, in addition, transactions between the same pair traders becomes rare (i.e., the probability of successive transaction between the same pair decays as the order of $N^{-2}$), 
			which implies quick disappearance of the two-body correlation between transaction pairs for $N\to \infty$. 
			This is our conjecture to validate the mean-field approximation for this model. 
			If this conjecture is correct, kinetic-like descriptions may be valid for various agent-based systems, 
			if agents are separated compulsorily to avoid successive interactions between the same pairs. 
			
			
		\subsection{Non-stationary property for price movements: power-law behavior}
			Financial markets are known to exhibit strong non-stationary properties statistically, such as the intraday activity patterns.  
			Here we discuss the impact of such non-stationary properties on the price movements and its relation to the celebrated power-law behavior for a long time. 
			
			Our theoretical model implies that the exponential law~{(\ref{eq:pricemovement_exp})} for the price movement as the basic statistical property. 
			This property was shown consistent with the real price movement in Ref.~\cite{Kanazawa2017} for a short time, by removing the non-stationary property in terms of the decay length $\kappa$. 
			The decay length $\kappa$ is related to the number of traders $N$ and the strength of trend-following $c$, both of which are expected to have non-stationary properties. 
			At least, indeed, the number of traders $N$ exhibits a trivial but strong non-stationary property with a correlation with the decay length. 
			
			\begin{figure}
				\centering
				\includegraphics[width=170mm]{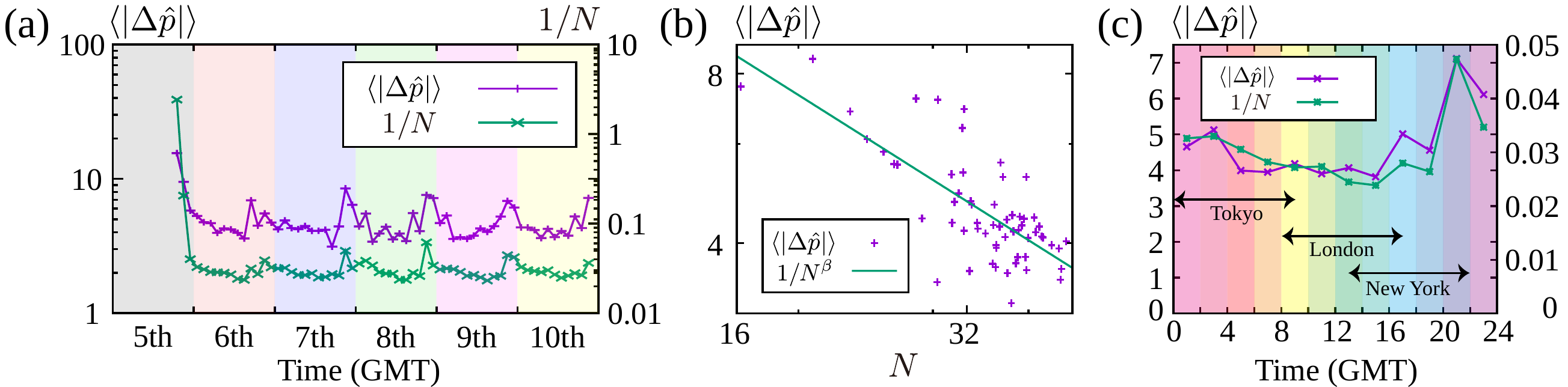}
				\caption	{
									(a)~Time series of the short-time market volatility $\la |\hDp|\ra$ and the inverse number of HFTs $1/N$. 
									Both $\la |\hDp|\ra$ and $1/N$ had a tendency to take large values during inactive hours of the EBS market, 
									such as (i) the time region just after the market opening (the 5th 18:00--22:00 GMT) and the end of the New York working hours (22:00--22:00 GMT). 
									This figure exhibits the correlation between $\la |\hDp|\ra$ and $1/N$ with Spearman's rank correlation coefficient of $0.63$. 
									(b)~Scattering plot between $\la |\hDp|\ra$ and $N$ in the log-log scales. 
									Regression analysis between $\log{\la |\hDp|\ra}$ and $\log{N}$ implies a power-law (almost linear) relation $\la |\hDp|\ra\propto 1/N^{\beta}$ with $\beta = 0.86\pm 0.1$. 
									In this analysis, we excluded the two samples after the market opening (the 5th 18:00--24:00 GMT) as outliers. 
									(c) Intraday patterns are studied for $\la |\hDp|\ra$ and $N$. 
									We took the averages of $\la |\hDp|\ra$ and $N$ conditional on two-hourly intraday time zones from Tuesday to Thursday, 
									with the arrow-type legends showing the working hours for Tokyo (0:00--9:00 GMT), London (8:00--17:00 GMT), and New York (13:00--22:00 GMT). 
									This figure shows that both $\la |\hDp|\ra$ and $1/N$ tended to take large values during the end of the New York working hours (22:00--22:00 GMT). 
								}
				\label{fig:DpVsNumHFTs}
			\end{figure}
			To illustrate this character, let us analyze the statistical relation between the mean absolute price movement $\la |\hDp|\ra$ and the number of HFTs $N$ in our dataset. 
			We measured $\la |\hDp|\ra$ as a representative of the market volatility for a short time and studied its correlation with $N$ every two hours in Fig.~\ref{fig:DpVsNumHFTs}a. 
			Spearman's rank correlation coefficient was $0.63$ between $\la |\hDp|\ra$ and $1/N$. 
			This result implies that the market volatility is relatively small when $N$ is large,
			which is qualitatively consistent with our theoretical prediction of $\la |\hDp|\ra \approx \kappa \sim 1/N^{\beta}$ 
			(e.g., $\beta=1$ if parameters are time-constant other than $N$). 
			The regression analysis between $\log{\la |\hDp|\ra}$ and $\log{N}$ implies $\beta = 0.86\pm 0.1$ as shown in Fig.~\ref{fig:DpVsNumHFTs}b. 
			We also note that both $\la \hDp\ra$ and $1/N$ had a tendency to become large during inactive hours of the EBS market (Fig.~\ref{fig:DpVsNumHFTs}c).
			
			The non-stationary property of the market volatility is related to the power-law behavior of the price movement for a long time. 
			In Ref.~\cite{Kanazawa2017}, the decay length $\kappa$ is shown to have a power-law distribution $P(\kappa)\propto \kappa^{-\alpha-1}$, 
			which implies the power-law price movement for a long time as the superposition of the short-time exponential distribution,
			\begin{equation}
				P_{\mathrm{long}}(\geq |\Delta p|) = \int d\kappa P(\kappa)P_{\mathrm{short}}(\geq |\Delta p|) \sim |\Delta p|^{-\alpha}
			\end{equation}
			with the complementary cumulative price movement distribution $P_{\mathrm{long}}(\geq |\Delta p|)$ and $P_{\mathrm{short}}(\geq |\Delta p|)\sim e^{-|\Delta p|/\kappa}$. 
			This result is consistent with previous empirical researches~\cite{Plerou1999,Lux1996,Guillaume1997,Longin1996}. 
			We thus concluded that both exponential law and power-law can consistently coexist at least in our dataset. 
	
			We here note that the FX market in our dataset was rather stable without any external shocks. 
			While the exponential-law was essential for a short time in our dataset, 
			we do not deny the possibility that the power-law may be essential even for a short time for unstable markets under external shocks. 
			We believe that that there would exist essentially different structures in unstable markets and 
			it would be interesting to study the statistics of traders' behavior in unstable markets under financial crisis for a future perspective. 

		\subsection{Non-stationary property for transaction interval: power-law behavior}
			As for the transaction interval, our theory predicts that the exponential-law~{(\ref{eq:MF_timeinterval})} is essential rather than the power-law. 
			This result is consistent with the previous report in Ref.~\cite{MTakayasu2002}, 
			showing that the exponential-law is essential for a short time but it superposition leads the power-law behavior of transaction interval for a long time.

		\subsection{Non-stationary property for order-book dynamics: stability of the order-book profile}
			We have discussed that both price movement and transaction interval are quite sensitive to non-stationary properties of the market. 
			On the other hand, the average order-book profile $f_{\mrA}(r)$ is relatively insensitive to such non-stationary properties, in contrast to the price movement and transaction interval. 
			Indeed, the average order-book profile $f_{\mrA}(r)$ is independent of the trend-following property $\tl{c}$.
			In addition, the order-book profile shows a convergence for $N\to \infty$, such that $\lim_{N\to \infty}f_{\mrA}(r)$ is an $L^2$-functions,
			which implies that large variation of $N$ does not have impact on the order-book profile. 
			
			Similar insensitivity does not exist for the price movement and transaction interval. 
			Indeed, they exhibit the strong divergence for $N\to \infty$ as $\lim_{N\to \infty}P(|\Delta p|)=\delta(|\Delta p|)$ and $\lim_{N\to \infty}P(\tau)=\delta(\tau)$,
			which implies the huge impact of large variation of $N$ on their statistics. 
			
			In this sense, the average order-book profile is a stable quantity to measure under non-stationary processes, whereas the price movement and transaction interval are unstable quantities. 
			Our theory provides the insight on the sensitivity of measured quantities to the non-stationary nature of the market.  
			We believe that developing systematic methods to remove such non-stationary nature is the key to understand not only the origin of power-laws in finance but also the essence of market microstructure. 
		
		\subsection{More is different: $N=2$ vs. $N\gg 1$}
			One of the most interesting features in statistical physics lies in the fact that many-body systems can exhibits essentially different characters from few-body systems, such as the critical phenomena and collective motion. 
			Though the current HFT model here does not exhibit critical phenomena, an essential difference can be shown between the cases of $N=2$ and $N\gg 1$.  
			To illustrate this point, let us consider the case of $\tl{c}=0$ without trend-following. 
			Our theory is applicable to solve the case of $N=2$ exactly, which leads the same solution presented in Ref.~\cite{Yamada2009}. 
			The price movement is then predicted to obey the exponential-law even without trend-following, which is qualitatively different from the Gaussian-law for $N\to \infty$. 
			This difference appears because the dynamics of the CM are not sufficiently slow for $N=2$. 
			 For $N=2$, indeed, one can show the absence of the zigzag noise term $\Delta \hxi[T]$ in the financial Langevin equation~{(\ref{eq:Financial_Brownian_Motion})},
			which leads the dominance of the random exponential noise $\hzeta[T]$. 
			For $N\gg 1$, on the other hand, the random noise $\hzeta[T]$ is negligibly small due to the slow CM dynamics, 
			and the trend-following effect becomes necessary to explain the exponential price movements statistics. 
			The model presented here thus exhibits essentially different characters as the number of traders increases. 
		
		\subsection{Does the trend-following effect break the random walk hypothesis?}
			Seemingly, the trend-following effect is strongly contradictory to the conventional assumption of the random walk hypothesis. 
			Our analysis however implies that the situation is not so simple: 
			In the absence of the trend-following, the market price exhibits the strong zigzag behavior, which is far from the pure random walks. 
			By adjusting the strength of trend-following appropriately (i.e., the marginal trend-following case), on the other hand, 
			the zigzag behavior is somewhat relieved and the market price time series rather approaches the random walks. 
			In this sense, the trend-following strategy might originate from the rational behavior of HFTs to equilibrate the strategies among traders. 
			It would be interesting to pursue the origin of trend-following behavior from economical viewpoints as future studies. 
			
			We also note that the real price time series exhibits slightly zigzag behaviors (i.e., the negative autocorrelation and the tendency for price movement to take different sign), which are consistent with our HFT model for the marginal trend-following case. 
			These different characters from the pure random walks have been well-known in finance and are obviously applicable to predict the direction of price movement in one-tick future. 
			It is not easy however to make profits over the market spread (i.e., the difference between the market best bid and ask prices) by utilizing only these properties. 
			While the real price time series slightly deviates from the pure random walks, it is not obvious whether these characters provide easy opportunities to statistically make profits. 
			Making profits requires us to predict price movements beyond the market spread, which is out of scope of this paper but is an interesting topic for a future study.  
		
		\subsection{Possible generalization 1: multiple-tick trend-following random walks and the PUCK model}
			In this manuscript, we have addressed the trend-following HFT model with one-tick memory. 
			It is straightforward to generalize the one-tick memory model toward a multiple-tick memory model, such that
			\begin{equation}
				\la \Delta \hz_i[T] \ra =c \tanh\frac{\hDp_{\EMA}[T-1]}{\Delta p^*}, \>\>\>\hDp_{\EMA}[T] \equiv \sum_{K=0}^\infty \frac{e^{-K/\tau_{\EMA}}}{Z_{\EMA}}\hDp[T-K],
			\end{equation}
			where $\hDp_{\EMA}[T]$ is the exponential moving average for the price movements $\{\hDp[T]\}_T$ with decay time $\tau_{\EMA}$ and renormalization constant $Z_{\EMA}\equiv 1/(1-e^{-1/\tau_{\EMA}})$. 
			In the authors' view, this model is more realistic because such an exponential moving average is a popular strategy among HFTs according to a detailed regression analysis for trend-following~\cite{Sueshige2018}.
			We then obtain a generalization of the financial Langevin equation as 
			\begin{equation}
				\hDp[T+1] =c\hat{\tau}[T]  \tanh \frac{\hDp_{\EMA}[T]}{\Delta p^*} + \Delta \hxi[T] + \hzeta[T]. \label{eq:generalizedFLangevin}
			\end{equation}
			
			The generalized financial Langevin equation~{(\ref{eq:generalizedFLangevin})} is equivalent to the potentials of unbalanced complex kinetics (PUCK) model~\cite{PUCK2006}, which was previously introduced by time-series data analyses. 
			Here we use an identity 
			\begin{equation}
				\hDp_{\EMA}[T] = \frac{e^{1/\tau_{\EMA}}}{Z_{\EMA}}\left\{\hp[T+1] - \hp_{\EMA}[T+1]\right\}, \>\>\>\hp_{\EMA}[T] \equiv \sum_{K=0}^\infty \frac{e^{-K/\tau_{\EMA}}}{Z_{\EMA}}\hp[T-K]
			\end{equation}
			for the exponential moving averages $\hDp_{\EMA}[T]$ and $\hp_{\EMA}[T]$, which leads the PUCK model
			\begin{equation}
				\hp[T+1]-\hp[T] =-\frac{\partial U(p)}{\partial p}\bigg|_{p = \hp[T]-\hp_{\EMA}[T]} + \Delta \hxi[T-1] + \hzeta[T-1]
			\end{equation}
			under a random potential $U(p) = - ce^{-1/\tau_{\EMA}}\Delta p^* Z_{\EMA}\hat{\tau}[T-1]\log \left\{\cosh (e^{1/\tau_{\EMA}}p/\Delta p^* Z_{\EMA})\right\}$. 
			In this sense, our theory is straightforwardly applicable to a derivation of the PUCK model.

		\subsection{Possible generalization 2: reduction to the random multiplicative processes}\label{sec:DiscussLinearTrend}
			In Sec.~\ref{sec:solution:Langevin}, we assume $\Delta \tl{p}^*\lesssim 1$ both for analytical simplicity and for consistency with the empirical report~\cite{Kanazawa2017}.  
			Here we discuss the case with $\Delta \tl{p}^*\gg 1$, whereby the hyperbolic trend-following reduces to the linear trend-following as $c\tanh(\hDp/\Dp^*) \approx c\hDp/\Dp^*$. 
			The financial Langevin equation~{(\ref{eq:Financial_Brownian_Motion})} is thus replaced with a linear financial Langevin equation
			\begin{equation}
				\hDp[T+1] = c\htau[T]\frac{\hDp[T]}{\Dp^*} + \Delta \hxi[T]+\hzeta[T].
			\end{equation}
			By introducing the second-order difference $\Delta^2\hp[T]\equiv \hDp[T+1]-\hDp[T]$, we obtain a similar equation to the conventional Langevin equation as 
			\begin{equation}
				\Delta^2 \hp[T] = -\hat \gamma[T]\hDp[T] + \Delta \hxi[T]+\hzeta[T]\label{eq:linearFinLangevin}
			\end{equation}
			with a random frictional coefficient $\hat \gamma[T]\equiv 1 - c\htau[T]/\Dp^*$, consistently with the simplified discussion in the supplementary material of Ref.~\cite{Kanazawa2017}. 
			Since Eq.~{(\ref{eq:linearFinLangevin})} belongs to the random multiplicative processes~\cite{TakayasuPRL1997}, 
			the price movement obeys the power-law statistics, consistently with the previous exact solution~\cite{Yamada2009} for the two-body case $N=2$. 
		
\section{Conclusion}\label{sec:conclusion}
		In this paper, we have presented a systematic solution for the trend-following trader model, which was empirically introduced in our previous work~\cite{Kanazawa2017}. 
		Starting from the microscopic dynamics of the individual traders, we have systematically reduced the multi-agent dynamics by generalizing the mathematical method developed in molecular kinetic theory. 
		We first introduce the phase space for our model and derive the dynamical equation for the phase space distribution function, 
		which corresponds to the Liouville equation in the conventional analytical mechanics. 
		On the basis of the Liouville equation for the trend-following trader model, we derive a hierarchy of reduced distributions in the parallel method to the BBGKY hierarchy.
		By introducing the mean-filed approximation, corresponding to the assumption of molecular chaos, we derive the mean-field dynamical equation for the one-body distribution function, similarly to the Boltzmann equation. 
		We then derive the analytic solution for the mean-field model, whose validity is numerically examined when the number of traders is sufficiently large. 
		We also derive the financial Langevin equation, governing the macroscopic dynamics of the financial Brownian motion, and study the macroscopic properties of the market price movements.  
		
		Here we have clarified the power of the kinetic frameworks in describing financial markets from microscopic dynamics.  
		In our conjecture, this success lies on the fact that the financial markets approximately satisfy the key assumptions of the binary interaction and molecular chaos (see Secs.~\ref{sec:Char_RealHFTs} and~\ref{sec:DiscussionMeanField} for related discussions); 
		the one-to-one transaction (i.e., the binary interaction) is the most basic interaction, and traders less likely transact with the same counterparty for $N\gg 1$. 
		We believe that the financial market is one of the best subjects to apply the kinetic theory, besides traffic flow and wealth distribution~\cite{Helbing,Nishinari,Prigogine,Pareschi}. 
		We also believe that generalization of kinetic theories would be a key to clarify various social systems from microscopic dynamics, since we have access to various microscopic data these days.

\begin{acknowledgements}
		We have greatly acknowledged to T. Ito, M. Katori, H. Hayakawa, M. Oshikawa, F. van Wijland, S. Ogawa, K. Yamada, T. Ooshida, D. Yanagisawa, S. Ichiki, K. Tamura, and J. Ozaki for fruitful discussions. 
		We also appreciate NEX for their provision of the EBS data on their foreign exchange platform. 
		This work was supported by Japan Society for the Promotion of Science KAKENHI (Grand No.~16K16016 and No.~17J10781) and Japan Science and Technology Agency, Strategic International Collaborative Research Program.
\end{acknowledgements}

\appendix

\section{Detailed Derivation of Financial Liouville Equation}\label{sec_app:Liouville_der}
		We here derive the financial Liouville equation for the trend-following trader model. 
		The dynamics of our model is given by 
		\begin{align}
			\frac{d\hz_i}{dt} = c\tanh \frac{\Delta p}{\Delta p^*} + \sigma \heta_{i;\ve}^{\mrR} + \heta_i^{\mrT}, \label{appeq:Dealer1}
		\end{align}
		where we have introduce the colored Gaussian noise $\heta_{i;\ve}^{\mrR}$ satisfying $\la\heta_{i;\ve}^{\mrR}\ra=0$ and $\la\heta_{i;\ve}^{\mrR}(t)\heta_{i;\ve}^{\mrR}(s)\ra=e^{-|t-s|/\ve}/2\ve$.
		For the mathematical convenience below, we finally take the white noise limit $\ve\to +0$: $\lim_{\ve\to0}\heta_{i;\ve}^{\mrR}=\heta_{i}^{\mrR}$. 
		We next consider the dynamics of the center of the mass $\bz$: 
		\begin{align}
			\frac{d\hzCM}{dt} = c\tanh \frac{\Delta p}{\Delta p^*} + \bar{\eta}, \>\>\>
			\bar{\eta} \equiv \frac{\sigma}{N}\sum_{i=1}^N \heta_{i;\ve}^{\mrR} + \frac{1}{N}\sum_{i,j}^{i<j}\sum_{k=1}^\infty (\Delta z_{ij}+\Delta z_{ji})\delta(t-\htau_{k;ij}).
		\end{align}
		
		Let next us consider the dynamics of an arbitrary function $f(\hGamma)$ for $\hGamma\equiv (\hz_1,\dots,\hz_N;\hzCM,\hp,\hDp)\in \mathcal{S}$. 
		The time-evolution of $f(\hGamma)$ is governed by the continuous movement by the continuous noise term $\heta_{i;\ve}^{\mrR}$ and the discontinuous jumps by the deterministic transaction term $\heta_i^{\mrT}$. 
		We then obtain
		\begin{align}
			\frac{df(\hGamma)}{dt} =& \sum_{i=1}^N\frac{\partial f}{\partial \hr_i} \left\{ c\tanh \frac{\hDp}{\Delta p^*} + \sigma\heta_{i;\ve}^{\mrR}\right\} 
					+ \frac{\partial f}{\partial \hzCM}\left\{ c\tanh \frac{\hDp}{\Delta p^*} + \frac{\sigma}{N}\sum_{i=1}^N\heta^{\mrR}_{i;\ve}\right\} \notag\\
			&+ \sum_{k=1}^\infty \sum_{i,j}^{i<j}[f(\hGamma+\Delta \hGamma_{ij})-f(\hGamma)]\delta(t-\htau_{k;ij})\label{appeq:PLEq:trans1}
		\end{align}
		where we have introduced the difference vector $\Delta \hGamma_{ij}$ induced by transactions defined by
		\begin{align}
			\Delta \hGamma_{ij}\equiv \left(0,\dots,0, \overbrace{-\frac{L_i}{2}\sgn(\hz_i-\hz_j)}^{i\rm{th}},0,\dots, 0, \overbrace{-\frac{L_j}{2}\sgn(\hz_j-\hz_i)}^{j\rm{th}},0,\dots, 0; -\frac{L_i-L_j}{2N}\sgn(\hz_i-\hz_j), \hp_{ij}^{\pst} - \hp, \hDp^{\pst}_{ij}-\hDp\right)
		\end{align}
		with $\hp_{ij}^{\pst}\equiv \hz_i-(L_i/2)\sgn(\hz_i-\hz_j)$ and $\Dp_{ij}^{\pst}\equiv \hp_{ij}^{\pst}-\hp$. 
		Let us decompose the sum of $\delta$-functions here as 
		\begin{align}
			\sum_{i,j}^{i<j}\sum_{k=1}^\infty \delta(t-\htau_{k;ij}) &= \sum_{i,j}^{i<j}\left[
					\sigma\delta\left(\hz_i-\hz_j-\frac{L_i+L_j}{2}\right)(\heta_{i;\ve}^{\mrR}-\heta_{j;\ve}^{\mrR}) - \sigma\delta\left(\hz_i-\hz_j+\frac{L_i+L_j}{2}\right)(\heta_{i;\ve}^{\mrR}-\heta_{j;\ve}^{\mrR})\right]\notag\\
			&= \sum_{i,j}\sigma\delta\left(\hz_i-\hz_j-\frac{L_i+L_j}{2}\right)(\heta_{i;\ve}^{\mrR}-\heta_{j;\ve}^{\mrR}). 
		\end{align}
		where we have used $\heta_{i;\ve}^{\mrR}-\heta_{j;\ve}^{\mrR}>0$ just before $\hr_i-\hr_j-(L_i+L_j)/2=0$ (or equivalently, $\heta_{i;\ve}^{\mrR}-\heta_{j;\ve}^{\mrR}<0$ just before $\hr_i-\hr_j+(L_i+L_j)/2=0$)
		by taking collision directions into account. 
		We then take the ensemble average of both hand side of Eq.~{(\ref{appeq:PLEq:trans1})} with the aid of the Novikov's theorem~\cite{Novikov} for an arbitrary functional $g[\heta_{i;\ve}^{\mrR}]$
		\begin{equation}
			\la \heta_{i;\ve}^{\mrR}(t) g[\heta_{i;\ve}^{\mrR}] \ra = \int_0^t ds\la \heta_{i;\ve}^{\mrR}(t)\heta_{i;\ve}^{\mrR}(s)\ra \left<\frac{\delta g[\eta_{i;\ve}^{\mrR}]}{\delta \eta_{i;\ve}^{\mrR}(s)}\right> 
		\end{equation}
		for the colored Gaussian noise $\heta_{i;\ve}^{\mrR}$. 
		Here we remark the following two important relations for the $\delta$-function for the phase space $\delta (\hGamma-\bGamma)\equiv \delta(\hzCM-\zCM)\delta(\hp-p)\delta(\hDp-\Delta p)\prod_{i=1}^N\delta(\hz_i-z_i)$: 
		\begin{align}
			 &\lim_{\ve\to 0}\la\delta (\hGamma-\bGamma)\delta(\hz_i-\hz_j-(L_i+L_j)/2)\heta_{i;\ve}^{\mrR}\ra =\delta(z_i-z_j-(L_i+L_j)/2)\lim_{\ve\to 0}\la\delta (\hGamma-\bGamma)\heta_{i;\ve}^{\mrR}\ra \notag\\
			=& \frac{\sigma}{2}\delta(z_i-z_j-(L_i+L_j)/2) \left< \left[\frac{\partial }{\partial \hz_i}+\frac{1}{N}\frac{\partial }{\partial \hzCM}\right]\delta (\hGamma-\bGamma)\right>
			=-\frac{\sigma}{2}\delta(z_i-z_j-(L_i+L_j)/2) \left[\partial_{i}+\frac{1}{N}\partial_{\CM}\right]P_t(\bGamma)
		\end{align}
		and 
		\begin{align}
			 &\lim_{\ve\to 0}\la\delta (\hGamma+\Delta \hGamma_{ij}-\bGamma)\delta(\hz_i-\hz_j-(L_i+L_j)/2)\heta_{i;\ve}^{\mrR}\ra =\delta(z_i-z_j)\lim_{\ve\to 0}\la\delta (\hGamma+\Delta \hGamma_{ij}-\bGamma)\heta_{i;\ve}^{\mrR}\ra \notag\\
			=& -\frac{\sigma}{2}\delta(z_i-z_j)\delta(p-z_i)\left[\partial_i+\frac{1}{N}\partial_{\CM}\right] \int d\Delta p' P_t (\bGamma-\Delta \bGamma'_{ij}) 
		\end{align}
		with the dummy variable
		\begin{equation}
			\Delta \bGamma'_{ij} \equiv \left(0,\dots,-\frac{L_i}{2},\dots,+\frac{L_j}{2},\dots,0;-\frac{L_i-L_j}{2N},\Dp,\Dp-\Dp'\right).
		\end{equation}
		By substituting $f(\hGamma)=\delta(\hGamma-\bGamma)$, we take the ensemble average for both hand-sides of Eq.~{(\ref{appeq:PLEq:trans1})} in the $\ve\to 0$ limit.
		We then obtain
		\begin{align}
			\frac{\partial P_t(\bGamma)}{\partial t} &= 
				\sum_{i=1}^N\left[-c\tanh \frac{\Delta p}{\Delta p^*} \left\{\partial_i + \frac{1}{N}\partial_{\CM}\right\}
				+\frac{\sigma^{2}}{2}\left\{\partial_i +\frac{1}{N}\partial_{\CM}\right\}^2\right] P_t(\bGamma)\notag\\
			&+ \sum_{i,j}\frac{\sigma^{2}}{2}\left\{-\delta(z_i-z_j)\delta(p-z_i)\int d\Dp'\tpartial_{ij}P_t(\bGamma-\Delta \bGamma'_{ij}) + \delta\left(z_i-z_j-\frac{L_i+L_j}{2}\right)\tpartial_{ij}P_t(\bGamma)\right\}\label{app:eq:BBGKY_trans}
		\end{align}
		with an abbreviation symbol $\tpartial_{ij}\equiv \partial_i-\partial_j$. 
		Here, let us pay attention to the signature of the derivatives. 
		Considering $P(\bGamma) \geq 0$ for all $\bGamma$ and $P(\bGamma) = 0$ for $z_i-z_j>(L_i+L_j)/2$,
		we obtain the signature of derivatives
		\begin{equation}
			\partial_{i}P_t(\bGamma)\bigg|_{z_i-z_j=(L_i+L_j)/2} \leq 0, \>\>\> \partial_{j}P_t(\bGamma)\bigg|_{z_i-z_j=(L_i+L_j)/2} \geq 0.
		\end{equation}
		Equation~{(\ref{app:eq:BBGKY_trans})} can be simplified into Eq.~{(\ref{eq:pseudo_Liouville})} in terms of signatures 
		by introducing the symmetric absolute derivative
		\begin{equation}
		 |\tpartial_{ij}|P_t(\bGamma) \equiv \left|\partial_{i}P_t(\bGamma)\right| + \left|\partial_{j}P_t(\bGamma)\right|.
		\end{equation}
		Note that Eq.~{(\ref{eq:pseudo_Liouville})} is a partial integro-differential equation because of the transaction jumps, though the conventional Liouville equation is a partial differential equation. 
		This implies that our financial Liouville equation~{(\ref{eq:pseudo_Liouville})} technically corresponds to the pseudo-Liouville equation~\cite{Resibois1977,Ernst1969,Beijeren1979,KanazawaTheses} rather than the Liouville equation. 
		
\section{Detailed Derivation of Financial BBGKY Hierarchy}\label{sec_app:BBGKY_der}
		We here derive the lowest BBGKY hierarchal equation for the reduced distribution function~{(\ref{eq:BBGKY_main})}, starting from the financial Liouville equation~{(\ref{eq:pseudo_Liouville})}. 
		We first introduce the relative price from the CM as $r_i\equiv z_i-z_{\CM}$. 
		By making transformation $\bGamma = (z_1,\dots,z_N;z_{\CM},p,\Dp) \to \bGamma_r\equiv (r_1,\dots,r_N;z_{\CM},p,\Dp)$,
		the financial Liouville equation can be rewritten as 
		\begin{align}
			\frac{\partial P_t(\bGamma_r)}{\partial t} &= \left[-c \tanh \frac{\Dp}{\Dp^*}\partial_{\CM}+\frac{\sigma^2}{2}\sum_{i=1}^N\left\{ \partial_i+ \frac{1}{N}\left(\partial_{\CM} - \sum_{k=1}^N\partial_k\right)\right\}^2\right]P_t(\bGamma_r)\\
			&+ \sum_{i,j}\frac{\sigma^{2}}{2}\left\{\delta(r_i-r_j)\delta(p-r_i-z_{\CM})\int d\Dp'|\tpartial_{ij}|P_t(\bGamma_r-\Delta \bGamma'_{ij;r}) - \delta\left(r_i-r_j-\frac{L_i+L_j}{2}\right) |\tpartial_{ij}|P_t(\bGamma_r)\right\},\notag
		\end{align}
		where we have used the chain rule for the variable transformation:
		\begin{equation}
			\frac{\partial }{\partial z_i} \to \frac{\partial }{\partial r_i}, \>\>\>\frac{\partial }{\partial z_{\CM}} \to \frac{\partial }{\partial z_{\CM}} - \sum_{k=1}^N\frac{\partial }{\partial r_i}.
		\end{equation}
		We have also introduced $\Delta \bGamma'_{ij;r} = \Delta \bGamma^{\prime (0)}_{ij;r} + \Delta \bGamma_{ij;r}^{\prime (1)}$ with 
		\begin{equation}
			\Delta \bGamma_{ij;r}^{\prime (0)} \equiv \left(0,\dots,-\frac{L_i}{2},\dots,+\frac{L_j}{2},\dots,0;0,\Dp,\Dp-\Dp'\right), \>\>\>
			\Delta \bGamma_{ij;r}^{\prime (1)} \equiv \frac{L_i-L_j}{2N}\left(+1,\dots, +1;-1,0,0\right). 
		\end{equation}
		According to the definition of the one-body, two-body, and three-body reduced distributions~{(\ref{eq:def:123bodydists})}, the lowest-order hierarchy is then derived as 
		\begin{align}
		\frac{\partial P_t^i(r_i)}{\partial t} &= \frac{\tilde{\sigma}^{2}}{2}\frac{\partial^2 P_t^{i}(r_i)}{\partial r_i^2} 
			+ \sum_{s=\pm 1}\sum_{j\neq i}\frac{\sigma^{2}}{2}\left[|\tpartial_{ij}|P^{ij}_t(r_i-\Delta r_{ij;s},r_j+\Delta r_{ji;s})\big|_{r_i=r_j} - |\tpartial_{ij}|P_t^{ij}(r_i,r_j)\big|_{r_i-r_j=s(L_i+L_j)/2}\right] \notag\\
		&+\sum_{s\pm 1} \sum_{j,k\neq i}\frac{\sigma^2}{2}\int dr_j  \left[|\tpartial_{jk}|P^{ijk}_t\left(r_i -\Delta r_{jk;s}^{(1)},r_j,r_k\right)
			- |\tpartial_{jk}|P_t^{ijk}(r_i,r_j,r_k)\right]\bigg|_{r_j-r_k=s(L_j+L_k)/2}.\label{eq:BBGKY}
		\end{align}
		with effective variance $\tilde{\sigma}^{2} \equiv \sigma^{2}(1-1/N)$ and jump size $\Delta r_{ij;s}\equiv \Delta r_{ij;s}^{(0)}+\Delta r_{ij;s}^{(1)}$ with 
		$\Delta r_{ij;s}^{(0)} \equiv -sL_i/2$, $\Delta r_{ij;s}^{(1)} \equiv s(L_i-L_j)/2N$. 
		Equation~{(\ref{eq:BBGKY_main})} is thus derived from Eq.~{(\ref{eq:BBGKY})} by introducing Liouville operators. 

\section{Detailed Derivation of Financial Boltzmann Equation~{(\ref{eq:BoltzmannEq})}}\label{sec:app:FBE_der}
		In this Appendix, we derive the financial Boltzmann equation~{(\ref{eq:BoltzmannEq})} from the financial BBGKY hierarchy~{(\ref{eq:BBGKY_main})}. 
		To simplify the hierarchal equation~{(\ref{eq:BBGKY_main})}, we use the symmetry among the traders in terms of the spread: 
		when the the spreads are equal for both $i$th and $j$th traders, their one-body distributions are also equal, namely,
		\begin{equation}
			L_i = L_j \Longrightarrow P^i_t(r) = P^j_t(r).
		\end{equation}
		Furthermore, there are also symmetries for the two-body and three-body distributions such that
		\begin{equation}
			L_i=L_k, \>\>\> L_j=L_l \Longrightarrow P^{ij}_t(r,r',r'') = P^{kl}_t(r,r',r'') 
		\end{equation}
		and
		\begin{equation}
			L_i=L_l, \>\>\> L_j=L_m, \>\>\> L_k=L_n \Longrightarrow P^{ijk}_t(r,r',r'') = P^{lmn}_t(r,r',r''). 
		\end{equation}
		On the basis of these symmetries, we introduce the conditional distributions on spreads.
		We denote the minimum and the maximum spreads among traders by $L_{\min}$ and $L_{\max}$, respectively. 
		We also assume that the number of the traders is so large that we can approximately regard spreads as continuously distributed. 
		In other words, the spread distribution $\rho_L \equiv \sum_{i=1}^N \delta (L-L_i)/N$ is an approximately continuous function. 
		We assume that $\rho_L=0$ for $L\not \in [L_{\min},L_{\max}]$. 
		The one-body and two-body distributions $\phi^L_t(r)$ and $\phi^{LL'}_t(r,r')$ are defined conditional on spreads $L$ and $L'$ by 
		\begin{equation}
			\phi^{L_i}_t(r)\equiv P_t^i(r), \>\>\> \phi^{L_iL_j}_t(r,r')\equiv P_t^{ij}(r,r'). 
		\end{equation}
		Here we make the following approximations for $N\to \infty$:
		\begin{enumerate}
			\item 	The effective variance $\tilde{\sigma}^{2} = \sigma^2(1-1/N)$ is approximated as the 
					\begin{equation}
						\tilde{\sigma}^{2} \approx \sigma^2.
					\end{equation}
			\item 	The discrete sum is approximated as continuous integrals:
					\begin{equation}
						\sum_{j}(\dots) \approx N\int_{L_{\min}}^{L_{\max}} dL'\rho_{L'}(\dots), \>\>\>
						\sum_{j,k}(\dots) \approx N^2 \int_{L_{\min}}^{L_{\max}} dL'dL''\rho(L') \rho(L'')(\dots).
					\end{equation}
			\item 	The relatively small displacement $\Delta r_{ij;s}^{(1)}$ is negligible:
					\begin{equation}
						\left|\Delta r_{ij;s}^{(1)}\right| \ll \left|\Delta r_{ij;s}^{(0)}\right|.\label{app:eq:assumption_Dr1_ignore}
					\end{equation}
		\end{enumerate}
		On the basis of these approximations, the lowest hierarchal equation~{(\ref{eq:BBGKY_main})} can be rewritten as
		\begin{align}
			\frac{\partial \phi^L_t(r)}{\partial t} \approx \frac{\sigma^2}{2}\frac{\partial^2 \phi^L_t(r)}{\partial r^2} 
			+ N\sum_{s=\pm 1}\int dL'\rho_{L'}\frac{\sigma^{2}}{2}\left[|\tpartial_{rr'}|\phi^{LL'}_t\left(r-\frac{sL}{2},r'+\frac{sL'}{2}\right)\bigg|_{r=r'} - |\tpartial_{rr'}|\phi^{LL'}_t(r,r')\big|_{r-r'=s(L+L')/2}\right], \label{eq:BBGKY2_app}
		\end{align}
		where correction terms of $O(N^{-1})$ are ignored. 
		We thus have obtained Eq.~{(\ref{eq:BBGKY2_app})}, which was derived in our previous paper~\cite{Kanazawa2017} by a heuristic argument. 
		The financial Boltzmann equation~{(\ref{eq:BoltzmannEq})} is then derived by making the mean-field approximation~{(\ref{eq:mlclr_chaos_MF_BE})} to Eq.~{(\ref{eq:BBGKY_main})}. 
		Note that the three-body correlation terms in Eq.~{(\ref{eq:BBGKY_main})} is finally irrelevant under these assumptions. 
		The consistency of this assumption is examined using the NLO solution of Eq.~{(\ref{eq:BoltzmannEq})} in Appendix.~{\ref{sec:app:revisited_consistency}}. 
		
\section{Boundary Condition}\label{sec_app:Boundary_Condition}
		We here note the boundary condition for the financial Boltzmann equation~{(\ref{eq:BoltzmannEq})}. 
		We introduce the cutoff for the boundaries at $r=\pm L_{\mrCUT}/2$ and assume the following four assumptions: 
		\begin{enumerate}
			\item Equation~{(\ref{eq:BoltzmannEq})} is valid only for $r\in [-L_{\mrCUT}/2,L_{\mrCUT}/2]$.
			\item The cutoff is taken sufficiently large: $L_{\mrCUT} > L_{\max}$.
			\item The probability is zero beyond the boundary: $\phi_t^L(r,t) = 0 \>\>\>\mathrm{for}\>\>\> r \not\in [-L_{\mrCUT}/2,L_{\mrCUT}/2].$
			\item The boundaries are the reflecting barriers, which ensure the conservation of the probability: 
					\begin{equation}
						\frac{\partial \phi_t^L(r,t)}{\partial r}\bigg|_{r=\pm L_{\mrCUT}/2} = 0 \Longrightarrow \frac{\partial }{\partial t}\int_{-L_{\mrCUT}/2}^{L_{\mrCUT}/2}dr\phi_t^L(r,t) = 0.\label{eq:Boundary_Condition}
					\end{equation}
		\end{enumerate}
		The probability conservation~{(\ref{eq:Boundary_Condition})} can be mathematically proved in Appendix.~\ref{sec_app:Prob_Conserve} under this boundary condition. 
		The cutoff parameter is finally taken infinity as $L_{\mrCUT}\to \infty$, and the main results in this paper do not depend on $L_{\mrCUT}$. 
		
		We here also note another related technical issue for the mean-field solution~{(\ref{eq:limit_triangular})}. 
		The large number limit $N\to \infty$ is taken before the limit for the boundary $L_{\mrCUT}\to \infty$. 
		We also note that, when the limit for the maximum spread $L_{\max}\to \infty$ is taken, the $L_{\max}\to \infty$ limit is taken in the last order in this paper to conserve the second assumption $L_{\mrCUT} > L_{\max}$. 
		Equation~{(\ref{eq:limit_triangular})} therefore technically implies  
		\begin{equation}
			\psi^L(r) \equiv \lim_{t\to \infty}\lim_{L_{\max}\to \infty}\lim_{L_{\mrCUT}\to \infty}\lim_{N\to \infty}\phi_t^L(r).
		\end{equation}

\section{Next-to-leading-order solution to the financial Boltzmann equation}\label{sec:app:subleading}
	\begin{figure}
		\centering
		\includegraphics[width=80mm]{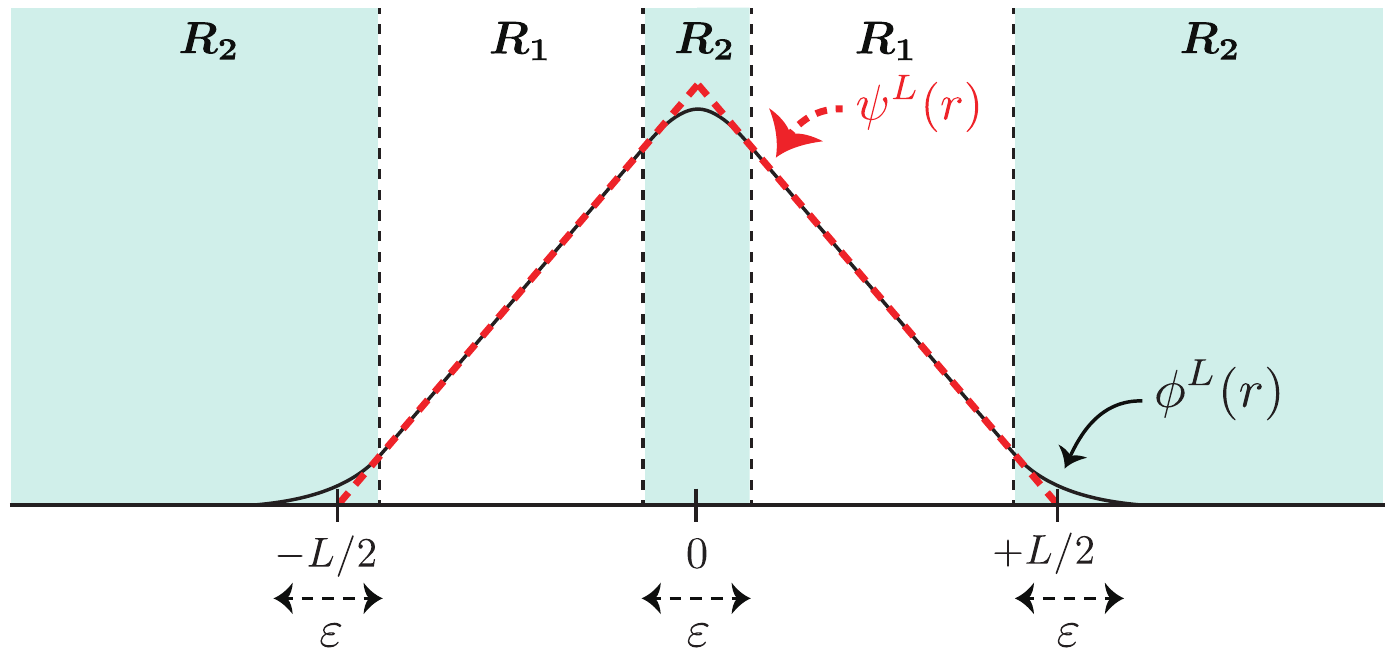}
		\caption	{
							Schematic of the NLO solution~{(\ref{eq:app:subleading_order_FBE})}. 
							There is a small deviation from the LO solution around the boundary layer $\bm{R}_2$ because of the finite number effect for $N$. 
							The deviation is studied within the NLO approximation for the financial Boltzmann equation~{(\ref{eq:BoltzmannEq})}. 
						}
		\label{fig:app:regimes}
	\end{figure}
	The LO solution for the financial Boltzmann equation~{(\ref{eq:BoltzmannEq})} is given by the tent function~{(\ref{eq:limit_triangular})}.
	Here we consider the NLO solution in the steady state, where the edges of tent functions are smooth because of the finite-number effect of traders. 
	We first derive the NLO solution by an intuitive asymptotic analysis, and will check that the solution satisfies the original financial Boltzmann equation~{(\ref{eq:BoltzmannEq})} by direct substitution. 
	We make the following three ansatzs (see Fig.~{\ref{fig:app:regimes}} as a schematic): 
	(i) There are two domains $\bm{R}_1\equiv (\ve/2,L/2-\ve/2)\cup (-\ve,-L/2+\ve/2)$ and $\bm{R}_2\equiv [-\infty,\infty]/\bm{R}_1$. 
	Here $\ve$ is the thickness of the boundary layers originating from the finite-number effect, with order $N^{-1/2}$ as will be shown later. 
	(ii) Out of the boundary layers $r\in \bm{R}_1$, the deviation from the LO solution $\psi^L(r)$ is negligible. 
	(iii) In the boundary layers $r\in \bm{R}_2$, the deviation from the LO solution $\psi^L(r)$ is not negligible.
	On the basis of these ansatzs, for $r>L/2$, we approximate 
	\begin{equation}
		\phi^{L'}(r')|_{r-r'=(L+L')/2} \simeq \frac{r'-L'/2}{L^{'2}/4}
	\end{equation}
	The financial Boltzmann equation~{(\ref{eq:BoltzmannEq})} is then approximated for $r > +L/2$ as
	\begin{align}
		0 &\simeq \frac{\sigma^{2}}{2}\frac{\partial^2 {\phi}^L(r)}{\partial r^2} - \frac{N\sigma^{2}}{2}\int_{L_{\min}}^{L_{\max}}dL'\rho_{L'}|\tpartial_{rr'}|\left\{{\phi}^L(r)\frac{r'-L'/2}{L'^2/4}\right\}\bigg|_{r-r'=(L+L')/2}\notag\\
		\Longleftrightarrow 
		0 &= \frac{\partial^2 \tl{\phi}^L(\tr)}{\partial \tr^2} + \frac{\tr}{L^{2}_\rho}  \frac{\partial \tl{\phi}^L(\tr)}{\partial \tr} - \frac{1}{L^{2}_\rho} \tl{\phi}^{L}(r)
	\end{align}
	where we have ignored the inflow flux $\tilde{J}^{LL'}(r+L/2)\propto \phi^L(r+L/2)$ on the basis of the ansatz and 
	we have introduced $\tr\equiv 2\sqrt{N}(r-L/2)$, $1/L_{\rho}^{*2} \equiv \int_{L_{\min}}^{L_{\max}}dL'\rho_{L'}/L'^{2}$, and $\tl{\phi}^L(\tr)\equiv \phi^L(L/2+\tr/2\sqrt{N})/2\sqrt{N}$. 
	The general solution around $\tr \sim 0$ is given by
	\begin{equation}
		\tl{\phi}^L(\tr) = C_1\tr + C_2 \mathcal{F}(\tilde{r}),\>\>\>\mathcal{F}(\tilde{r})\equiv \frac{\tr}{2}\erfc\left(\frac{\tr}{\sqrt{2}L_\rho^*}\right)-\frac{e^{-\tr^2/2L^{*2}_{\rho}}}{\sqrt{2\pi/L^{*2}_{\rho}}}
	\end{equation}
	with arbitrary coefficients $C_1$ and $C_2$.  
	Under the boundary condition $\lim_{\tr \to \infty}\phi^L(\tr)=0$, we obtain $C_1=0$.
	Considering the asymptotic relation $\tl{\phi}^L(\tr)\sim C_2\tr$ for $\tr\to-\infty$, 
	we obtain $C_2=-1/NL^2$ for the asymptotic connection to the LO solution in $\bm{R}_{1}$. 
	We thus obtain the NLO solution around the boundary $|r|\sim L/2$,
	\begin{equation}
		\phi^L(r) = \frac{1}{L^2/4}\left[\frac{e^{-N(|r|-L/2)^2/L^{*2}_{\rho}}}{2\sqrt{2N\pi/L^{*2}_{\rho}}}-\frac{|r|-L/2}{2}\erfc\left(\frac{\sqrt{2N}(|r|-L/2)}{L_\rho^*}\right)\right],\label{eq:app:subleading_order_FBE}
	\end{equation}
	which is consistent with the LO solution~{(\ref{eq:limit_triangular})} for $N\to \infty$: $\lim_{N\to \infty}\phi^L(r)=\psi^L(r)$. 

	We have obtained the NLO solution~{(\ref{eq:app:subleading_order_FBE})} rather intuitively,
	but we can check that the solution satisfies the original Boltzmann equation~{(\ref{eq:BoltzmannEq})} up to the order of $N^{-1/2}$ by direct substitution. 
	Around $r\sim L/2$, indeed, we obtain
	\begin{align}
		&\frac{\sigma^2}{2}\frac{\partial^2 \phi^L(r)}{\partial r^2} - \frac{N\sigma^2}{2}\int dL'\rho_{L'}|\tpartial_{rr'}|\phi^L(r)\phi^{L'}(r')\bigg|_{r-r'=(L+L')/2}\notag\\
		=&\frac{\sigma^2}{2}\left[-\frac{8\sqrt{N}}{L^2}\frac{\partial^2 \mathcal{F}(\tr)}{\partial \tr^2} - \frac{8\sqrt{N}}{L^2}\int \frac{dL'\rho_{L'}}{L^{\prime 2}}|\tpartial_{\tr\tr'}|\mathcal{F}(\tr)\mathcal{F}(\tr')\bigg|_{\tr'=-\tr}\right]\notag\\
		=&-\frac{4\sqrt{N}\sigma^2}{L^2}\left[\frac{\partial^2 \mathcal{F}(\tr)}{\partial \tr^2} + \frac{1}{L^{*2}_{\rho}}|\tpartial_{\tr\tr'}|\mathcal{F}(\tr)\mathcal{F}(\tr')\bigg|_{\tr'=-\tr}\right] = 0,
	\end{align}
	where we have ignored the inflow $\tilde{J}^{LL'}(r+L/2)\propto \phi^L(r+L/2)=O(\exp(-NL^2/4L^{*2}_\rho))$ around $r\sim L/2$. 
	This implies that the solution~{(\ref{eq:app:subleading_order_FBE})} satisfies the financial Boltzmann equation~{(\ref{eq:BoltzmannEq})} directly. 
	We also note that the NLO correction is the order of $N^{-1/2}$ and is consistent with the assumptions in Appendix.~{\ref{sec:app:FBE_der}},
	where correction terms of $O(N^{-1})$ are ignored for the derivation of Eq.~{(\ref{eq:BBGKY2_app})}. 

\section{Numerical simulation of the microscopic model}\label{sec:app:MonteCarlo}
		Here we explain the numerical implementation of the trend-following HFT model. 
		We focused on two type of buy-sell spread distributions given by the $\delta$-distributed spread~{(\ref{eq:avg_order_book_1})} and the $\gamma$-distributed spread~{(\ref{eq:avg_order_book_2})}. 
		The length and time units of this system are taken by $L^*$ and $L^{*2}/(\sigma^2N)$, respectively. 
		We performed the Monte Carlo simulation for various number of traders $N$ and trend-following parameters $(\tl{c},\Delta \tl{p}^*)$ under a fixed discretization time $\Delta t= 0.01 L^{*2}/(\sigma^2N)$. 
		For initialization, we first run the simulation for the time interval of $10L^{*2}/\sigma^2$ and then run the simulation again to take samples. 
		The simulation time was set to be $10^5$ ticks except for the MSD plots in Fig.~{\ref{fig:WeakTrend}c,d}, Fig.~{\ref{fig:StrongTrend}d}, and Fig.~{\ref{fig:MarginalTrend}c,d}. 
		For Fig.~{\ref{fig:WeakTrend}c,d}, Fig.~{\ref{fig:StrongTrend}d}, and Fig.~{\ref{fig:MarginalTrend}c,d}, the simulation time was set to be $10^6$ ticks. 

\section{Detailed derivation of the master-Boltzmann equation~{(\ref{eq:main_MB_macro})} for the macroscopic description}\label{sec:app:MB_macro}
	In this Appendix, the master-Boltzmann equation~{(\ref{eq:main_MB_macro})} is derived from the BBGKY hierarchal equation~{(\ref{eq:BBGKY_main_2})} for macroscopic variables.
	By applying a mean-field approximation~{(\ref{eq:mlclr_chs_macro3})} to Eq.~{(\ref{eq:BBGKY_main_2})}, 
	we obtain 
	\begin{align}
		\frac{\partial P_t(z_{\CM},p,\Delta p)}{\partial t} =& 
			\left[-c\tanh\frac{\Delta p}{\Delta p^*}\partial_{\CM}+\frac{\sigma^{2}}{2N}\partial_{\CM}^2\right] P_t(z_{\CM},p,\Delta p) + \frac{N^2\sigma^{2}}{2}\int dLdL'\rho_L\rho_{L'}\times\notag\\
		&\bigg\{\int d\Dp'|\tpartial_{rr'}|\phi^L\left(r+\frac{L}{2}-z_{\CM}\right)\phi^{L'}\left(r'-\frac{L'}{2}-z_{\CM}\right)\bigg|_{r=r'=p}P_t\left(z_{\CM}+\frac{L-L'}{2N},p-\Dp,\Delta p'\right)\notag\\
		&- P(z_{\CM},p,\Delta p)\int dr|\tpartial_{rr'}|\phi^L(r)\phi^{L'}(r')\big|_{r=r'+(L+L')/2}\bigg\}.
	\end{align}
	Using the NLO solution~{(\ref{eq:app:subleading_order_FBE})}, 
	we deduce a closed master-Boltzmann equation for the macroscopic dynamics~{(\ref{eq:main_MB_macro})}. 
	Equation~{(\ref{eq:app:MasterBoltzmann_der})} can be rewritten into the standard form of the master equation~\cite{GardinerB},
	\begin{align}
		\frac{\partial P_t(\bm{Z})}{\partial t} = \left[- c\tanh \frac{\Delta p}{\Delta p^*} \partial_{\CM}+\frac{\sigma^{2}}{2N}\partial_{\CM}^2\right] P_t(\bm{Z})
		+ \int d\bm{X} \left\{W(\bm{Z}|\bm{X})P_t(\bm{X})-W(\bm{X}|\bm{Z})P_t(\bm{Z})\right\}\label{eq:app:MasterBoltzmann_der2}
	\end{align}
	with macroscopic-variable vectors $\bm{Z}\equiv (z_{\CM},p,\Delta p)$, $\bm{X}\equiv (z'_{\CM},p',\Delta p') \equiv (z_{\CM}-y,p-\Delta p,\Delta p')$ and volume element $d\bm{X}\equiv dz'_{\CM}dp'd\Delta p'$. 
	Here, the transition rate $W(\bm{Z}|\bm{X})$ is specified by
	\begin{equation}
		W(\bm{Z}|\bm{X}) \equiv \frac{1}{\tau^*}\mathcal{N}\left(p-z_{\CM};\frac{L^{*2}_{\rho}}{4N}\right) w_N(z_{\CM}-z_{\CM}')\delta (\Delta p-p+p').
	\end{equation}
	Because the master equations corresponds one-to-one with SDEs~\cite{BreuerB}, 
	a set of SDEs
	\begin{subequations}\label{eq:set:SDE_macro}
		\begin{align}
			\frac{d\hz_{\CM}}{dt}	&= c\tanh \frac{\hDp}{\Delta p^*} + \frac{\sigma}{\sqrt{N}}\hxi^{\mrG} + (\hz_{\CM}^{\pst}-\hz_{\CM})\hxi^{\mrP}_{\tau^*}\\
			\frac{d\hp}{dt}			&= (\hp^{\pst}-\hp)\hxi^{\mrP}_{\tau^*} \\
			\frac{d\hDp}{dt}		&= (\hDp^{\pst}- \hDp)\hxi^{\mrP}_{\tau^*}
		\end{align}
	\end{subequations}
	is derived from the master-Boltzmann equation~{(\ref{eq:app:MasterBoltzmann_der2})}. 
	Here $\hxi^{\mrG}$ is the white Gaussian noise with unit variance and $\hxi^{\mrP}_{\tau^*}$ is the white Poisson noise with mean interval $\tau^*$. 
	The post-collisional states are given by $(\hz_{\CM}^{\pst},\hp^{\pst},\hDp^{\pst})\equiv (\hz_{\CM} + \hat{\nu}[T]/N, \hz_{\CM}^{\pst} + (L^{*}_{\rho}/2\sqrt{N})\hxi[T], \hp^{\pst} - \hp)$ at the tick time $T$. 
	$\hat{\nu}[T]$ is a discrete-time white noise term obeying $P(\nu)=\tilde{w}(\nu)$ with an $N$-independent distribution $\tilde{w}(\nu)=w_N(\nu/N)/N$ and $\hxi[T]$ is a discrete-time white Gaussian noise with unit variance. 

\section{Detailed derivation of the financial Langevin equation~{(\ref{eq:Financial_Brownian_Motion})}}\label{sec:app:F_Langevin_der}
	Here we simplify the three SDEs~{(\ref{eq:set:SDE_macro})} for $(\hz_{\CM},\hp,\hDp)$ in continuous time $t$
	into a single SDE~{(\ref{eq:Financial_Brownian_Motion})} for price movement $\hDp$ in discrete time. 
	According to the set of SDEs~{(\ref{eq:set:SDE_macro})}, $\hz_{\CM}$ exhibits random walks with constant drift in the absence of transaction, 
	without updates for $\hp$ and $\hDp$. 
	The movement of the CM between transactions (i.e., the time interval $[\hat{t}[T]+\ve,\hat{t}[T+1]-\ve]$ with infinitesimal positive number $\ve$) is then basically determined by the time interval $\hat{\tau}[T]\equiv \hat{t}[T+1]-\hat{t}[T]$ as 
	\begin{equation}
		\hz_{\CM}(\hat{t}[T+1]-\ve) - \hz_{\CM}(\hat{t}[T]+\ve) = c\hat{\tau}[T]\tanh \frac{\hDp}{\Delta p^*} + \sqrt{\frac{\sigma^2 \hat{\tau}[T]}{N}}\hat{\mu}[T] + O(\ve)
	\end{equation}
	with Gaussian random noise $\hat{\mu}[T]$ with unit variance. 
	Within the mean-field approximation, $\hat{\tau}[T]$ is an exponential random number with mean interval $\tau^*$. 
	At the instance of the transaction at time $\hat{t}[T]$, there is a jump originating from the Poisson noise term $(\hz^{\pst}_{\CM}-\hz_{\CM})\hxi^{\mrP}_{\tau^*}$,
	\begin{equation}
		\hz_{\CM}(\hat{t}[T+1]+\ve) - \hz_{\CM}(\hat{t}[T+1]-\ve) = \frac{1}{N}\hat{\nu}[T]+O(\ve)
	\end{equation}
	with random number $\hat{\nu}[T]$ obeying a probability distribution $P(\nu)=\tl{w}(\nu)$ with an $N$-independent $\tl{w}(\nu)\equiv w(\nu/N)/N$. 
	In summary, we obtain the following stochastic dynamics in tick time: 
	\begin{subequations}\label{eq:app:financialLangevin_der_trans}
	\begin{align}
		\hz_{\CM}[T+1] &= \hz_{\CM}[T] + c\hat{\tau}[T]\tanh \frac{\hDp[T]}{\Delta p^*} + \sqrt{\frac{\sigma^{2}\hat{\tau}[T]}{N}}\hat{\mu}[T] + \frac{1}{N}\hat{\nu}[T]\\
		\hp [T+1] &= \hz_{\CM}[T+1] + \sqrt{\frac{L^{*2}_{\rho}}{4N}}\hxi[T]\\
		\hDp [T+1] &= \hp[T+1] - \hp[T]
	\end{align}
	\end{subequations}
	with Gaussian random number $\hxi[T]$ with unit variance. 
	To be precise, $\hz_{\CM}[T]\equiv \lim_{\ve \to +0}\hz_{\CM}(\hat{t}[T]+\ve)$, $\hp[T]\equiv \lim_{\ve \to +0}\hp(\hat{t}[T]+\ve)$, and $\hDp[T]\equiv \lim_{\ve \to +0}\hDp(\hat{t}[T]+\ve)$. 
	By eliminating the two variables $(\hz_{\CM},\hp)$ from Eq.~{(\ref{eq:app:financialLangevin_der_trans})},
	we obtain Eq.~{(\ref{eq:Financial_Brownian_Motion})} as a single stochastic difference equation in tick time. 

	\paragraph*{Case 1: Single spread.}
		For the $\delta$-distributed spread $\rho_L=\delta(L-L^*)$,
		we obtain
		\begin{equation}
			\tl{w}(y) = \delta(y),\>\>\>
			\tau^* = \frac{L^{*2}}{2N\sigma^{2}}, \>\>\>
			L^{*2}_{\rho}=L^{*2},
		\end{equation}
		which implies the absence of $\hat{\nu}(T)$ is absent for the $\delta$-distributed spread. 
		This is natural because the CM is conserved during transaction for this special case. 

	\paragraph*{Case 2: $\gamma$-distribution.}
		For the $\gamma$-distributed spread $\rho_L=L^3e^{-L/L^*}/6L^{*4}$ , 
		we obtain
		\begin{equation}
			\tl{w}(y) = \frac{L^*+2|y|}{2L^{*2}}e^{-2|y|/L^*},\>\>\>
			\tau^* = \frac{3L^{*2}}{N\sigma^{2}}, \>\>\>
			L^{*2}_{\rho}=6L^{*2}.
		\end{equation}

\section{Derivation of the diffusion equation~{(\ref{eq:diffusion_WT})} for weak trend-following case}\label{sec:app:Diffusion_WT}
		In this Appendix, we derive the diffusion equation~{(\ref{eq:diffusion_WT})} for weak trend-following case $\tl{c}\ll 1$. 
		By ignoring the trend-following term, we integrate Eq.~{(\ref{eq:LinearBL_macro})} over $p$ and $\Dp$ to obtain
		\begin{align}
			\frac{\partial P_t(z_{\CM})}{\partial t} = \frac{\sigma^2}{2N}\frac{\partial^2 P_t(z_{\CM})}{\partial z_{\CM}^2} + \frac{1}{\tau^*}\int_{-\infty}^\infty dy w_N(y)[P_t(z_{\CM}-y)-P_t(z_{\CM})].
		\end{align}
		Given that $w_N(y)$ satisfies the scaling of the system size expansion~\cite{VanKampen}
		\begin{equation}
			\tilde{w}(y) \equiv \frac{1}{N}w_N\left(\frac{y}{N}\right)
		\end{equation}
		with an $N$-independent non-negative function $\tilde{w}(y)$, 
		we apply the Kramers-Moyal expansion~\cite{VanKampen,GardinerB}
		\begin{equation}
			\frac{1}{\tau^*}\int_{-\infty}^\infty dy w_N(y)[P_t(z_{\CM}-y)-P_t(z_{\CM})] = \frac{2\sigma^2}{L_{\rho}^{*2}}\sum_{k=1}^\infty \frac{(-1)^k}{N^{k-1}}\frac{\alpha_k}{k!} \frac{\partial^k P_t(z_{\CM})}{\partial z_{\CM}^k}
		\end{equation}
		with $N$-independent Kramers-Moyal coefficient $\alpha_k\equiv \int_{-\infty}^\infty dyy^n\tilde{w}(y)$. 
		By taking the series up to the order of $N^{-1}$, we finally obtain the diffusion equation for the CM~{(\ref{eq:diffusion_WT})}. 
		In the case of $\gamma$-distributed spread, we obtain $\alpha_2 = L^{*2}.$

\section{Theoretical probability of the successive same sign for $\hDp$ without trend-following}\label{sec:app:successive_sign_MF}
	Here we study the probability of taking the successive same sign for price movement $\hDp$ in the absence of trend-following within the mean-field approximation~{(\ref{eq:Financial_Brownian_Motion})}. 
	Let us introduce three Gaussian random variable $\hat{x}$, $\hat{y}$, $\hat{z}$ with unit variance and study $\hat w\equiv \hat y - \hat x$ and $\hat u \equiv \hat z - \hat y$. 
	To analyze the probability of the successive same sign for the mean-field model~{(\ref{eq:Financial_Brownian_Motion})}, it is sufficient to study the probability of taking the same sign for $\hat w$ and $\hat u$ as 
	\begin{equation}
		P(\hat w>0 \cap \hat u>0) + P(\hat w<0 \cap \hat u<0) = 2P(\hat w>0 \cap \hat u>0) = \int_{0}^{\infty}dwdu P(w,u).
	\end{equation}
	Here we obtain
	\begin{align}
		&P(\hat w>0 \cap \hat u>0) 	= \int_{0}^{\infty}dwdu \int_{-\infty}^\infty dxdydz\delta (w-y+x)\delta (u-z+y)\frac{e^{-(x^2+y^2+z^2)/2}}{(2\pi)^{3/2}} \notag \\
									&= \int_{0}^{\infty}dwdu \int_{-\infty}^\infty \frac{dz}{(2\pi)^{3/2}}\exp\left[-\frac{1}{2}(z-u-w)^2-\frac{1}{2}(z-u)^2-\frac{1}{2}z^2\right] = \frac{1}{6}.
	\end{align}
	We thus conclude that the probability of the successive same sign for the mean-field model~{(\ref{eq:Financial_Brownian_Motion})} is given by $1/3$.

\section{Consistency of the solution~{(\ref{eq:app:subleading_order_FBE})} with the financial BBGKY hierarchy~{(\ref{eq:BBGKY2_app})}}\label{sec:app:revisited_consistency}
	We have obtained the NLO solution~{(\ref{eq:app:subleading_order_FBE})} to the financial Boltzmann equation~{(\ref{eq:BoltzmannEq})}. 
	The financial Boltzmann equation~{(\ref{eq:BoltzmannEq})} is derived from the financial BBGKY hierarchal equation~{(\ref{eq:BBGKY_main})} with three-body correlation terms assumed irrelevant in Appendix.~\ref{sec:app:FBE_der}. 
	Here we check that the consistency between the NLO solution~{(\ref{eq:app:subleading_order_FBE})} and this assumption directly. 
	Let us introduce the three-body distribution function $\phi^{LL'L''}_T(r,r',r'')$ for the relative prices as
	\begin{equation}
		\phi^{L_iL_jL_k}_t(r,r',r'') = P_t^{ijk}(r,r',r''). 
	\end{equation}
	On the basis of the assumption~{(\ref{app:eq:assumption_Dr1_ignore})}, the following remaining term $R$ is ignored for $N\to \infty$
	\begin{equation}
		R \equiv N^2\int dr'dL'dL''\rho_{L'}\rho_{L''}\left[|\tpartial_{r'r''}|\phi^{LL'L''}_t\left(r-\frac{s(L'-L'')}{2N},r',r''\right) - |\tpartial_{r'r''}|\phi^{LL'L''}_t\left(r,r',r''\right)\right]\Bigg|_{r'-r''=s(L'+L'')/2}.
	\end{equation}
	This term can be shown irrelevant for $N\to \infty$ under the assumption of molecular chaos for three-body distribution:
	\begin{equation}
		\phi^{LL'L''}_t(r,r',r'') \approx \phi^{L}_t(r)\phi^{L'}_t(r')\phi^{L''}_t(r'').
	\end{equation}
	Indeed, we obtain
	\begin{equation}
		R \approx N^2\sum_{s=\pm 1}\int dL'dL''\rho_{L'}\rho_{L''}\int dr'|\tpartial_{r'r''}|\phi^{L'}_t(r')\phi^{L''}_t(r'')\bigg|_{r'-r''=s(L'+L'')/2}  \left[\phi^{L}_t\left(r-\frac{s(L'-L'')}{2N}\right) - \phi^{L}_t\left(r\right)\right]. 
	\end{equation}
	By substituting the NLO solution~{(\ref{eq:app:subleading_order_FBE})}, 
	we obtain
	\begin{equation}
		\int dr'|\tpartial_{r'r''}|\phi^{L'}_t(r')\phi^{L''}_t(r'')\bigg|_{r'-r''=s(L'+L'')/2} \approx \frac{4L_{\rho}^{*2}}{NL^{\prime 2}L^{\prime \prime 2}}, \>\>\>
		\frac{1}{L^{*2}_\rho}\equiv \int \frac{dL\rho_L}{L^2}.
	\end{equation}
	$R$ is then shown to be the order of $O(N^{-1})$ as 
	\begin{equation}
		R \approx \sum_{k=1}^\infty \frac{(2N)^{-2k+1}}{(2k)!}\int dL'dL''\rho_{L'}\rho_{L''} (L'-L'')^{2k}\frac{\partial^{2k}\phi^L_t(r)}{\partial r^{2k}} = O(N^{-1}),
	\end{equation}
	where we have used the Taylor expansion
	\begin{equation}
		\phi^{L}_t\left(r-\frac{s(L'-L'')}{2N}\right) - \phi^{L}_t\left(r\right) = \sum_{k=1}^\infty \frac{(-s)^k}{k!}\left(\frac{L'-L''}{2N}\right)^k\frac{\partial^k\phi^L_t(r)}{\partial r^k}. 
	\end{equation}
	The ignorance of the three-body correlation term $R$ is thus validated on the basis of the NLO solution~{(\ref{eq:app:subleading_order_FBE})}.

\section{Proof of probability conservation}\label{sec_app:Prob_Conserve}
		We explain the detail of the calculation to derive the conservation of the probability for the reflecting boundary condition. 
		The total probability where the order exists in the range $[-L_{\mrCUT}/2,L_{\mrCUT}/2]$ is given by $\int_{-L_{\mrCUT}/2}^{L_{\mrCUT}/2}dr\phi_t^L(r)$.
		The time-derivative of the total probability obeys the following equation: 
		\begin{align}
			\frac{\partial }{\partial t}\int_{-L_{\mrCUT}/2}^{L_{\mrCUT}/2}dr\phi_t^L(r)
			&= \int_{-L_{\mrCUT}/2}^{L_{\mrCUT}/2}dr\left\{\frac{\sigma^2}{2}\frac{\partial^2 \phi_t^L(r)}{\partial r^2} + N\sum_{s=\pm1}\int_{0}^\infty dL'\rho_{L'}\left[J^{LL'}_{t;s}(r+sL/2)-J^{LL'}_{t;s}(r)\right]\right\}\notag\\
			&= \frac{\sigma^2}{2}\left[\frac{\partial \phi_{t}^{L}(r)}{\partial r}\right]_{-L_{\mrCUT}/2}^{L_{\mrCUT}/2} \!\!+\! N\!\sum_{s=\pm 1} \int_{-L_{\mrCUT}/2}^{L_{\mrCUT}/2}dr\int_{L_{\min}}^{L_{\max}}dL' \rho_{L'}\left[J_{t;s}^{LL'}(r+sL/2)-J_{t;s}^{LL'}(r)\right].
		\end{align}
		Considering the following identity for the integrals
		\begin{align}
			\int_{-L_{\mrCUT}/2}^{L_{\mrCUT}/2}dr J_{t;s=+1}^{LL'}(r+sL/2)
			=&\frac{\sigma^2}{2}\int_{-L_{\mrCUT}/2}^{L_{\mrCUT}/2}dr\left[\phi^{L'}_{t}(r-L'/2)|\partial\phi^L_t(r+L/2)| + |\partial\phi^{L'}_t(r-L'/2)|\phi^L_t(r+L/2)\right]\notag\\
			=&\frac{\sigma^2}{2}\int_{-(L_{\mrCUT}-L')/2}^{(L_{\mrCUT}-L)/2}dr\left[\phi^{L'}_t(r-L'/2)|\partial\phi^L_t(r+L/2)| + |\partial\phi^{L'}_t(r-L'/2)|\phi^L_t(r+L/2)\right]
		\end{align}
		and
		\begin{align}
			\int_{-L_{\mrCUT}/2}^{L_{\mrCUT}/2}dr J_{t;s=+1}^{LL'}(r)
			=&\frac{\sigma^2}{2}\int_{-L_{\mrCUT}/2}^{L_{\mrCUT}/2}dr     \left[\phi^{L'}_t(r-(L+L')/2)|\partial\phi^L_t(r)| + |\partial\phi^{L'}_t(r-(L+L')/2)|\phi^L_t(r)\right]\notag\\
			=&\frac{\sigma^2}{2}\int_{(-L_{\mrCUT}+L+L')/2}^{L_{\mrCUT}/2}dr\left[\phi^{L'}_t(r-(L+L')/2)|\partial\phi^L_t(r)| + |\partial\phi^{L'}_t(r-(L+L')/2)|\phi^L_t(r)\right]\notag\\
			=&\frac{\sigma^2}{2}\int_{-(L_{\mrCUT}-L')/2}^{(L_{\mrCUT}-L)/2}dr\left[\phi^{L'}_t(r-L'/2)|\partial\phi^L_t(r+L/2)| + |\partial\phi^{L'}_t(r-L'/2)|\phi^L_t(r+L/2)\right],
		\end{align}
		we obtain
		\begin{equation}
			\int_{-L_{\mrCUT}/2}^{L_{\mrCUT}/2}dr\left[J_{t;s=+1}^{LL'}(r+L/2)-J_{t;s=+1}^{LL'}(r)\right] = 0.
		\end{equation}
		Here, the assumption $L_{\mrCUT}>L_{\max}$ is used in changing the integral interval. 
		In a parallel calculation, 
		we obtain
		\begin{equation}
			\int_{-L_{\mrCUT}/2}^{L_{\mrCUT}/2}dr\left[J_{t;s=-1}^{LL'}(r-L/2)-J_{t;s=-1}^{LL'}(r)\right] = 0.
		\end{equation}
		These relations imply 
		\begin{equation}
			\frac{\partial }{\partial t}\int_{-L_{\mrCUT}/2}^{L_{\mrCUT}/2}dr\phi^L_t(r) = \frac{\sigma^2}{2}\left[\frac{\partial \phi^L_t(r)}{\partial r}\right]_{-L_{\mrCUT}/2}^{L_{\mrCUT}/2}.
		\end{equation}
		We then show the conservation of the probability~{(\ref{eq:Boundary_Condition})} for the reflecting boundary condition.


\begin{thebibliography}{99}

\bibitem{KuboB}
	R. Kubo, M. Toda, and N. Hashitsume, {\it Statsitical Physics II} (Springer-Verlag, Berlin, 1991), 2nd ed.

\bibitem{Chapman1970}
	S. Chapman and T.G. Cowling, 
	{\it The Mathematical Theory of Non-Uniform Gases} (Cambridge University Press, Cambridge, 1970).
\bibitem{Broeck2004}
	C. Van den Broeck, R. Kawai, and P. Meurs, 
	Phys. Rev. Lett. \tb{93}, 090601 (2004). 
\bibitem{Broeck2006}
	C. Van den Broeck and R. Kawai, 
	Phys. Rev. Lett. \tb{96}, 210601 (2006). 
\bibitem{Brilliantov}
	N.V. Brilliantov and T. P\"oschel, 
	{\it Kinetic Theory of Granular Gases} (Oxford Univ. Press, New York, 2004).
\bibitem{Bertin2006}
	E. Bertin, M. Droz, and G. Gr\'egoire, 
	Phys. Rev. E \tb{74}, 022101 (2006); 
	J. Phys. A \tb{42}, 445001 (2009).
\bibitem{Helbing}
	D. Helbing,
	Rev. Mod. Phys. \tb{73}, 1067 (2001). 
\bibitem{Nishinari}
	A. Schadschneider, D. Chowdhury, and K. Nishinari, 
	{\it Stochastic Transport in Complex Systems: From Molecules to Vehicles} (Elsevier, Amsterdam, 2010).
\bibitem{Prigogine}
	I. Prigogine and R. Herman, 
	{\it Kinetic Theory of Vehicular Traffic} (American Elsevier, New York, 1971). 
\bibitem{Cai2004}
	D. Cai, L. Tao, M. Shelley, and D.W. McLaughlin, Proc. Natl. Acad. Sci. U.S.A. \tb{101}, 7757 (2004).
\bibitem{Buice2013}
	M.A. Buice and C.C. Chow, PLoS Comput. Biol. 9, e1002872 (2013).
\bibitem{Pareschi}
	L. Pareschi and G. Toscani, {\it Interacting Multiagent Systems} (Oxford University Press, Oxford, 2014).

\bibitem{GardinerB}
	C.W. Gardiner, {\it Handbook of Stochastic Methods}, 4th ed. (Springer, Berlin, 2009).
\bibitem{Resibois1977}
	P. R\'esibois and M. de Leener, {\it Classical Kinetic Theory of Fluids} (Wiley, New York, 1977).

\bibitem{McDonald}
	J.-P. Hansen and I. McDonald, {\it Theory of Simple Liquids}, 3rd ed. (Academic Press, Amsterdam, 2006).
\bibitem{VanKampen}
	N.G. van Kampen, 
	{\it Stochastic Processes in Physics and Chemistry}, 3rd ed. (Elsevier, Amsterdam, 2007);
	N.G. van Kampen, Can. J. Phys. \tb{39}, 551 (1961).
\bibitem{Spohn1980}
	H. Spohn, Rev. Mod. Phys. \tb{52}, 569 (1980).

\bibitem{Mantegna1999}
	R.N. Mantegna and H.E. Stanley, {\it Introduction to econophysics: correlations and complexity in finance} (Cambridge University Press, Cambridge, 1999).  
\bibitem{Slanina2014}
	F. Slanina, {\it Essentials of Econophysics Modelling} (Oxford University Press, Oxford, 2014). 

\bibitem{Bachelier1900}
	L. Bachelier, Annales Scientifiques de l'\' Ecole Normale Sup\' erieure \tb{17}, 21 (1900).
\bibitem{Einstein1905}
	A. Einstein, Ann. Phys.-Berlin \tb{322}, 549 (1905).


\bibitem{Mantegna1995}
	R.N. Mantegna and H.E. Stanley, Nature (London) \tb{376}, 46 (1995). 
\bibitem{Lux1996}
	T. Lux, Appl. Financ. Econ. \tb{6}, 463 (1996). 
\bibitem{Plerou1999}
	V. Plerou, P. Gopikrishnan, L.A. Nunes Amaral, M. Meyer, and H.E. Stanley, 
	Phys. Rev. E \tb{60}, 6519 (1999). 
\bibitem{Guillaume1997}
	D.M. Guillaume, M.M. Dacorogna, R.R. Dav\'e, U.A. M\"uller, R.B. Olsen, and O.V. Pictet, Finance Stoch. \tb{1}, 95 (1997). 
\bibitem{Longin1996}
	F.M. Longin, J. Business \tb{69}, 383 (1996).

\bibitem{JDHamilton}
	J.D. Hamilton, {\it Time Series Analysis} (Princeton University Press, Princeton, 1994).
\bibitem{Engle1982}
	R.F. Engle, Econometrica \tb{50}, 987 (1982). 
\bibitem{PUCK2006}
	M. Takayasu, T. Mizuno, and H. Takayasu, Physica A \tb{370}, 91 (2006). 

\bibitem{Kyle1985}
	A.S. Kyle, Econometrica \tb{53}, 1315 (1985). 
\bibitem{Takayasu1992}
	H. Takayasu, H. Miura, T. Hirabayashi, and K. Hamada, Physica A \tb{184}, 127 (1992). 
\bibitem{Bak1997}
	P. Bak, M. Paczuski, and M. Shubik, Physica A \tb{246}, 430 (1997). 
\bibitem{Lux1999}
	T. Lux and M. Marchesi, Nature (London) \tb{397}, 498 (1999). 
\bibitem{SatoTakayasu1998}
	A.-H. Sato and H. Takayasu, Physica A \tb{250}, 231 (1998).
\bibitem{Yamada2007}
	K. Yamada, H. Takayasu, and M. Takayasua, Physica A, \tb{382}, 340 (2007). 
\bibitem{Yamada2009}
	K. Yamada, H. Takayasu, T. Ito, and M. Takayasu, Phys. Rev. E \tb{79}, 051120 (2009).
\bibitem{Yamada2010}
	K. Yamada, H. Takayasu, and M. Takayasu, J. Phys.: Conf. Ser. \tb{221}, 012015 (2010). 

\bibitem{Maslov2000}
	S. Maslov, Physica A \tb{278}, 571 (2000). 
\bibitem{Daniels2003}
	M.G. Daniels, J.D. Farmer, L. Gillemot, G. Iori, and E. Smith, Phys. Rev. Lett. \tb{90}, 108102 (2003).
\bibitem{Smith2003}
	E. Smith, J.D. Farmer, L. Gillemot, and S. Krishnamurthy, Quantitative Finance \tb{3}, 481 (2003). 
\bibitem{Bouchaud2002}
	J.-P. Bouchaud, M. M\'ezard, and M. Potters, Quantitative Finance \tb{2}, 251 (2002). 
\bibitem{Farmer2005}
	J.D. Farmer, P. Patelli, and I. Zovko, Proc. Natl. Acad. Sci. U.S.A. \tb{102}, 2254 (2005).
\bibitem{Toth2011}
 	B. T\'oth, Y. Lemp\'eri\`ere, C. Deremble, J. de Lataillade, J. Kockelkoren, and J.-P. Bouchaud, Phys. Rev. X \tb{1}, 021006 (2011).
 \bibitem{Donier2015}
	J. Donier, J. Bonart, I. Mastromatteo, and J.-P. Bouchaud, Quantitative Finance \tb{15}, 1109 (2015). 
 	
\bibitem{Yura2014}
	Y. Yura, H. Takayasu, D. Sornette, and M. Takayasu, Phys. Rev. Lett. \tb{112}, 098703 (2014); Phys. Rev. E \tb{92}, 042811 (2015). 
\bibitem{Kanazawa2017}
	K. Kanazawa, T. Sueshige, H. Takayasu, and M. Takayasu, to appear in Phys. Rev. Lett. (2018); arXiv:1703.06739. 

\bibitem{Evans2008}
	D.J. Evans and G. Morriss, {\it Statistical Mechanics of Nonequilibrium Liquids}, 2nd ed. (Cambridge University Press, Cambridge, 2008).
	 
\bibitem{Ernst1969}
	M.H. Ernst, J.R. Dorfmann, W.R. Hoegy, and J.M.J. van Leeuwen, Physica \tb{45}, 127 (1969).
\bibitem{Beijeren1979}
	H. van Beijeren and M. H. Ernst, J. Stat. Phys. \tb{21}, 125 (1979).
\bibitem{KanazawaTheses}
	K. Kanazawa, {\it Statistical Mechanics for Athermal Fluctuation: Non-Gaussian Noise in Physics} (Springer, Berlin, 2017). 

\bibitem{Cercignani1994}
	C. Cercignani, R. Illner, and, M. Pulvirenti, {\it The Mathematical Theory of Dilute Gases} (Springer, New-York, 1994). 
	
\bibitem{EBS_Schmit}
	A.B. Schmidt, Ecology of the Modern Institutional Spot FX: The EBS Market in 2011, 
	SSRN 1984070 (2011). 
\bibitem{EBSRule}
	EBS Dealing Rules -- Appendix -- EBS Market (at the time of June 2016).


\bibitem{HakenB}
	H. Haken, {\it Synergetics: Introduction and Advanced Topics} (Springer, Berlin, 2004).
\bibitem{David2003}
	H.A. David and H.N. Nagaraja, {Order Statistics}, 3rd ed. (Wiley, New York, 2003).	
	
	
\bibitem{MTakayasu2002}
	M. Takayasu, H. Takayasu, and M.P. Okazaki, {\it Empirical Science of Financial Fluctuations} (Springer, Tokyo, 2002), p.18-25.
	
\bibitem{Sueshige2018}
	T. Sueshige, K. Kanazawa, H. Takayasu, and M. Takayasu, in preparation.  

\bibitem{TakayasuPRL1997}
	H. Takayasu, A.-H. Sato, and M. Takayasu, Phys. Rev. Lett. \tb{79}, 966 (1997). 


\bibitem{Novikov}
	E.A. Novikov, Sov. Phys. JETP \tb{20}, 1290 (1965); 
	W. Horsthemke and R. Lefever, {\it Noise-Induced Transitions: Theory and Applications in Physics, Chemistry, and Biology} (Springer-Verlag, Berlin, 1984); 
	R.F. Fox, Phys. Rev. A \tb{33}, 467 (1986). 

\bibitem{BreuerB}
	H.P. Breuer and F. Petruccione, {\it The theory of open quantum systems} (Oxford University Press, Oxford, 2002). 

\end{thebibliography}
\end{document}